 \documentclass[preprint2]{aastex}

\usepackage{graphicx}
\usepackage{natbib}
\usepackage{rotating}
\usepackage{rotfloat}
\usepackage[flushleft]{threeparttable}

\shorttitle{Observed CO and dust emission in post-AGB stars}
\shortauthors{He, Szczerba, Hasegawa \& Schmidt}

\begin{document}


\title{Observed relationship
    between CO\,2-1 and dust emission during post-AGB phase. \thanks{The observation
    project was funded by Academia Sinica, Institute of Astronomy and
    Astrophysics, Taiwan}}


\author{J. H. He\altaffilmark{1}, 
        R. Szczerba\altaffilmark{2},
        T. I. Hasegawa\altaffilmark{3}, and
        M. R. Schmidt\altaffilmark{2}}
\email{jinhuahe@ynao.ac.cn}

\altaffiltext{1}{Key Laboratory for the Structure and Evolution of Celestial Objects, 
  Yunnan Observatories, Chinese Academy of Sciences; 
  P.O. Box 110, Kunming, Yunnan Province, PR China}

\altaffiltext{2}{N. Copernicus Astronomical Center, Rabianska 8,  87-100 Torun, POLAND}

\altaffiltext{3}{Academia Sinica, Institute of Astronomy and Astrophysics,
  P.O. Box 23-141, Taipei 10617}




\begin{abstract}

A CO\,2-1 line survey is performed toward a sample of 58 high
Galactic latitude post-AGB (pAGB) stars. To complement the observations, a 
compilation of
literature CO\,2-1 line data of known pAGB stars is done. After combining the datasets, CO\,2-1 line data are available for 133 pAGB stars (about 34 per cent of known pAGB stars) among which 44 are detections.
The CO line strengths are compared with infrared dust emission for these pAGB stars by defining 
a ratio between the integrated CO\,2-1 line flux and {\it IRAS} $25\micron$ flux
density (CO-IR ratio). The relationship between the CO-IR ratio and the {\it IRAS} color C23
(defined with the 25 and 60\,$\micron$ flux densities) 
is called here the CO-IR diagram. The pAGB objects are found to be located between AGB stars and planetary nebulae (PNe), and segregate into 
three distinctive groups (I, II and III) on the CO-IR diagram. 
By analyzing their various properties such as chemical types, spectral types, 
binarity, circumstellar envelope expansion velocities, and pAGB sub-types on the CO-IR diagram, it is argued that the
group-I objects are mainly intermediate mass C-rich pAGB stars in early pAGB stage (almost all of 
the considered carbon rich `$21\micron$' stars belong to this group); 
the group-II objects are massive or intermediate mass pAGB stars which already follow the profound trend of PNe; and 
the group-III objects are mainly low mass binary pAGB stars with very weak CO\,2-1 line emission (almost all of the considered RV\,Tau variables belong to this group). 
The CO-IR diagram is proven to be a powerful tool to investigate the co-evolution of circumstellar gas and dust 
during the short
pAGB stage of stellar evolution. 

\end{abstract}


\keywords{
stars: AGB and post-AGB --- 
(stars:) circumstellar matter --- 
stars: mass-loss --- 
(ISM:) planetary nebulae: general --- 
infrared: stars ---
radio lines: stars
}



\section{INTRODUCTION}
\label{intro}

After the super-wind has ceased, the evolution 
of the remnant circumstellar envelope (CSE) around a single post-Asymptotic Giant Branch
(post-AGB, or pAGB) star
is dominated first by its expansion, 
and later by
photochemical processes when the central star temperature quickly rises up. 
The expansion of the remnant CSE produces a 
detached circumstellar shell and the object is usually characterized by a double peak spectral 
energy distribution (SED) due to the lack of hot
dust~\citep{kowk93}. CO molecules in the CSE are protected from photodissociation 
mainly by self-shielding and line shielding by atomic and molecular H$_2$~\citep[see e.g., ][]{mamo88}.

There have been few dedicated works on the relationship between dust and CO in pAGB
stars.
\citet{1991AA...245..499A} had noticed that some pAGB stars of RV Tauri
type are usually very deficient in CO line emission and proposed
peculiar elemental abundances that prevent the formation of CO
molecules as
the possible interpretation of the phenomenon.
\citet{1992AA...257..701B} observed $^{12}$CO and $^{13}$CO\,1-0, 2-1 lines toward several young pAGB stars. Combined
with literature data, their results revealed a correlation
between CO\,1-0 integrated intensity and {\it IRAS} $60\,\micron$ ($F_{60}$) flux density. However,
there are some exceptional pAGB stars that show either too
strong or too weak CO\,1-0 line intensities relative to $F_{60}$ flux densities.
They also noticed the trend that the CO line is
stronger in some AGB carbon stars than in 
investigated pAGB stars. They compared the CO line of
the young pAGB stars with other pAGB stars from literature and tentatively concluded
that younger pAGB stars might have stronger CO lines. With more observational 
CO data accumulated to date, it is possible to extend such studies in a more systematic way.

Here we present a new observational study of relationship between
dust and CO in the evolving circumstellar envelopes of pAGB objects using 
the compilation of such objects from the Torun Catalog of pAGB
stars \citep{szcz07, szcz11}. After 
description of our observations and data reduction in 
Sect.~\ref{obs}, the results are presented in Sect.~\ref{results}. 
To augment the data set for analysis, we have also performed as complete as possible compilation of literature CO\,2-1 data
for all known pAGB stars in Sect.~\ref{litdata} (with details presented in the Appendix). 
Then, the observed relation between the integrated CO\,2-1 line fluxes and IR dust
emission flux densities is investigated and compared to that of AGB stars and PNe in Sect.~\ref{analysis}. 
To better understand the observed relationship between CO and dust emission, some other properties 
of the pAGB stars, such as CSE chemical types, spectral types, binarity, spectral energy distribution types, 
CSE expansion velocities and pAGB sub-types, are discussed in Sect.~\ref{discuss}.
At the end, the main points of this work are summarized in Sect.~\ref{summary}.

\section{SAMPLE SELECTION, NEW OBSERVATION AND DATA REDUCTION}
\label{obs}

In order to explore the relationship between dust and CO in the pAGB 
circumstellar envelopes with smaller contamination from interstellar CO
emission,
a sample of 58 high Galactic latitude pAGB stars (with $|b|\ge 15^{\circ}$) that are accessible from the Arizona Radio 
Observatory 10\,m submillimetre telescope (AROSMT, for objects with
a declination $\ge -38^{\circ}$) were selected from
the Torun Catalog of pAGB stars \citep{szcz07}. However, the discussions in this paper will be based on the second version of the catalogue
\citep{szcz11}\footnote{The Torun
pAGB star catalogue is publicly available at:
http://www.ncac.torun.pl/postagb2}. We note that, although great efforts have been made in building the catalog, 
the evolutionary status of some objects (such as RV Tau type and R CrB stars) is still uncertain in the catalog. 
Almost all of these 
pAGB objects have optical
and/or infrared photometry and/or spectroscopy data, which provides a reliable 
basis for the analysis in this work.

The ALMA band-6 sideband-separating receiver on the AROSMT telescope
was used for our $^{12}$CO\,J=2-1 line survey from 2007 November to 2008 April. Sky subtraction was made 
by beam-switch mode with a 2
arcmin throw at 1\,Hz wobbling in the azimuth direction. Two filter banks (FFBs,
1\,GHz width, 1024 channels) were used for the two linear
polarizations. 
The beam width was about $32\arcsec$ in this line.
A nominal factor of 44.4 Jy/K can be used to derive line peak fluxes. 

The GILDAS/CLASS package was used to reduce and analyze the data. The two
polarizations were combined to improve the S/N. An on-plus-off
integration time of $20$\,min per object yielded an typical root mean square (RMS) of about 15\,mK at a spectral
resolution of 1\,MHz in the
polarization-averaged spectral baseline. 
A linear baseline was removed from each spectrum. 

\section{RESULTS}
\label{results}

The CO\,2-1 line were detected toward only six sources among the 58
observed in this survey. 
The baseline RMS noise of all observed data and the line parameters for
the detected sources are given in Tables\,\ref{nondet} and \ref{detect},
respectively. A main beam efficiency of 0.75 is used to convert the antenna temperature 
$T_{\rm A}^*$  into main beam temperature  $T_{\rm mb}$. Also given in Table~\ref{detect} are integrated line flux, $F_{\rm int}$, 
and its ratio to the {\it IRAS} 25\,$\mu$m flux density 
\begin{equation}
R_{\rm CO25}=F_{\rm int}/F_{25}. 
\label{eq1}
\end{equation} 
Only good quality {\it IRAS} flux densities (with $Q=2$ or $3$) are used 
in this work. The {\it IRAS} 25\,$\mu$m flux density is considered because it is the representative  
wavelength of the cool dust emission from the circumstellar envelopes of all evolved stars considered 
here: AGB stars, pAGB stars, and PNe. In addition, 25\,$\mu$m flux densities have smaller interstellar 
contamination than 60\,$\mu$m flux densities for low galactic latitude objects which will be 
compiled from literature in the next section. 

Among the six detected sources, there is only one new detection of CO\,2-1 line in
IRAS\,07430+1115 (only its CO\,1-0 line was detected before). The CO\,2-1 line was observed
but not detected toward this object by \citet{2005ApJ...624..331H}, which was
perhaps due
to the wrong source position used by them (the position they observed
was away from the {\it IRAS} position by about $30\arcsec$, comparable to the beam size). 
The very narrow CO 2-1 line toward IRAS\,19437-1104 is also new, but it is possibly a contamination from a high Galactic latitude 
molecular cloud, because: 
1) so narrow CO line ($V_{\rm exp}=0.7$\,km\,s$^{-1}$) is very rare among pAGB stars, and 
2) the LSR velocity of the narrow CO line (at $V_{\rm LSR}=3.58$\,km\,s$^{-1}$) is the
same as an interstellar cloud toward the same direction as measured by
\citet{dame01}. Therefore, this object will not enter our discussions hereafter. 
%
\begin{deluxetable}{l@{ }r@{ }r@{}c@{ }l@{ }r@{ }r@{}c}
\tabletypesize{\scriptsize}
\tablewidth{0pc} 
\tablecaption{The 58 observed high Galactic latitude post-AGB stars.
\label{nondet}}
\tablehead{ 
\colhead{Name\tablenotemark{a}} &
\colhead{RA(2000)}              &
\colhead{DEC(2000)}             &
\colhead{RMS($T_{\rm mb}$)}     &
\colhead{Name\tablenotemark{a}} &
\colhead{RA(2000)}              &
\colhead{DEC(2000)}             &
\colhead{RMS($T_{\rm mb}$)}     \\
\colhead{}                      &
\colhead{hh:mm:ss.sss}          &
\colhead{dd:mm:ss.sss}          &
\colhead{mK}                    &
\colhead{}                      &
\colhead{hh:mm:ss.sss}          &
\colhead{dd:mm:ss.sss}          &
\colhead{mK}                    }
\startdata 
01005$+$7910          &01:04:45.514  &$ $79:26:46.312 &12   &NGC\,6254         &16:57:11.743  &$-$04:03:59.670 & 9 \\
LX\,And               &02:19:44.091  &$ $40:27:22.208 &22   &F17015$+$0503     &17:04:00.085  &$ $04:59:00.816 & 8 \\
05208$-$2035          &05:22:59.424  &$-$20:32:53.116 &11   &PG\,1704+222      &17:06:46.171  &$ $22:05:52.090 & 9 \\
RV\,Col               &05:35:44.206  &$-$30:49:35.490 &13   &UZ\,Oph           &17:21:59.260  &$ $06:54:42.217 &10 \\
06338$+$5333          &06:37:52.422  &$ $53:31:02.017 & 9   &V453\,Oph         &17:26:49.115  &$-$02:23:36.146 &12 \\
07430$+$1115          &07:45:51.390  &$ $11:08:19.626 & 8   &17436$+$5003      &17:44:55.470  &$ $50:02:39.465 &14 \\
09371$+$1212          &09:39:53.963  &$ $11:58:52.604 &13   &F17495$+$0757     &17:52:01.248  &$ $07:56:29.194 &10 \\
10158$-$2844          &10:18:07.590  &$-$28:59:30.840 &22   &17534$+$2603      &17:55:25.182  &$ $26:02:59.964 &16 \\
DN\,Leo               &10:38:55.230  &$ $10:03:48.500 &18   &18062$+$2410      &18:08:20.083  &$ $24:10:43.323 &11 \\
11157$+$5257          &11:18:33.583  &$ $52:40:54.606 & 8   &18095$+$2704      &18:11:30.670  &$ $27:05:15.547 &15 \\
11472$-$0800          &11:49:48.040  &$-$08:17:20.420 &17   &V443\,Lyr         &18:29:31.508  &$ $33:58:40.897 &12 \\
BD$+$13 2491          &12:07:10.820  &$ $12:59:07.750 &15   &V534\,Lyr         &18:37:58.775  &$ $37:26:05.676 &15 \\
CD$-$31 9638          &12:20:44.940  &$-$32:33:26.190 &20   &19132$-$3336      &19:16:32.759  &$-$33:31:20.434 &30 \\
12538$-$2611          &12:56:30.140  &$-$26:27:36.950 &19   &F19179$-$3336     &19:21:07.327  &$-$31:23:50.323 &22 \\
BPS\,CS\,22877$-$0023 &13:25:39.470  &$-$08:49:19.060 &18   &19437$-$1104      &19:46:30.528  &$-$10:56:54.934 &13 \\
13467$-$0141          &13:49:17.120  &$-$01:55:44.820 &16   &19500$-$1709      &19:52:52.692  &$-$17:01:50.358 &22 \\
F15160$+$0215         &15:18:36.150  &$ $02:04:16.280 &18   &19590$-$1249      &20:01:49.831  &$-$12:41:17.779 &14 \\
F15240$+$1452         &15:26:20.820  &$ $14:41:36.340 &16   &20023$-$1144      &20:05:05.413  &$-$11:35:57.915 &13 \\
BT\,Lib               &15:31:15.890  &$-$23:21:36.220 &20   &RX\,Cap           &20:14:55.215  &$-$12:56:34.702 &25 \\
15465$+$2818          &15:48:34.407  &$ $28:09:24.264 &13   &V590\,Aql         &20:17:08.554  &$-$04:03:06.991 &22 \\
BD$+$33 2642          &15:51:59.882  &$ $32:56:54.373 &15   &FQ\,Aqr           &20:51:21.341  &$ $02:18:46.360 &18 \\
BD$+$26 2763          &15:58:58.275  &$ $26:08:04.646 &15   &20547$+$0247      &20:57:16.284  &$ $02:58:44.554 &30 \\
LSE\,29               &16:09:24.560  &$-$27:13:38.160 &22   &PHL\,1580         &21:30:25.244  &$-$19:22:34.374 &32 \\
V2205\,OPH            &16:28:35.373  &$-$09:19:31.814 &20   &NGC\,7089         &21:33:32.413  &$ $00:49:05.793 &11 \\
F16277$-$0724         &16:30:30.020  &$-$07:30:52.050 &18   &PHL\,174          &21:50:48.657  &$-$19:42:00.270 &20 \\
NGC\,6205             &16:41:33.680  &$ $36:26:07.706 &14   &22006$-$1652      &22:03:19.699  &$-$16:37:35.245 &34 \\
V652\,Her             &16:48:04.693  &$ $13:15:42.282 &16   &22327$-$1731      &22:35:27.528  &$-$17:15:26.956 &32 \\
TT\,Oph               &16:49:35.888  &$ $03:37:54.242 & 9   &BD$+$39 4926      &22:46:11.231  &$ $40:06:26.294 &15 \\
LS\,IV\,$-$04\,1      &16:56:27.731  &$-$04:47:23.708 &10   &DS\,Aqr           &22:53:17.037  &$-$18:35:30.984 &32 \\\hline 
\enddata 
\tablecomments{Positions and baseline RMS noise level of the observed CO\,2-1 spectra.}
\tablenotetext{a}{IRAS or FIRAS prefix has been omitted}
\end{deluxetable} 
%
\begin{deluxetable}{l@{ }l@{ }l@{ }l@{}c@{}r@{}r@{}r@{}r@{}c@{}r}
\tabletypesize{\scriptsize}
\tablewidth{0pc} 
\tablecaption{CO line parameters of the six detected high Galactic latitude post-AGB stars.
\label{detect}}
\tablehead{ 
\colhead{{IRAS}                         } &
\colhead{{Other}                        } &
\colhead{$l$                            } &
\colhead{$b$                            } &
\colhead{Sp.Type                        } &
\colhead{$T_{\rm mb}$ \tablenotemark{a} } & 
\colhead{$V_{\rm exp}$\tablenotemark{a} } & 
\colhead{$I_{\rm int}$\tablenotemark{a} } &
\colhead{$F_{\rm int}$\tablenotemark{a} } &
\colhead{$V_{\rm LSR}$                  } &
\colhead{$R_{\rm CO25}$\tablenotemark{b}} \\
\colhead{                               } &
\colhead{Name                           } &
\colhead{deg                            } &
\colhead{deg                            } &
\colhead{                               } &
\colhead{mK                             } &
\colhead{km\,s$^{-1}$                   } &
\colhead{K.km\,s$^{-1}$                 } &
\colhead{Jy\,MHz                        } &
\colhead{km\,s$^{-1}$                   } &
\colhead{MHz                            } }
\startdata 
07430$+$1115 &             &208.9312 &$+$17.0670 &G5Ia  & 74( 8) & 6.5(0.6) &1.03(0.05) & 35.0(1.7) &$ $15.7 & 1.17( 0.24) \\ 
09371$+$1212 &Frosty Leo   &221.8895 &$+$42.7269 &K7II  &166(13) &22.8(0.8) &8.04(0.14) &273.7(4.8) &$-$15.6 &59.62(11.97) \\ 
17436$+$5003 &V814 Her     &077.1331 &$+$30.8696 &F3Ib  &256(14) & 8.6(0.4) &4.71(0.10) &160.3(3.4) &$-$35.3 & 0.87( 0.18) \\ 
17534$+$2603 &89 Her       &051.4341 &$+$23.1876 &F2Ibe &233(16) & 3.5(0.3) &1.75(0.07) & 59.6(2.4) & $-$8.4 & 1.09( 0.22) \\ 
19437$-$1104 &DY Aql       &029.1826 &$-$17.1308 &G5    &122(13) & 0.7(1.3) &0.17(0.03) &           &$ $ 3.6 &             \\ 
19500$-$1709 &V5112 Sgr    &023.9837 &$-$21.0361 &F2/   &259(22) &10.4(1.2) &5.77(0.24) &196.4(8.2) &$ $24.4 & 1.19( 0.24) \\ 
             &             &         &           &F3Iab &        &          &           &           &        &             \\                                                                                                                                        
\enddata                                                                  
\tablenotetext{a}{The values in the parentheses are $1\,\sigma$ noise level.}
\tablenotetext{b}{$R_{\rm CO25}$ is defined by Eq.~\ref{eq1} in Sect.~\ref{results}.} 
\end{deluxetable} 

\begin{figure}
\epsscale{1.0}
\plotone{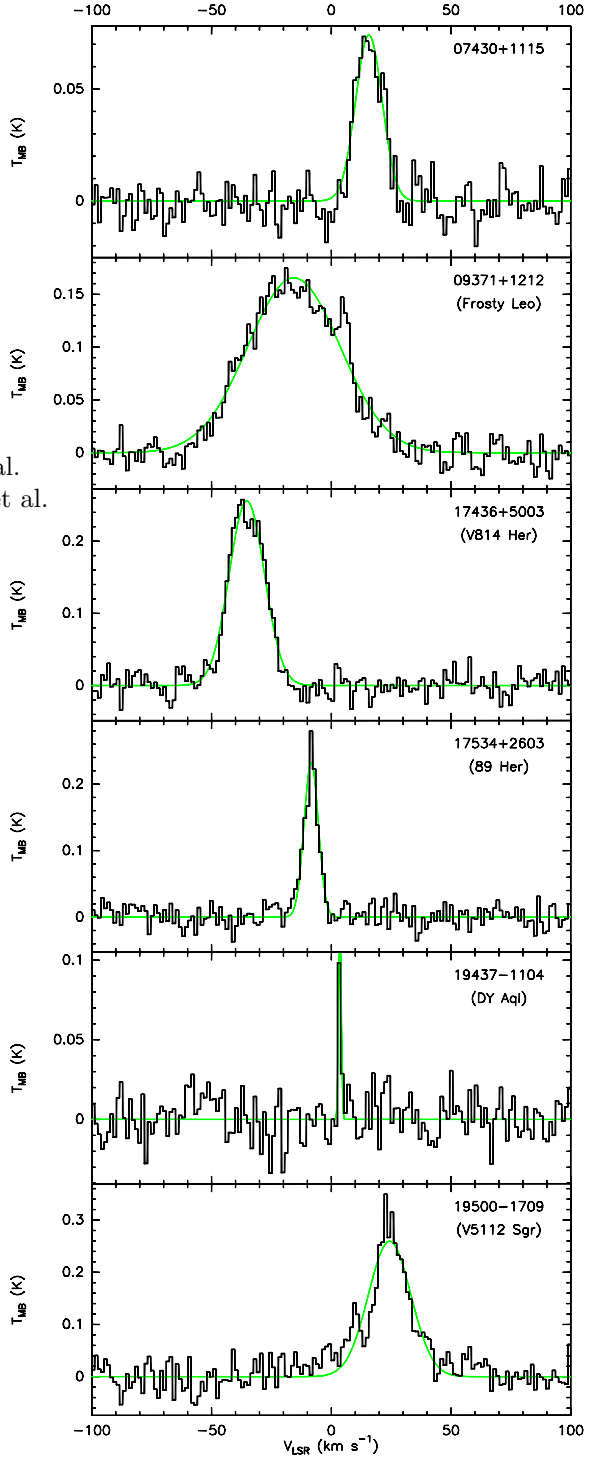}
\caption{The observed CO\,2-1 line spectra of the detected high Galactic latitude
post-AGB sample stars. Overplotted smooth curves are Gaussian fittings. 
The very narrow line of IRAS\,19437-1104 is possibly a contamination 
from an interstellar cloud (see details in Sect.~\ref{results}).}
\label{hglsgspec}
\end{figure}

The resulting CO line profiles of the six detected
sources are shown in Fig.~\ref{hglsgspec}. Here we give 
notes for some interesting
spectral line features in these profiles. 
\begin{enumerate}
\item{A common feature of these
CO\,2-1 lines is the presence of high velocity line wings. Four of
the five detected circumstellar CO lines show evidence of such line wings
(with the only exception being IRAS\,07430+1115, see 
Fig.~\ref{hglsgspec}). The Gaussian line wings are usually produced 
by fast (bipolar) winds from the central pAGB star.} 
\item{The CO\,2-1 line of IRAS\,19500-1709 is a composite profile with a strong narrow feature
superimposed at the center of a broader flat pedestal. Similar composite line
profiles had been found in some other AGB or pAGB stars, such as
RS\,Cnc \citep{gera03,libe10}, EP\,Aquarii
\citep{wint07}, and more such sources in \citet{1998ApJS..117..209K} and \citet{wint03}. The
interpretation of such profile is still controversial, however. }
\item{The sharp peak CO line profile of IRAS\,17534
$+$2603 had been nicely
interpreted by \citet{buja07} on the basis of their CO line
interferometry data with a bipolar hourglass
shaped outflow model.}
\item{ A small narrow emission like feature can be recognized around $V_{\rm LSR}=5$\,km\,s$^{-1}$ 
on the top of the broad line profile of Frosty Leo. A plateau was seen near the same velocity in a better quality line profile 
obtained with IRAM (with smaller beam) by \citet{buja01}. The change of profile shape of this feature 
with different beam sizes indicates that it should originate from the circumstellar envelope instead of from 
interstellar clouds. }
\item{Absorption like features are present in the CO\,2-1 line
  profile of two objects. An absorption like feature is seen in the blue line wing of
  IRAS\,19500-1709 at $V_{\rm LSR}=14.4$\,km\,s$^{-1}$, or at a
  Doppler shift of $\approx -10$\,km\,s$^{-1}$ with respect to the
  systemic velocity of 24.4\,km\,s$^{-1}$. The same object was observed with 
  IRAM 30m by \citet{buja01} and clearer absorption like features 
  had been seen around similar velocities on top of broad line wings in their better 
  profile data. As argued
  by \citet{1992AA...257..701B}, it could be the signpost of still deeply
embedded small-size fast winds. Similar absorption like features
also appear in
the blue line wing of IRAS\,17436+5003 at Doppler shifts of
-33, -23 and -14\,km\,s$^{-1}$ with respect to the systemic velocity of -35.3\,km\,s$^{-1}$ in
Fig.~\ref{hglsgspec}. The bluest of the three absorption features even
shows negative strength. Although the three
absorption features in this object are weak, their appearance can be nicely confirmed in this object by
comparing our data with previous
observation of the same line with IRAM 30m telescope by
\citet{1992AA...257..701B} more than 16 years ago. These absorption features were also briefly mentioned by
\citet{cast04} with their more recent observations. Part of these weak but stable
absorption features could be similarly interpreted by 
embedded small-size fast winds, as for the case of
IRAS\,19500-1709. The absorption with negative flux could also be
interpreted similarly given that continuum emission is strong enough in the compact
outflow regions. However, it is hard for this scenario to explain
all the three absorption features in this line profile with simple
CSE structures. Similar absorption line features have
been found in the CO line profiles of other well-known AGB or
pAGB stars, e.g., CRL\,2688 \citep{kawa87,cox00,buja01}, 
CRL\,618 \citep{1996ApJ...467..341H,sanc04}, and some other sources from \citet{cast10}.}
\end{enumerate}

\section{LITERATURE DATA}
\label{litdata}
 
Because the detection rate of CO\,2-1 line in our high Galactic latitude pAGB star sample 
turns out to be low, we complement our sample with compilation of literature 
CO\,2-1 line data for all 393 
known pAGB stars (including likely, RCrB-eHe-LTP and RV\,Tau types) 
from the second version of the Torun Catalog \citep{szcz11}. 
Here eHe means extreme helium stars, while LTP stands for Late Thermal Pulse objects. The details of the data compilation 
are given in Appendix~\ref{app-lit-data}.  Only the summary of the literature data are given here.

In total, the compilation by October 26, 2011 includes 175 literature CO\,2-1 line data entries 
(see Tables~\ref{lit-pAGB-CO} and \ref{lit-pAGB} in Appendix~\ref{app-lit-data}) of 87 pAGB stars.
The repeated observations have been averaged together to yield the most representative CO\,2-1 line flux for each of these 87 objects. They are presented in Table~\ref{lit-pAGB-obj} in Appendix~\ref{app-lit-data}, 
together with IRAS flux densities, IRAS colors and the ratio between integrated CO\,2-1 line flux and IRAS 25\,$\mu$m flux density.

After combining this with our high Galactic latitude sample and removing duplicated objects, the total number of observed pAGB stars is 133,
among which 44 objects 
were detected in CO\,2-1 line. Thus, about 34 per cent of the known pAGB stars have been observed in the CO\,2-1 line,
and the detection rate is also about 34 per cent. 

According to the analysis of the statistical properties of the observed sample (see 
Appendix~\ref{app-statistics}), 
it is found that the available CO\,2-1 observations have the following bias effects: 
1) pAGB stars in the Galactic Center direction are under-represented; 
2) the available CO\,2-1 line observations are biased to strong IR emitters and the detectability of CO\,2-1 line is sensitivity limited.

In addition, 1117 CO\,2-1 line data entries are collected for 
751 other objects (AGB stars and PNe) from the following papers for the purpose of comparison with pAGB stars: the compilation of \citet{loup93}; some S stars from \citet{jori98,rams09}; some 
PNe from \citet{hugg05,hugg96}; and some O-rich semiregular and irregular variables from \citet{kers99}. Among them are
569 AGB stars (28 OH/IR stars plus another 155 O-rich AGB stars, 56 S stars, and 330 C-rich AGB stars) 
and 182 PNe. Repeated CO observations for these objects are also averaged together, as it was done for pAGB objects.  We do not present the detailed data in this paper, however, since they are not complete compilations for any of these type of objects.

The IRAS point source flux densities of these objects are also extracted from the IRAS Point Source Catalog 
and in a few cases from the IRAS Faint Source Catalog. Only those good quality IRAS data (with the quality factor Q$>$1) 
are used in this work.

\section{DATA ANALYSIS}
\label{analysis}

\subsection{CO-IR relation of high and low Galactic latitude post-AGB stars}
\label{pagb}

Although the relation between CO line strength and infrared dust 
emission strength has been studied before \citep[e.g., ][]{1992AA...257..701B}, 
it is still lack of in-depth investigation.
In this work, 
we do not try to compare the dust and gas mass loss rates. Instead, we
directly compare the observed CO line fluxes with {\it IRAS} flux densities  
($R_{\rm CO25}$ ratio - see Eq.~\ref{eq1}).
The greatest advantage of the direct comparison is that it allows to exclude the
uncertainties introduced by distances and empirical formulas for mass loss rate.
The traditional {\it IRAS} color-color (C-C) diagram of the two colors,
\begin{eqnarray*}
{\rm C12}&=&2.5\log (F_{25\micron}/F_{12\micron}),\\
{\rm C23}&=&2.5\log (F_{60\micron}/F_{25\micron}),
\end{eqnarray*} 
is also used. Here, $F_{12}$, $F_{25}$ and $F_{60}$ are the IRAS flux densities. 
The C23 color is a better tracer of the cool dust emission around pAGB stars than C12. 
Thus, we will concentrate on $R_{\rm CO25}$-C23 relationships of our sample stars. 

\begin{figure*}  
\epsscale{1.0}
\plotone{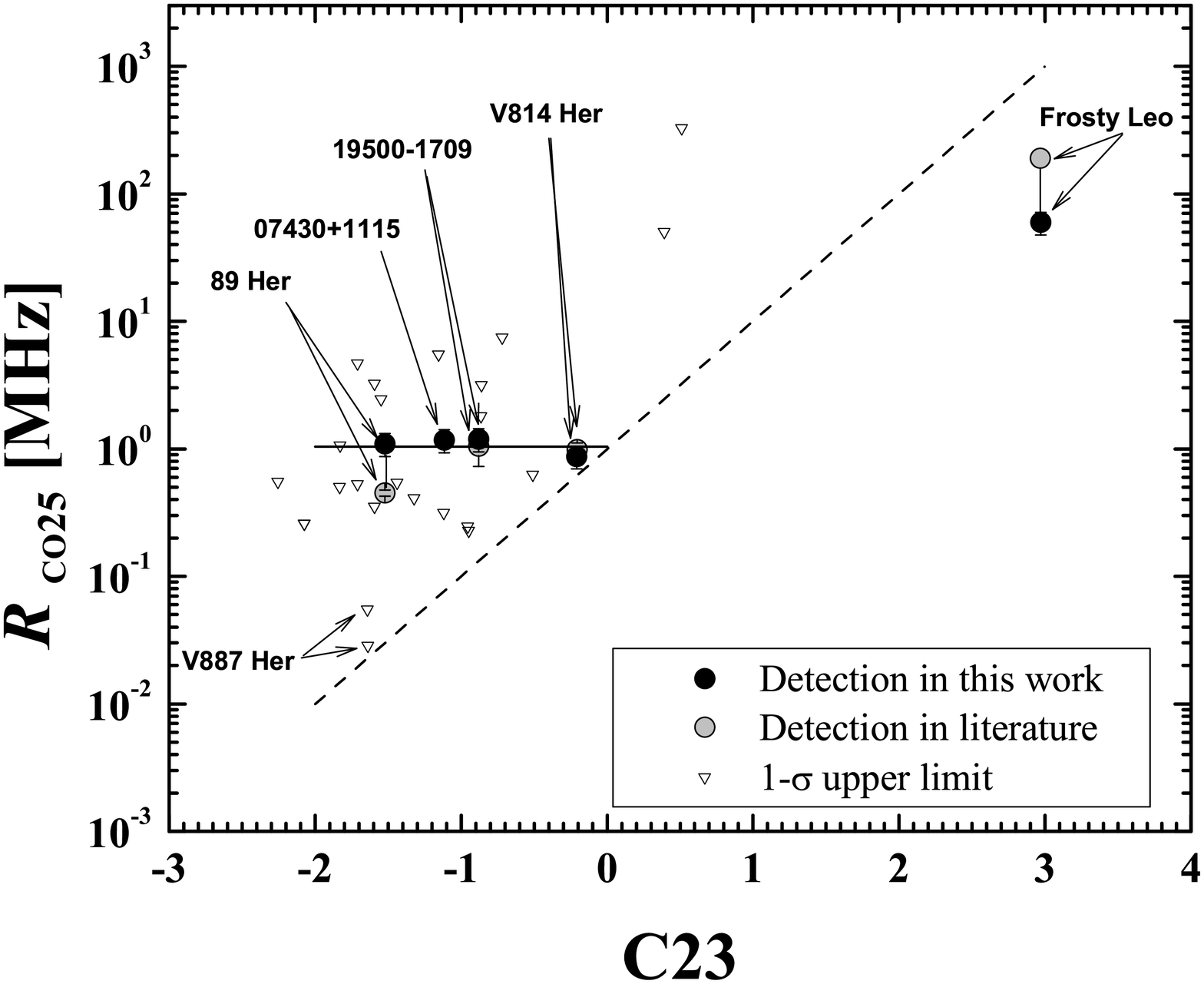}
\plotone{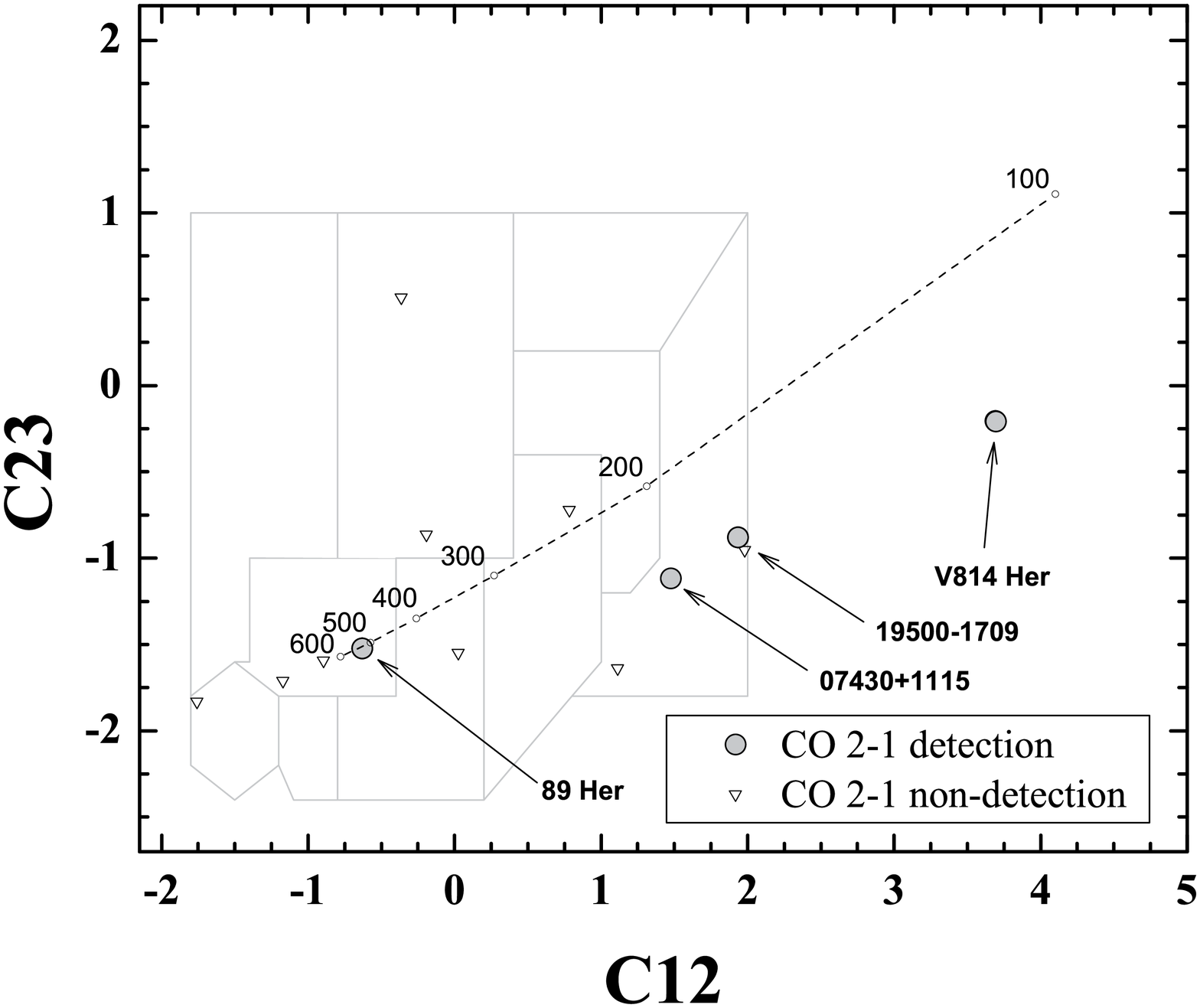}
\plotone{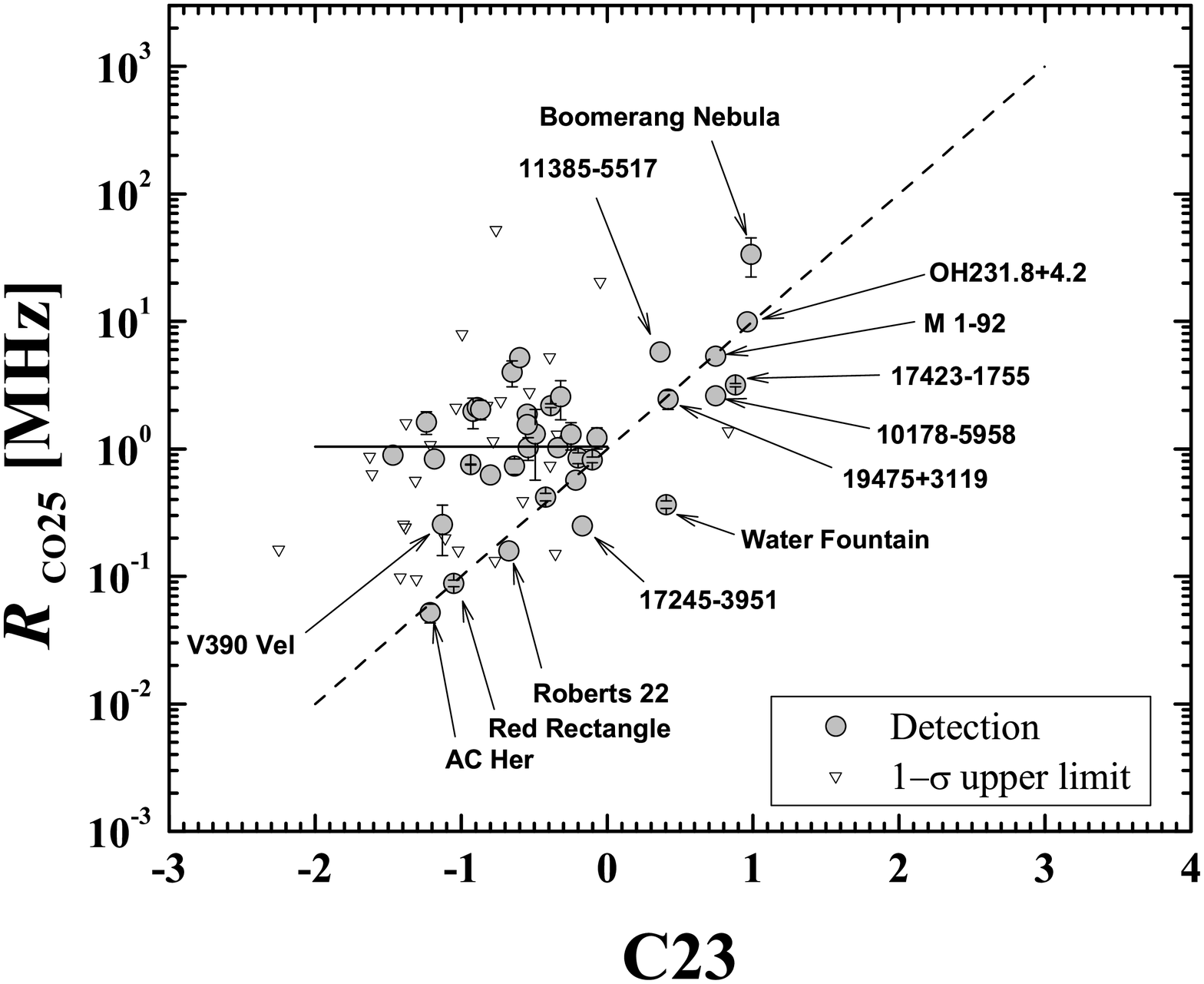}
\plotone{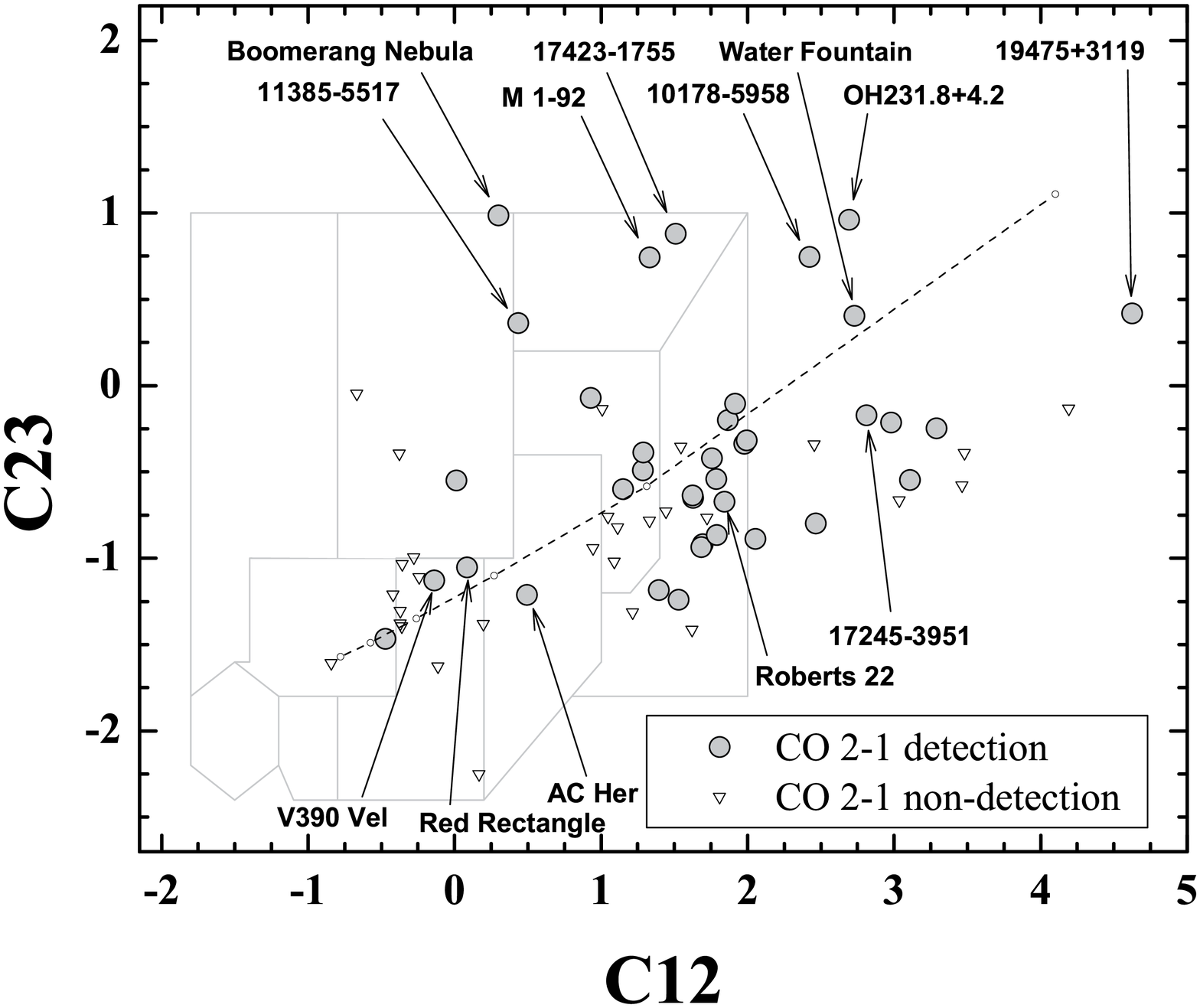}
\caption{The integrated CO\,2-1 line flux to {\it IRAS} $25\,\micron$ flux density ratios ($R_{\rm CO25}$ - see Eq.~\ref{eq1} in Sect.~\ref{results})
are plotted against {\it IRAS} color C23 (the so called CO-IR diagram defined in Sect.~\ref{pagb}) for the
observed high and low Galactic latitude pAGB stars in the left top and bottom panels, respectively. 
Also shown, in the right panels, are the traditional IRAS color-color diagrams for the same groups of pAGB objects. 
The error bars are the $20$ per cent nominal uncertainty of the CO line fluxes. The horizontal solid
line and slant dashed line on the CO-IR diagrams represent the major trends of the pAGB star 
distribution. The short dashed line on the IRAS C-C diagrams is the black body (BB) line with integer numbers being the BB temperatures. 
The regions in the IRAS C-C are those defined 
by \citet{vand89} for evolved stars.}
\label{fighglppn}
\end{figure*}

Our new observational results of the high Galactic latitude (h.g.l.) pAGB stars are combined with literature 
data. The h.g.l. objects (we adopt $|b|>15^{\circ}$) should be mainly local stars or low mass halo stars 
statistically less massive than the low Galactic latitude (l.g.l.) counterparts. 
Thus, we plot their $R_{\rm CO25}$-C23 and IRAS C-C diagrams separately in Fig.~\ref{fighglppn}. 
Compared with our new observations, there is no 
CO\,2-1 line detections in literature toward other h.g.l. objects. However, the average 
literature CO line fluxes show clear discrepancies with our new measurements for Frosty Leo and 89 Her
(see the two symbols linked to the same object names in the figure). 
The lower literature CO line flux of 89 Her could be due to the fact that it has been partially resolved  
by the small beam of the IRAM telescope which was used to obtain two of the three 
available literature CO datasets. For Frosty Leo, the literature data was obtained by a 12m telescope 
(similar as in this work) more than 18 years ago. The reason for the discrepancy is unclear.
The 1-$\sigma$ upper limits of non-detection sources are estimated by assuming a typical CSE
expansion velocity of 10\,km\,s$^{-1}$ ($\approx 7.7$\,MHz). The $R_{\rm CO25}$-C23 diagram 
will be called CO-IR diagram hereafter. 

Shown in the upper panels of Fig.~\ref{fighglppn} are h.g.l. pAGB stars. 
Both the literature data (gray circles) and our new observations (black dots for the detected objects) are shown, as indicated 
by double arrows in the figure. In the upper left panel, almost all the detected pAGB stars, 
excluding the only exceptional case of Frosty\,Leo, show very similar CO-IR flux ratios of 
\begin{equation}
\label{eq-I-25}
R_{\rm CO25} \approx 1.04(\pm 0.08) {\rm [MHz]}
\end{equation}
as delineated by the horizontal line in the figure. This relation is valid in the color range of 
$-1.6<{\rm C23}<0.0$ for the considered pAGB stars. The slant dashed line will be explained below.

On the IRAS C-C diagram in the upper right panel, Frosty\,Leo is not shown because of the poor 
quality of its {\it IRAS} $12\micron$ flux density data. The remaining four objects are distributed 
along an elongated region below the black body line. Particularly, the C12 colors vary more than 
the C23 colors. This trend of increasing colors could be the result of fast weakening 
of the $12\micron$ flux density when the inner CSE of the pAGB stars is gradually evacuated. 
Thus, the increasing IR color sequence of the detected h.g.l. pAGB stars could be an evolutionary sequence. 
Two special objects are commented in the footnote\footnote{
Frosty\,Leo stands out with its very red C23 color and 
very high $R_{\rm CO25}\approx 60$ in the upper left panel of Fig.~\ref{fighglppn}.
Although the very red C23 color can be partially attributed to its
very strong water ice emission band around $46\micron$ \citep{forv87},
the very large $R_{\rm CO25}$ ratio needs other mechanisms or explanation. This object also has
other peculiar properties such as a high initial stellar mass of 4.23\,M$_\odot$ \citep{mura08}, a
massive compact expanding ring seen in CO\,2-1 and 1-0 lines and in near-IR polaro-imaging data 
\citep{cast05,Doug90} around a binary system \citep{rodd95}, 
and frenzied equatorial jets \citep{saha00}. }
\footnote{
V887\,Her is also peculiar in that the 1-$\sigma$ upper limit of the $R_{\rm CO25}$ ratio 
is more than 10 times smaller than that of the five CO detections 
in the upper left panel of Fig.~\ref{fighglppn}. 
It is an RV Tauri variable \citep{Sahin11} showing depletion of refractory
elements such as Al, Y and Zr.}.

In the two lower panels of Fig.~\ref{fighglppn}, the l.g.l. pAGB stars are shown. 
The CO-IR diagram in the lower left panel shows that many stars with CO detection (gray circles) 
are crowded around a similar region as 
the high latitude stars (with $R_{\rm CO25} \approx 1.04$ and ${\rm C23}<0.0$), while 
the rest objects seem to be located in an elongated region, extending from blue C23 color and 
small $R_{\rm CO25}$ ratio to red C23 color and large $R_{\rm CO25}$ ratio. The object names are marked out 
in the figure for the latter group of stars. We find that a slant straight line (the dashed line) of
\begin{equation}
\label{eq-II-25}
\log (R_{\rm CO25})\approx{\rm C23}.
\end{equation}
can represent this trend reasonably well. 
This log-linear trend holds in the color range of $-1.2<{\rm C23}<1$. Eq.~\ref{eq-II-25} is not a fit to the data, 
but a simple function recognized by eyes. The lower right corner of this CO-IR diagram is devoid of objects, 
hinting that this log-linear trend could represent a border of pAGB star distribution on the CO-IR diagram. 

The l.g.l. pAGB objects, which are distributed along the slant line and have C23$<$0 are located on or below BB line on the IRAS C-C diagram. 
On the other hand, the rest objects that have C23$>$0 (with the exception of IRAS\,19475$+$3119) are located above BB line 
(see the lower panels of Fig.~\ref{fighglppn}). In addition, it is seen that
the l.g.l. pAGB stars, which 
appear in a similar region as the h.g.l. ones on the CO-IR diagram (compare the two left panels) 
also occupy a similar region on the IRAS C-C diagrams (compare the two right panels). This
indicates that these h.g.l. and l.g.l. objects share a similar combination of dust and CO gas characteristics.

\subsection{CO-IR diagrams in the context of AGB-pAGB-PN sequence}
\label{compare}

\begin{figure*}
\epsscale{2.0}
\plotone{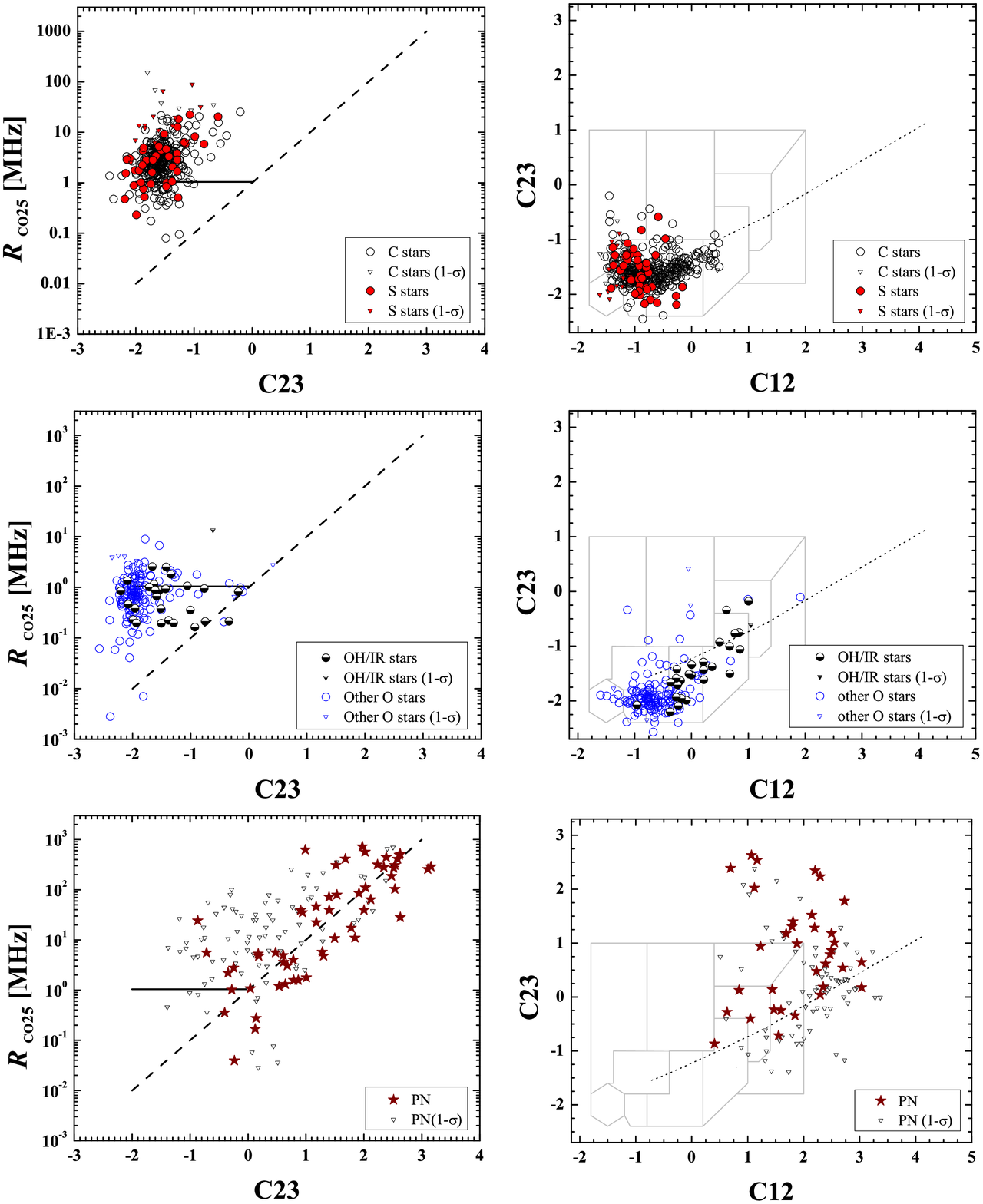}
\caption{The same as Fig.~\ref{fighglppn} but for C- and O-rich AGB stars and PNe. 
  The trends of pAGB stars distribution are represented by the two lines on the CO-IR diagrams. A color plot 
  is provided in the online version.}
\label{figcompare}
\end{figure*}

Although some regularity has begun to emerge on the previously discussed $R_{\rm CO25}$-C23 
diagrams of the pAGB stars, the discussed trends are still murky due to the limited number of objects involved. 
In this section, we try to verify these trends in a broader context of the AGB-pAGB-PN evolutionary sequence. 
The CO\,2-1 line data collected from literature for some AGB stars
and PNe make this comparison possible. 
Thus, the CO-IR diagrams of these AGB stars and PNe are plotted 
and compared with the pAGB stars (represented by the two straight lines that delineate the major trends) 
in Fig.~\ref{figcompare}. The traditional IRAS C-C diagrams are also shown for them. 
For clarity, the C-rich and O-rich AGB stars are plotted in separate panels. 

\subsubsection{AGB stars and PNe on the CO-IR diagrams}
\label{AGB-PN}

We briefly discuss the distribution of AGB stars and PNe on the CO-IR and IRAS C-C diagrams, while more 
comparison among these objects is presented in Appendix~\ref{app-AGB-stars}.

In the top left panel of Fig.~\ref{figcompare}, C-rich AGB stars (empty circles) mainly concentrate 
in a compact region on the $R_{\rm CO25}$-C23 diagram, and S stars (gray filled circles)
mainly scatter in a similar region as the C stars. On the IRAS C-C diagram (the top right panel), 
most of these stars are distributed slightly below the black body line (the dotted line) or in regions 
with red C23 but blue C12 colors. Their mean IR color is C23$=-1.54\pm0.35$ and their mean CO-IR flux ratio 
is  $\log(R_{\rm CO25})=0.41\pm0.40$.
%

In the middle left panel of Fig.~\ref{figcompare}, many O-rich AGB stars (empty circles) concentrate in another compact 
region different from that of C stars on the CO-IR diagram, whilst an extreme subsample of them --- OH/IR stars (half-shaded black circles) ---
scatter in regions with similar or smaller $R_{\rm CO25}$ ratios but usually redder C23 colors than the other O-rich AGB stars. 
Their mean IR color is C23$=-1.70\pm0.76$ and their mean CO-IR flux ratio 
is  $\log(R_{\rm CO25})=-0.13\pm0.58$, which means that they are slightly bluer and have 3.5 times smaller mean $R_{\rm CO25}$ ratios
than C-rich AGB stars on average. 
On the IRAS C-C diagram in the middle right panel of the figure, most of these O-rich AGB stars 
concentrate in a region below the black body line, while the OH/IR stars stretch out to much redder regions, as expected.

In the bottom left panel of Fig.~\ref{figcompare}, PNe show a pronounced increasing log-linear trend on the CO-IR diagram. 
The trend spans more than 4 orders of magnitude of the $R_{\rm CO25}$ ratio and are in close agreement to 
the log-linear trend of pAGB stars (the dashed line).  On the IRAS C-C diagram in the bottom right panel, 
most of the PNe appear in a very red region above the black body line, as expected for cold 
dusty CSEs.

\subsubsection{
The pAGB trends on the CO-IR diagram in the context of AGB-pAGB-PN evolution}
\label{compare-all}

By comparing the pAGB stars (represented by the two straight lines) with AGB stars and PNe in Fig.~\ref{figcompare},
it is now clear that the pAGB stars are distributed in a transitional region between the AGB stars and PNe on the CO-IR diagram.
The subgroup of pAGB stars represented by the horizontal solid line in the figure have similar CO-IR ratios (with a mean $\log(R_{\rm CO25})=0.02\pm0.03$) as AGB stars, but have redder IR colors (with C23$=-1.6\sim 0.0$).
The log-linear trend followed by the rest group of pAGB stars on the 
CO-IR diagram agree quite well with the trend of PNe. Thus those pAGB stars with very red C23 colors ($>0$) could be 
the precursor of the shown PNe. The other end of the log-linear trend (with ${\rm C23}<0$) is populated by some pAGB stars that 
have exceptionally small $R_{\rm CO25}$ (very weak CO line emission). 

The distinctive characteristics of these pAGB stars allow us 
to divide them into three subgroups, which, as we show later, 
have different properties.

\subsection{Grouping of post-AGB stars}
\label{grouping}

\begin{figure}
\epsscale{1.0}
\plotone{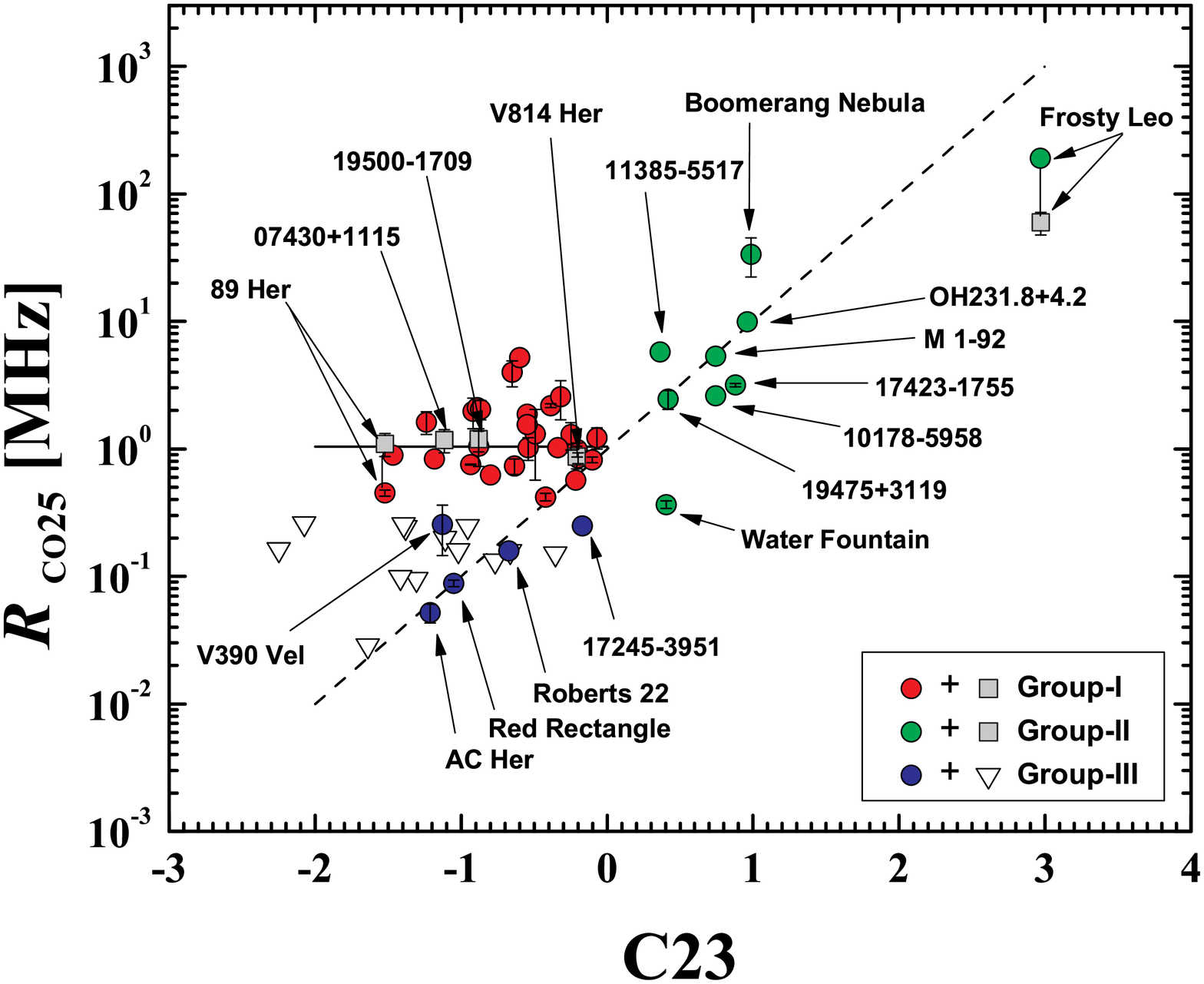}
\plotone{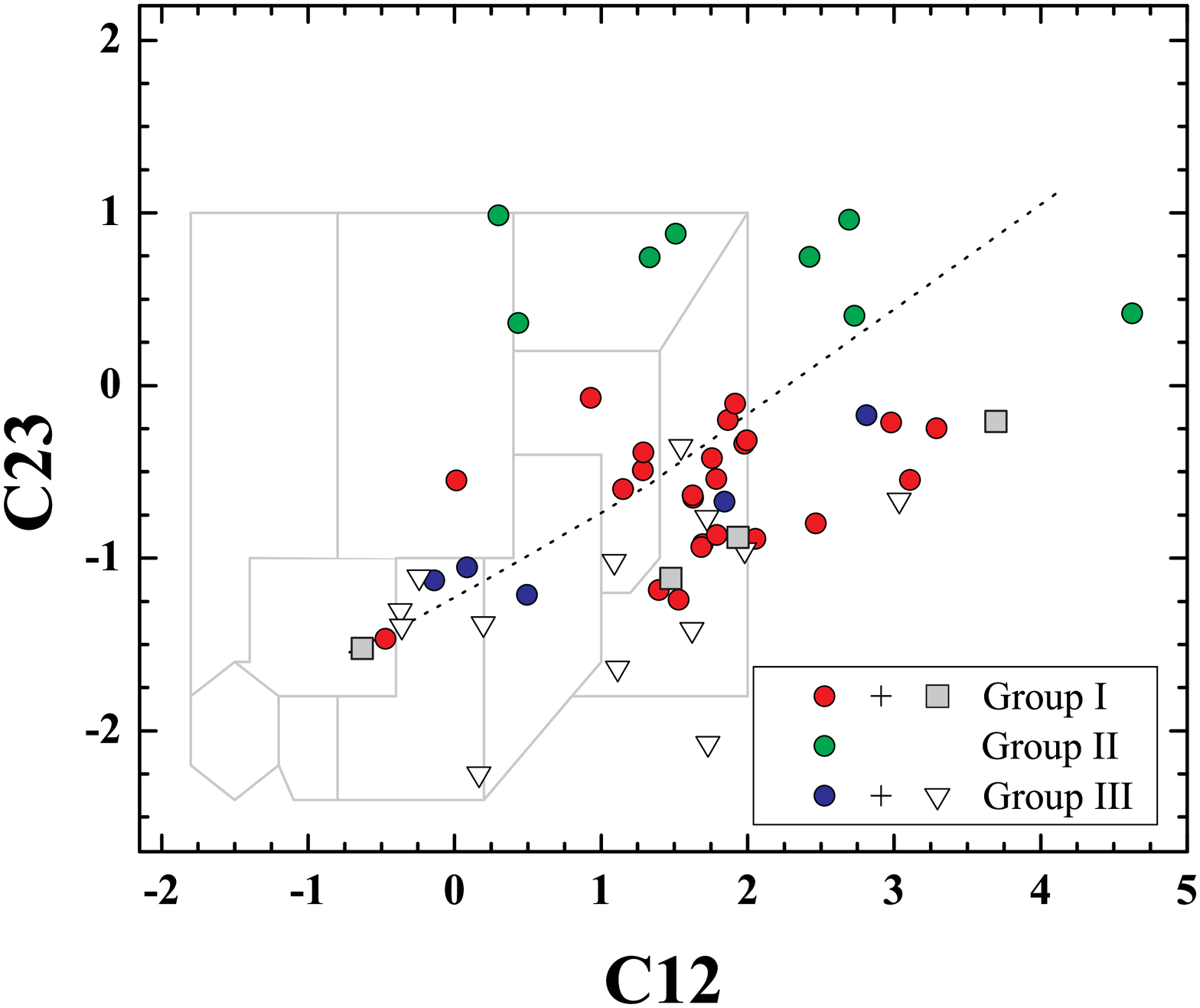}
\caption{Grouping of all pAGB stars on the CO-IR diagram and the resulting groups on the {\it IRAS} C-C diagram. The lines and regions are the same as in Fig.~\ref{fighglppn}. 
  Object names are labelled out for our five high Galactic latitude pAGB stars
  and the CO-detected pAGB stars that belong to groups-II and III.}
\label{figppn}
\end{figure}

Our sample of pAGB stars aggregate in different regions on the CO-IR diagram, which is not so well seen on the 
traditional IRAS C-C diagram. As we will see below, the aggregation on the CO-IR diagram just reflects 
different nature (e.g., mass, binarity, chemistry, etc.), 
and the stage of pAGB evolution of these stars through the contrast of dust and CO line emission. 
Here, we merge the h.g.l. and l.g.l. pAGB stars and divide them
into three CO-IR groups, as shown in the upper panel of Fig.~\ref{figppn}. Only significant, from point of view of performed source classification, CO non-detections
are plotted in Fig.~\ref{figppn}.

{\it group-I} --- those pAGB stars with $R_{\rm CO25} > 1/3 $ and ${\rm C23}<0$ (the red filled circles, or gray filled squares 
in Fig.~\ref{figppn}). They compose the largest group of pAGB stars, which is
distributed in a horizontally elongated region on the CO-IR diagram, with similar $R_{\rm CO25}$ ratio of $\sim 1$\,MHz 
(actually in the range of 0.42-5.2\,MHz, with a median of $\sim 1$\,MHz). 
Their distribution can be roughly represented by the horizontal line that 
was determined for our h.g.l. pAGB stars in Sect.~\ref{pagb}.

{\it group-II} --- those pAGB stars with 
${\rm C23}>0$ (the green filled circles and one gray filled 
square in Fig.~\ref{figppn}). They are the reddest group 
with usually enhanced CO\,2-1 emission relative to their IR dust emission. They occupy the red 
part of the log-linear 
trend as delineated by the dashed line in the figure.

{\it group-III} --- those pAGB stars with $R_{\rm CO25} < 1/3 $ and ${\rm C23}<0$ (the blue filled circles in Fig.~\ref{figppn}). 
They have exceptionally weak CO\,2-1 line emission, although their C23 colors are similar to that of group-I stars. 
They occupy the blue part of the log-linear trend (the dashed line) and the region immediately above it. This group also includes some 
CO\,2-1 non-detections whose 1-$\sigma$ upper limit of $R_{\rm CO25}$ ratios are smaller than 1/3 and thus they are also CO deficient pAGB stars. 
We did not plot the other non-detections, since they could be members of different groups.

All of the three CO-IR groups of pAGB stars are listed in Table~\ref{tabgrouping}, with different groups in separate sub-sections of the table. 
The vicinity of intersection between the solid and dashed lines in the top panel of Fig.~\ref{figppn} do no 
allow us to be sure to which group a given object belong. Therefore, the classification  
of the objects near the group borders (here we have assumed somewhat arbitrarily a rectangular border region as $R_{\rm CO25}>1/3$ and  $-0.5<{\rm C23}<0$)
 is tentative, and we labeled them with a leading `*' symbol before their object names in Table~\ref{tabgrouping}.
The table is organized in a way to enable easy comparison between
the CO-IR and IRAS C-C diagrams in Fig.~\ref{figppn}. Objects in groups-I and II are sorted 
with the increasing C23 colors, while those of group-III 
are ordered in decreasing C23 colors. 
Also listed in the table are some properties of the pAGB stars that will 
be explained and discussed later in Sect.~\ref{discuss}.

In the lower panel of Fig.~\ref{figppn}, the three groups are displayed also on the traditional dust emission diagnostic tool,  the IRAS C-C diagram.
Group-II pAGB stars are clearly separated from the other stars by their large C23 colors.
However, the groups-I and III pAGB stars are well mixed on the IRAS C-C diagram, which demonstrates that
the involvement of gas emission (the CO\,2-1 line) in the grouping does have brought us 
a new information about the pAGB stars.

Group-I pAGB stars distribute in an elongated region on the IRAS C-C diagram,
with their C23 colors varying only by a factor about 1.5, 
while their C12 colors varying by a larger factor of about 4.0. 
The larger variation range of the C12 colors may be the natural consequence of fast weakening of the $12\micron$ dust emission 
in the expanding detached CSE. 
Thus, group-I pAGB stars are possibly still in early pAGB stage when the detached CSE is still actively developing, 
which is also supported by the fact that their $R_{\rm CO25}$ ratios are similar to that of AGB stars. 
Their similar $R_{\rm CO25}$ ratios hints that the thermal balance in the gas (represented by the CO\,2-1 line)
and dust (represented by the $25\micron$ emission) could be still tightly coupled during the early pAGB stage.

The red C23 color and richness of CO\,2-1 emission of the group-II stars indicate that they could have massive 
and cool CSEs. To the opposite, the weakness of CO\,2-1 emission and the blue C23 colors of the group-III stars 
suggest that they could be lower mass pAGB stars with more transparent spherical component in their CSEs where CO molecules have been 
partially or totally destroyed by penetrating UV photons.
The well known low mass binary disc system 
Red\,Rectangle \citep{mens02,witt09} that shows narrow CO line \citep{1995ApJ...453..721J,1996ApJ...472..703D} 
is just a member of the group-III. However, it is not clear why the distinct groups-II and III 
share the same log-linear relationship on the CO-IR diagram.

\section{DISCUSSION}
\label{discuss}
As we showed in Sect.~\ref{analysis}, the AGB, pAGB and PN CO-emitters segregate into different regions on the CO-IR diagram, which sets up a new platform for discussing the various aspects of the pAGB stars evolution. In this section, we consider chemical and central star spectral types, spectral energy distribution (SED) types, binarity, CSE expansion velocities of the pAGB stars, and their pAGB sub-types to investigate evolution and diversity of these objects using CO-IR diagram.

To facilitate the discussion, we collect the 
important properties of the three groups of pAGB stars 
in Table~\ref{tabgrouping}. Altogether, excluding 32 non-detections with upper limit of $R_{\rm CO25} > 1/3$, we have gathered information for 55 objects.
 There are following columns in the table: 
Object name, C23 -- IRAS color, 
$R_{\rm CO25}$ -- CO to IR flux ratio, 
$V_{\rm exp}$ -- CSE expansion velocity, 
SED -- spectral energy distribution type, 
Binarity, 
Chem.type (*/CSE) -- central star and CSE chemical type,
Spectral type -- central star spectral type, 
pAGB class, and 
Chem.ref (*/CSE) -- literature for chemical type of the star and CSE.
The references for chemical and spectral types are given in footnotes to Table~\ref{tabgrouping}.
Most of the data are already collected in the Torun Catalog, but not yet for chemistry of central stars and their circumstellar envelopes.   Therefore, we have performed dedicated literature study to determine  
chemical types (but also to fine-tune spectral types) of pAGB objects, so credits to the original works or compilations are given below Table~\ref{tabgrouping}. Note that chemistry of CSE is also inferred from dust features seen in Infrared Space Observatory (ISO) spectra (if available in several cases) which can be found in the Torun catalog.

\begin{deluxetable}{l@{}l@{}r@{ }r@{}r@{}c@{}c@{ }c@{}l@{}r@{}c}
\tablecolumns{11} 
\tabletypesize{\tiny}
\tablewidth{0pc}
\tablecaption{Properties of the three CO-IR groups of post-AGB stars.
\label{tabgrouping}}
\tablehead{
\colhead{                          }  &
\colhead{Object                    }  &
\colhead{C23                       }  &
\colhead{RCO25                     }  &
\colhead{$V_{\rm exp}$             }  &
\colhead{SED\tablenotemark{a}      }  &
\colhead{Binarity\tablenotemark{a} }  &
\colhead{Chem.type                 }  &
\colhead{Spectral\tablenotemark{s} }  &
\colhead{pAGB\tablenotemark{a}     }  &
\colhead{Chem.ref\tablenotemark{c} }  \\
\colhead{                          }  &
\colhead{name                      }  &
\colhead{                          }  &
\colhead{                          }  &
\colhead{${\rm km\,s}^{-1}$        }  &
\colhead{                          }  &
\colhead{                          }  &
\colhead{*/CSE                     }  &
\colhead{type                      }  &
\colhead{class                     }  &
\colhead{*/CSE                     }  }
\startdata
 \multicolumn{11}{l}{\it ------------ Group I ------------}\\
$ $ & 89 Her          & -1.52 & $ $  1.09\tablenotemark{b} &   3.5\tablenotemark{b} & 0      & B & O/O    & F2Ibe           & UU Her         & 24/10   	\\
$ $ & AI Sco          & -1.47 & $ $  0.89                  &   2.8                  & 0      & B & C?/O   & G4              & RV Tau         & 9/8     	\\
$ $ & 22272+5435      & -1.24 & $ $  1.62                  &  10.1                  & IVa    &   & C/C    & G5Ia            & 21 micron      & 1       	\\
$ $ & 04296+3429      & -1.18 & $ $  0.83                  &  12.8                  & IVa    &   & C/C    & F3I$^{(s1)}$    & 21 micron      & 1       	\\
$ $ & 07430+1115      & -1.12 & $ $  1.17\tablenotemark{b} &   6.5\tablenotemark{b} & IVa    &   & C/C    & G5Ia            & 21 micron      & 1       	\\
$ $ & 20000+3239      & -0.94 & $ $  0.75                  &  12.2                  & IVa    &   & C/C    & G8Ia            & 21 micron      & 1       	\\
$ $ & 07134+1005      & -0.92 & $ $  1.97                  &   9.9                  & IVb    &   & C/C    & F7Ie$^{(s2)}$   & 21 micron      & 1       	\\
$ $ & 16594-4656      & -0.89 & $ $  2.10                  &  14.0                  & III    &   & C/C    & Ae$^{(s2)}$     & 21 micron      & 1       	\\
$ $ & 19500-1709      & -0.88 & $ $  1.19\tablenotemark{b} &  10.4\tablenotemark{b} & IVb    &   & C/C    & F0Ie$^{(s2)}$   & 21 micron      & 1       	\\
$ $ & 23304+6147      & -0.87 & $ $  2.04                  &  12.1                  & IVa    &   & C/C    & F6I$^{(s3)}$    & 21 micron      & 1       	\\
$ $ & 14488-5405      & -0.80 & $ $  0.62                  &  15.0                  & IVb    &   & ?/?    & A0Ie$^{(s2)}$   & possible       & -/-     	\\
$ $ & 06530-0213      & -0.65 & $ $  3.99                  &  11.3                  & IVa    &   & C/C    & G1I$^{(s2)}$    & 21 micron      & 1       	\\
$ $ & 17441-2411      & -0.63 & $ $  0.73                  &  12.5                  & III    &   & ?/?    & F4I$^{(s2)}$    & IRASsel        & -/-     	\\
$ $ & 08005-2356      & -0.60 & $ $  5.20                  & 100.0                  & I      &   & C/O    & F5Ie$^{(s2)}$   & IRASsel        & 2/15     \\
$ $ & 23541+7031      & -0.55 & $ $  1.55                  &  54.0                  & II     &   & ?/O    & Be              & hotpAGB        & -/29     \\
$ $ & 22223+4327      & -0.55 & $ $  1.87                  &  14.0                  & IVb    &   & C/C    & F7I$^{(s2)}$    & 21 micron      & 1        \\
$ $ & 19454+2920      & -0.54 & $ $  1.02                  &  15.6                  & III    &   & ?/C    & ~               & IRASsel        & -/11     \\
$*$ & 19480+2504      & -0.49 & $ $  1.31                  &  17.5                  & III    &   & C/C    & ~               & IRASsel        & 7/19     \\
$*$ & 20028+3910      & -0.42 & $ $  0.42                  &  14.5                  & III    &   & ?/O    & F3-7I$^{(s1)}$  & IRASsel        & -/10     \\
$*$ & 22574+6609      & -0.39 & $ $  2.18                  &  19.8                  & II     &   & C/C    & ~               & 21 micron      & 1        \\
$*$ & 09032-3953      & -0.34 & $ $  1.02                  &  25.0                  & II     &   & ?/?    & ~               & IRASsel        & -/-      \\
$*$ & 23321+6545      & -0.32 & $ $  2.56                  &  16.1                  & III    &   & ?/C    & ~               & IRASsel        & -/11     \\
$*$ & 19114+0002      & -0.25 & $ $  1.29                  &  31.5                  & IVa    &   & O/O    & F7I$^{(s2)}$    & IRexc          & 24/15    \\
$*$ & 17106-3046      & -0.21 & $ $  0.57                  &   1.0                  & III    &   & ?/O    & F5I$^{(s2)}$    & IRASsel        & -/28     \\
$*$ & V814 Her        & -0.21 & $ $  0.87\tablenotemark{b} &   8.6\tablenotemark{b} & IVb    &   & O/O    & A7I$^{(s2)}$    & UU Her         & 24/15    \\
$*$ & 17150-3224      & -0.20 & $ $  0.85                  &  14.8                  & III    &   & ?/O    & F3-7I$^{(s1)}$  & OHmaser        & -/12     \\
$*$ & 18276-1431      & -0.10 & $ $  0.82                  &  12.2                  & IVa    &   & ?/O    & K2/K3:II/III:   & OHmaser        & -/30     \\
$*$ & CRL 618         & -0.07 & $ $  1.22                  &  16.7                  & II     &   & C/C    & B0              & refneb         & 4        \\
 \multicolumn{11}{l}{\it ------------ Group II ------------}\\                                                                               
$ $ & 11385-5517      &  0.36 & $ $  5.74                  &  45.0                  & 0      & B & C/O    & F0Iape          & IRexc          & 24/31    \\
$ $ & Water Fountain  &  0.41 & $ $  0.37                  &  46.0                  & II     &   & ?/O    & ~               & Water fountain & -/14     \\
$ $ & 19475+3119      &  0.42 & $ $  2.45                  &  15.1                  & IVb    &   & O/O    & F3Ib            & IRASsel        & 24/32    \\
$ $ & M 1-92          &  0.74 & $ $  5.34                  &  32.5                  & III    & B & ?/O    & B0.5IV;F2       & refneb         & -/26     \\
$ $ & 10178-5958      &  0.75 & $ $  2.61                  &  13.0                  & III    &   & ?/C    & BIe$^{(s2)}$    & hotpAGB        & -/20     \\
$ $ & 17423-1755      &  0.88 & $ $  3.17                  &  26.5                  & II/III & B & ?/O    & Be              & IRASsel        & -/17     \\
$ $ & OH 231.8+4.2    &  0.96 & $ $  9.94                  &  21.4                  & II/III &   & O/O    & M6              & possible       & 16/18    \\
$ $ & Boomerang Neb   &  0.99 & $ $ 33.60                  &  25.5                  & II     &   & ?/?    & ~               & refneb         & -/-      \\
$ $ & Frosty Leo      &  2.97 & $ $ 59.62\tablenotemark{b} &  22.8\tablenotemark{b} & IVb    & B & ?/O    & K7II            & hglsg          & -/32     \\
 \multicolumn{11}{l}{\it ------------ Group III ------------}\\                                                                              
$ $ & 17245-3951      & -0.17 & $ $  0.25                  &  15.0                  & III    &   & ?/O    & F6I$^{(s2)}$    & OHmaser        & -/27     \\
$ $ & 19374+2359      & -0.35 & $<$  0.15                  &                        & II     &   & ?/O    & B3-6I           & IRASsel        & -/13     \\
$ $ & 19306+1407      & -0.66 & $<$  0.16                  &                        & IVa    &   & ?/C,O  & B0-1I$^{(s1)}$  & IRASsel        & -/10     \\
$ $ & Roberts 22      & -0.67 & $ $  0.16                  &  31.0                  & II     &   & ?/C,O  & A0Ie$^{(s2)}$   & OHmaser        & -/10     \\
$ $ & 19477+2401      & -0.77 & $<$  0.13                  &                        & II     &   & C/C    & G0I$^{(s1)}$    & 21 micron      & 1        \\
$ $ & 01005+7910      & -0.95 & $<$  0.25                  &                        & IVb    &   & ?/C    & B2Iab:e         & hglBsg         & -/5      \\
$ $ & 19386+0155      & -1.02 & $<$  0.16                  &                        & I      &   & O/O    & F5I$^{(s2)}$    & IRASsel        & 22       \\
$ $ & Red Rectangle   & -1.05 & $ $  0.09                  &   4.9                  & II     & B & C/C,O  & B8V             & refneb         & 2/3,21 	\\
$ $ & RV Tau          & -1.11 & $<$  0.20                  &                        & 0      & B & C/?    & K3pvar          & RV Tau         & 24/-     \\
$ $ & V390 Vel        & -1.13 & $ $  0.25                  &   5.6                  & I      & B & O/?    & F3e             & RV Tau         & 24/-     \\
$ $ & AC Her          & -1.21 & $ $  0.05                  &   1.5                  & 0      & B & O/O    & F4Ibpvar        & RV Tau         & 24/10    \\
$ $ & U Mon           & -1.30 & $<$  0.09                  &                        & 0      & B & O/?    & K0Ibpvar        & RV Tau         & 24/-     \\
$ $ & DY Ori          & -1.38 & $<$  0.24                  &                        & 0      & B & O/O    & ~               & RV Tau         & 24/8     \\
$ $ & AR Pup          & -1.39 & $<$  0.25                  &                        & III    & B & O/?    & F0Iab:...       & RV Tau         & 24/-     \\
$ $ & AI CMi          & -1.41 & $<$  0.10                  &                        & I      &   & O/O    & K3/5I$^{(s2)}$  & RV Tau         & 23/19    \\
$ $ & V887 Her        & -1.64 & $<$  0.03                  &                        & IVa    &   & O/O    & F3Ib            & OHmaser        & 25/6    	\\
$ $ & V886 Her        & -2.07 & $<$  0.26                  &                        & IVb    &   & O/O    & B3e$^{(s2)}$    & IRASsel        & 24/5     \\
$ $ & 20004+2955      & -2.25 & $<$  0.16                  &                        & 0      &   & ?/O    & G7Ia            & IRexc          & -/33     \\
\enddata
\tablecomments{$*$ These objects appear near the joint region between the three CO-IR groups and 
thus their group identities are tentative. (Objects are sorted by C23 color within each group.) }
\tablenotetext{a}{SED types, binarity and the sub-types of post-AGB stars are taken from the Torun Catalog.}
\tablenotetext{b}{ These values are preferentially taken from this work, instead of from literature.}
\tablenotetext{c}{The circumstellar/central star chemical types were originally compiled from the CO data papers and SIMBAD database, 
but are later refined with dedicated literature study. Here is the list of the dedicated references and relevant comments: 
 1) Carbon rich star with $21\,micron$ dust feature;
 2) \citet{Bak97} (optical spectrum);
 3) \citet{wat98} (ISO spectrum);
 4) \citet{CGCS};
 5) \citet{Cerr09} (Spitzer: PAH/Silicates);
 6) \citet{Eder88} (OH maser);
 7) \citet{GCGCS};
 8) \citet{Giel11} (Spitzer: Silicate);
 9) \citet{Giri05} (optical spectrum);
10) \citet{Hodg04} (ISO spectrum);
11) \citet{Hony02} ($30\,\micron$ feature);
12) \citet{1993AA...273..185H} (OH maser);
13) \citet{Kwok87} (CFHT IR photometry);
14) \citet{Lik88} (OH \& H2O masers);
15) \citet{Lik89} (OH maser);
16) Spectral type of M6 in Simbad;
17) \citet{Mant11} (Spitzer: Silicate \& H2O ice absorption);
18) \citet{Morr80} (OH maser);
19) \citet{1993AA...267..515O} (IR + HCN/CO);
20) \citet{Part01} (ISO: PAH);
21) \citet{Pee02} (ISO: PAH);
22) \citet{Pere04} (Optical and IR spectra);
23) \citet{Rao12} (optical spectrum);
24) \citet{SSSS} (abundance compilation);
25) \citet{Sahin11} (optical spectrum);
26) \citet{Seaq91} (OH maser);
27) \citet{Sev02} (OH maser);
28) \citet{Silva93} (OH maser);
29) \citet{Volk87} (IRAS/LRS);
30) \citet{teLin91} (OH maser);
31) \citet{teLin92} (OH maser);
32) \citet{teLin96} (OH maser);
33) \citet{vand02} (ISO spectrum).
}
\tablenotetext{s}{The central star spectral types are mostly obtained from SIMBAD database, while several of them are updated with following literature:
(s1) = \citet{Sanc08};
(s2) = \citet{Suar06};
(s3) = \citet{Pere07}.}
\end{deluxetable}

\subsection{Chemical and spectral types of post-AGB objects}
\label{chem-sp-type}

For discussion in this paper, we assume that the chemical type of a pAGB object is C-rich if it has a C-rich central star and/or C-rich dust in its CSE. Also stars with dual dust chemistry in their CSE (simultaneous presence of C- and O-rich dust features and/or molecules) are treated as having C-rich chemical type. Knowledge of chemical type is important since it allows to roughly estimate mass of the progenitor. Single carbon stars are formed only in a limited progenitor mass range. 
For solar metallicity this happens for $\sim 1.5$\,M$_\odot < M_{\rm ZAMS}<\sim 5$\,M$_\odot$, 
and the mass range shrinks and shifts towards somewhat smaller values for lower metallicities \citep{Piovan03}. For progenitor mass lower and higher than the above mass range a star will remain O-rich. Note, however, that in close binary systems, mass transfer to a companion star may reduce the star's AGB lifetime and the star may remain O-rich even if its progenitor was of intermediate mass.

\begin{figure*}
\epsscale{2.0}
\plotone{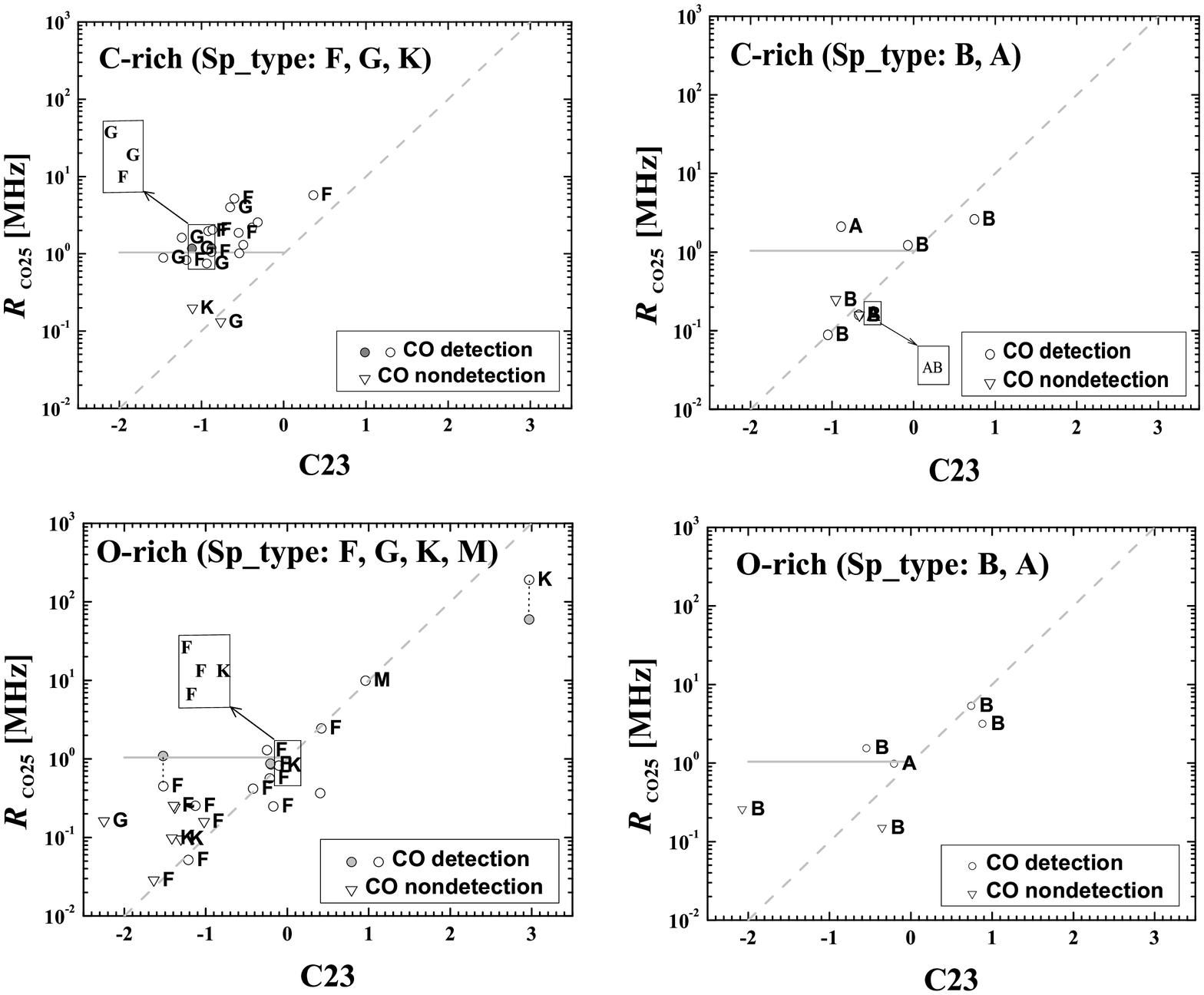}
\caption{The distribution of chemical types and central star spectral types for pAGB stars on the $R_{\rm CO25}$-C23 diagram. The C- and O-rich pAGB objects are plotted in upper and lower panels, while central stars of late spectral types (F, G, K and M) and of early ones (B and A) are plotted
in the left and right panels, respectively. 
The solid and dashed lines are the same as in the left panels of Fig.~\ref{fighglppn}.  Crowded regions have been enlarged in rectangles. Gray filled circles are new results from this work. They are connected by vertical line with the corresponding literature CO data, if their differences are obvious.}
\label{fig-chem-spec}
\end{figure*}

The distribution of the chemical types from Table~\ref{tabgrouping} is visualized 
together with central star spectral types on the CO-IR diagram in Fig.~\ref{fig-chem-spec}. 
C-rich and O-rich pAGB stars are plotted in separate rows, while those with late and early spectral types
in different columns.

In the left two panels of Fig.~\ref{fig-chem-spec}, the C- and O-rich pAGB stars 
with late spectral types (F, G, K and M) show distinctive distributions. C-rich pAGB stars of late spectral types are mostly group-I sources, with only few objects, like Red Rectangle, Roberts\,22 and IRAS\,19477+2401\footnote{This object shows the `21 micron' feature, but its $R_{\rm CO25}$ is uncertain due to interstellar contamination of the CO\,2-1 emission \citep{2005ApJ...624..331H}.}, belonging to group-III.  On the other hand, O-rich pAGB stars with similar spectral types belong predominantly to group-III\footnote{As we show in the next subsection this group contains a significant number of known binary systems.}, and to the transition region between different groups. Group-II objects with late spectral types are not numerous, in general, but among them there is only one C-rich star (IRAS\,11385$-$5517) with SiC emission at 11.3\,$\mu$m seen in its ISO spectrum, but also with OH emission detected from its shell (see Torun catalog and references in Table~\ref{tabgrouping}). 

Post-AGB stars of earlier spectral types (right panels of Fig.~\ref{fig-chem-spec}) are also not numerous in our sample but are located mostly along the log-linear track (dashed line in the figure), with a clear exception being V886\,Her (C23 = -2.07)  - a massive \citep{Ryans:2003eu} fast evolving O-rich pAGB star \citep{Arkhipova:1999nx}.  

From point of view of the progenitor mass, a feasible interpretation of the discussed trends is that group-I objects, being predominantly C-rich, are intermediate mass pAGB stars; the group-III sources that are CO-deficient objects are the lowest mass pAGB stars; and the group-II sources are intermediate or high mass stars, which already follow the PNe trend on the CO-IR diagram (see bottom left panel in Fig.~\ref{figcompare}).
This simplified interpretation is supported by the significantly large percentage of O-rich sources (about 70 per cent, see Table~\ref{tabgrouping}) among groups-III and II. 
On the other hand, the paucity of O-rich pAGB stars in group-I region (see the two bottom panels of  Fig.~\ref{fig-chem-spec}) seems to suggest that O-rich pAGB stars do not evolve into this region.

From point of view of a single star evolution, we expect that during pAGB stage, the C23 color almost continuously increases with time, while $F_{25}$ flux generally decreases after a short-lasting increase at the transition phase between AGB and pAGB
\citep[see, e.g.,][]{Szczerba:1993fk, stef98}. Such behavior is due to cooling of circumstellar shell, which is moving away from the central star. However, evolution of the CO\,2-1 line flux, another key factor determining the position of a source on the CO-IR diagram, is neither simple nor investigated theoretically to our knowledge. In this respect the CO-IR diagram serves as an observational tool that allows us to  put constraints on the CO\,2-1 line flux behavior during pAGB phase of stellar evolution.   

PNe are well concentrated along the log-linear region (see bottom left panel in Fig.~\ref{figcompare}) to which ultimately  pAGB stars (with exception of the low mass ones that could disperse their circumstellar shells before the onset of photoionization) should evolve. In the frame of single star evolution, the existence of such trend can be understood if the CO\,2-1 line flux does not change much during late pAGB (pre-PNe) and PNe phase of evolution, while the $F_{25}$ $\mu$m flux density monotonically decreases. However, in the earlier stages of pAGB evolution, the $F_{25}$ continuum (see above) and probably also CO\,2-1 line fluxes could change non-monotonically, resulting in a relatively complex distribution in Fig.~\ref{fig-chem-spec} (e.g., the lack of O-rich pAGB sources in group-I and the presence of C-rich ones among group-III).
\begin{figure}
\epsscale{0.8}
\plotone{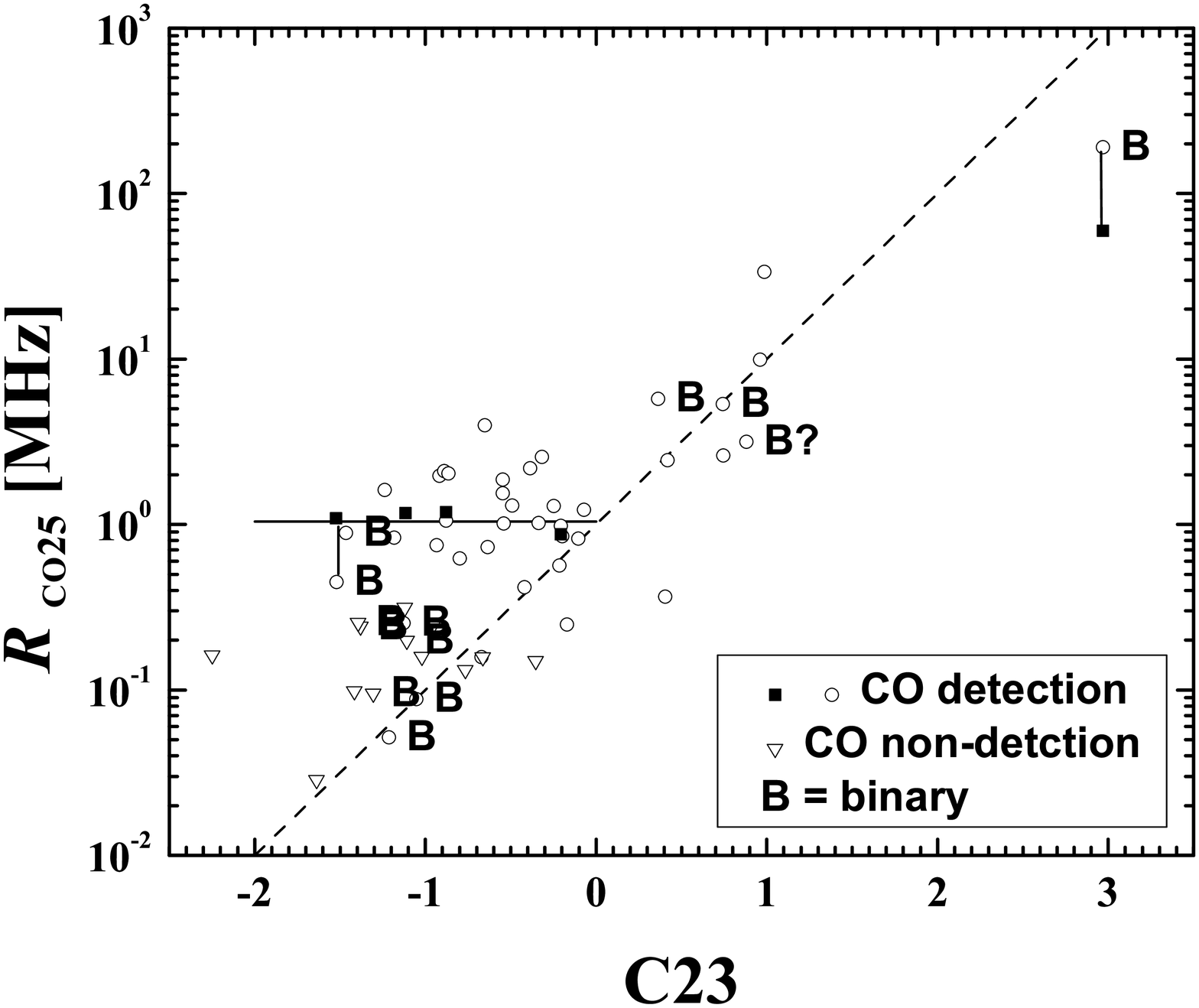}
\caption{The distribution of binary pAGB stars (marked as `B') on the 
CO-IR diagram. Only group-III contains some CO non-detections. The solid and dashed lines represent 
the distribution trends of groups-I, II and III.}
\label{fig-binarity}
\end{figure}

\subsection{SED and Binarity}
\label{binarity}
The SEDs of pAGB stars have been classified according to the scheme introduced by \citet{vand89}, with 
addition of a type 0 \citep{szcz11}. There are six SED types in total: types 0, I, and II, which show significant near infrared (NIR) excess, and types III, IVa, and IVb showing cold dust emission together with a second peak at shorter wavelengths from central star emission. \citet{deru06} proposed that the NIR excess, which is seen in the first three SED types (0, I and II) is a signature of gravitationally bounded circumbinary discs perhaps formed during strong binary interaction \citep{vanw09}. The other three SED types (III, IVa and IVb), are signatures of a detached shells and/or expanding tori, which are formed due to mass loss on AGB, or by interaction of AGB star with its companion \citep{zijl07}, respectively.  

Among our CO-IR group-I sources, three quarters (13 out of 17) have SED types of III, IVa or IVb, indicating that their dust emission is dominated by a detached shell/expanding torus. On the other hand, two third (12 out of 18) of the CO-IR group-III pAGB stars have SED types of 0, I or II, showing that they have excess emission from hot dust, a signature of a disc. It is natural to expect that formation of a disc is due to interaction among stars, when primary star was a giant. 
It is also interesting to note that the pAGB sources that are located in the transition region (those marked by `*' in Table~\ref{tabgrouping}), are showing mostly (8 out of 11) emission from cold (detached) CSEs (with SEDs of type III, IVa, and IVb). The situation in the CO-IR group-II is less evident, as about the same fraction of sources show presence of hot$+$cold or only cold dust.

Information about binarity is directly obtained from the Torun Catalog of pAGB stars. Although such information is by no means complete in the catalog, some interesting features still can be recognized in the distribution of known binary pAGB stars on the CO-IR diagram, as shown in Fig.~\ref{fig-binarity}. 
First, known binaries appear in all three regions of pAGB stars on CO-IR diagram. 
However, most of them appear in the CO-deficient region of group-III. 
In the current sample, about 39 per cent (7 out of 18) group-III pAGB stars are known 
binaries. Among the six C-rich stars that belong to group-III, two are known binary systems (Red Rectangle and RV\,Tau). Binaries are also common in the group-II region (about 40 per cent). What is also striking on Fig.~\ref{fig-binarity} is dichotomy in C23 color distribution of binaries. One, more numerous group, have blue C23 colors (no cold dust), while the second group have red C23 colors (significant amount of cold dust, with Frosty Leo being the most extreme example).

\subsection{CSE expansion velocity}
\label{vexp}

The CSE expansion velocity of all the 42 pAGB stars with detection of CO\,2-1 line show profound trend: $V_{\rm exp}$ is smallest among group-III pAGB stars, intermediate among group-I ones and largest among group-II objects. Excluding few exceptional objects, which will be discussed below, we obtain average velocities from the CO line widths for each group: $12\pm 2$\,km\,s$^{-1}$ for group-I; $28\pm 12$\,km\,s$^{-1}$ for group-II and 
$4\pm 2$\,km\,s$^{-1}$ for group-III. Note that the vast majority of objects located in the transition region (those marked with `*' in Table~\ref{tabgrouping}.) have velocities in between those characteristic for group-I and group-II, with clear exception being IRAS\,17106$-$3046, which have very low  $V_{\rm exp}$ that is more typical for group-III objects. Because the AGB wind velocity is expected to be higher for more massive AGB stars \citep[see, e.g., the discussions of ][]{nyma92}, the trend we found suggests that group-II objects are statistically more massive than group-I pAGB stars. On the other hand, so low expansion velocities for group-III objects suggest that CO is observed from circumstellar disks (rotating and/or expanding) rather than from outflows \citep{Bujarrabal:2005yq, Buja13}. It is very likely that such disks are formed in binary systems \citep[e.g.][]{van-Winckel:2003rt}.

There are two objects in group-I (89\,Her and AI\,Sco), which have very low expansion velocities of their CSE derived from CO lines. Both of them are known to be binaries with 0-type SED, O-rich chemistry, and a blue C23 color close to that typical for O-rich AGB stars. They share characteristics of group-III pAGB objects, but have at the same time  a relatively strong CO 2-1 emission in comparison with their 25$\mu$m continuum flux.

The other two group-I objects (IRAS\,08005$-$2356 and IRAS\,23541$+$7031), on the other hand, have very high expansion velocities, much larger even than those typical for group-II pAGB objects. IRAS\,08005$-$2356 has a broad CO line, which was only tentatively detected by \citet{1994AAS..103..301H}. However, similarly high velocities are also detected in optical spectrum of this star by \citet{slij91}, and \citet{Sanc08}, which may be interpreted as a signature of ongoing fast wind (or jet) from this source. Its type-I SED indicates significant near infrared emission, which may be interpreted as emission from hot dust in a disc, but its 
$R_{\rm CO25}=5.20$ is the largest among group-I pAGB stars. IRAS\,23541$+$7031 has a disk or torus of molecular gas, which seems to be expanding  \citep{Castro-Carrizo:2002ly}. It has very rich and complex wind activity and rapidly evolving shocked material \citep{Sanchez-Contreras:2010gf}. In this respect, a broad CO 2-1 line detected by \cite{1996ApJ...467..341H} is not surprising.

Roberts\,22 and IRAS\,17245-3951 (Walnut Nebula) are another two special objects with normal or high CSE expansion velocities (31 and 15 \,km\,s$^{-1}$, respectively), but this time being members of our group-III objects ($R_{\rm CO25}=0.16$ and 0.25, respectively).  Again, the broad CO lines may be interpreted by the contribution from bipolar outflows, because both show typical bipolar nebulae. We note that Roberts\,22 was known to show broadened $H_\alpha$ emission line and was thus suspected to be a Wolf-Rayet star \citep{robe62}. 

\subsection{Post-AGB sub-types}
\label{pagb-type}

Finally, we would like to put attention to the fact that, among sources from Table~\ref{tabgrouping}, there are two pAGB classes \citep{szcz07} that are the most abundant.  They are `21 micron' sources (12 objects) and RV\,Tau stars (8 sources). These two groups have quite well defined properties. The `21 micron' sources are a group of C-rich intermediate mass stars with the still unidentified 21\,$\mu$m dust feature \citep{Kwok:1989ul}, which also show s-process elements enhancement \citep{Van-Winckel:2000qf}, and have periodic variability well correlated with their effective temperature \citep{Hrivnak:2010pd}. RV\,Tau stars are luminous variable stars, which show alternating deep and shallow minima with periods between 30 and 150 days, and spectral types F, G and K \citep[e.g.][]{Preston:1963bh}. Most of them show IR excess, which is interpreted as a remnant of AGB mass loss \citep{Jura:1986lq}. RV\,Tau stars with near-IR excess (very likely due to radiation from disks) are probably binaries \citep{de-Ruyter:2005dq}. 

These two groups have well established positions on the CO-IR diagram. Almost all members of the `$21\,\micron$' type sources from our sample are in our group-I, with one belonging to the transition region and only one to group-III (IRAS\,19477$+$2401). It is remarkable that 10 (out of 17, or 59 per cent) objects in group-I are `$21\,\micron$' sources. Their IR colors are in the range $-1.3<{\rm C23}<-0.55$. All but one have a G or F spectral type, 9 (among 12) have a SED of IVa or IVb type, none of them is known to be binary, their outflow velocities are around 10\,km\,s$^{-1}$ (with an average of $12\pm 3$\,km\,s$^{-1}$). 

On the other hand, 7 (out of 8) RV\,Tau stars from our sample belong to group-III objects, with only one source (AI Sco) being a member of group-I on the CO-IR diagram. This is not so surprising since RV\,Tau variables are well-known to be CO-deficient stars \citep{1991AA...245..499A}. RV\,Tau stars from our sample consist of the majority of the blue part of the group-III objects, with IR colors in a narrow range of $-1.5<{\rm C23}<-1.1$. They have O-rich chemistry,  7 (out of 8) have SED classified as 0 or I (with AR\,Pup being the only exception with SED of type-III), 7 of them are also known binaries (with the only exception of AI\,CMi, which do not have near-IR excess - see e.g. SED in the Torun catalog), and the three RV\,Tau variables with detected CO\,2-1 line all show very narrow CO lines (with $V_{\rm exp}=1.5 \sim 5.6$\,km\,s$^{-1}$).

\subsection{Overall properties of the CO-IR groups of post-AGB stars}
\label{interpret}

\begin{figure}
\epsscale{0.8}
\plotone{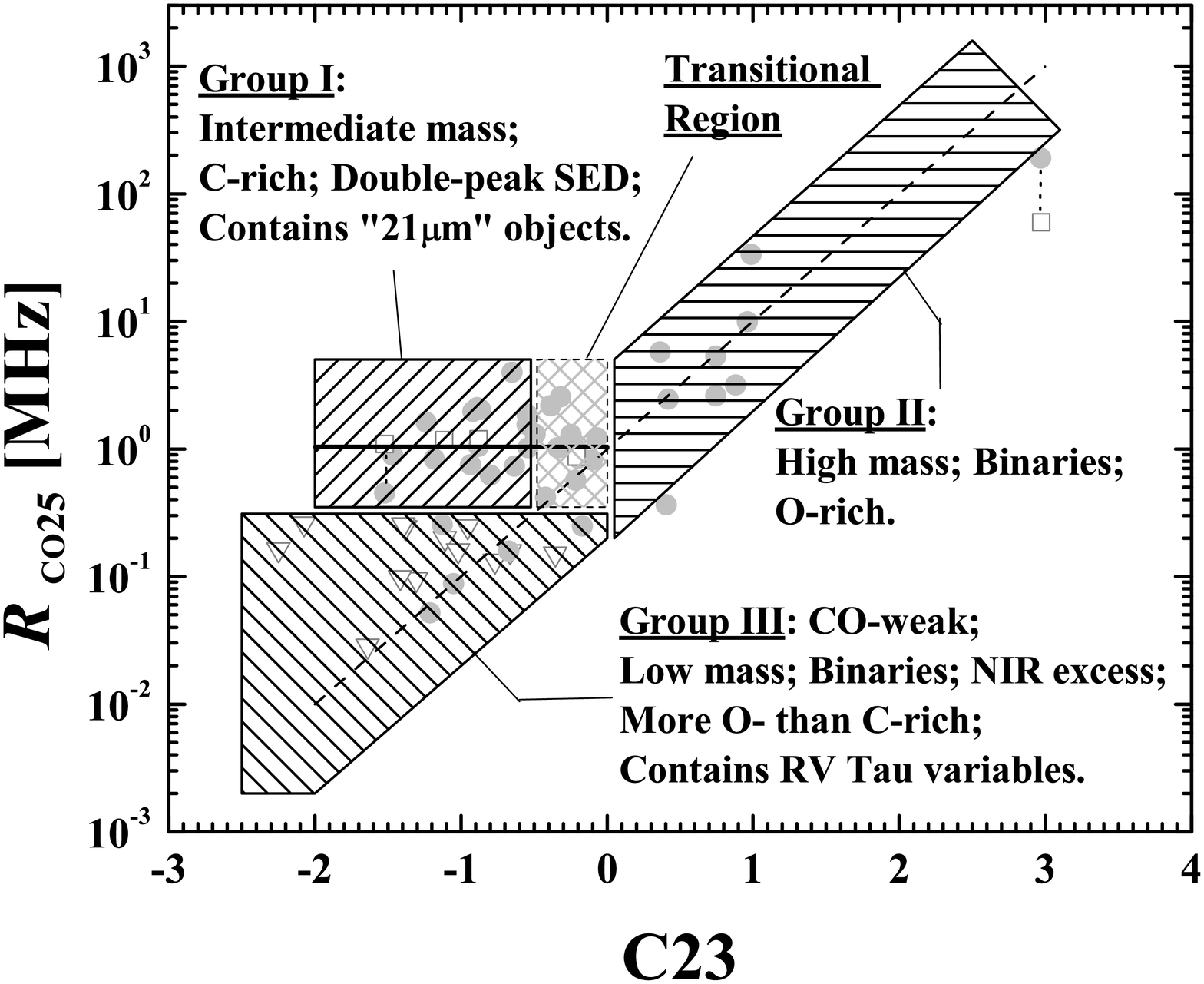}
\caption{The cartoon illustrates the main regions and the key features of the three pAGB CO-IR groups.
  Observed pAGB stars are shown as gray points.}
\label{cartoon}
\end{figure}

We summarize below the properties of the different groups of pAGB objects, as inferred from the above discussion. The key features of each CO-IR group are collected in Table~\ref{group-summary} and illustrated in a cartoon (Fig.~\ref{cartoon}). 
\begin{deluxetable}{lllclrlll}
\tabletypesize{\scriptsize}
\tablewidth{0pc} 
\tablecaption{Summary of the bulk properties of the three CO-IR groups.
\label{group-summary}}
\tablehead{ 
\colhead{Group} &\colhead{C23} &\colhead{$R_{\rm CO25}$} &\colhead{Chemistry} &\colhead{Sp.Type} &\colhead{Binary} &\colhead{SED.Type} &\colhead{$V_{\rm exp}$} &Special objects \\
\colhead{}      &\colhead{}    &\colhead{}               &\colhead{}           &\colhead{}        &\colhead{}       &\colhead{}         &\colhead{km\,s$^{-1}$}  &\colhead{}    }
\startdata 
I          &$-1.52\sim -0.5$   &$0.42\sim 5.2$           &C                    &F, G              &$7\%$            &III,IVa,IVb        &$12\pm 2$               &$21\,\micron$ objects \\
II         &$+0.36\sim +2.97$  &$0.37\sim 59.62$         &O or C               &K to B            &$44\%$           &0,II,III,IVb       &$28\pm 12$              &                      \\
III        &$-2.25\sim -0.17$  &$<0.26$                  &O                    &M to B            &$39\%$           &0,I,II             &$4\pm 2$                &RV\,Tau variables     \\
\hline 
\enddata 
\end{deluxetable} 

\subsubsection{Group-I}
\label{groupI}

The group-I pAGB stars have a relatively narrow range of $R_{\rm CO25}$ ratios from 0.42 to 5.2\,MHz, with a median of $\sim 1$\,MHz and C23 color varying between -1.5 and -0.5. 
They represent the early pAGB stage of intermediate mass pAGB stars,
because of their F to G spectral types and
C-rich chemistry.
This group contains almost all of the considered `$21\,\micron$' objects.

\subsubsection{Group-II}
\label{groupII}

The group-II pAGB stars are the reddest objects in our sample. They are distributed along
a log-linear trend on the CO-IR diagram, which coincides with a similar trend of PNe. 
On average, group-II objects are more massive pAGB stars than group-I sources, due to their 
higher $V_{\rm exp}$. However, this region should also contain evolved intermediate mass pAGB stars that are approaching PNe phase.

\subsubsection{Group-III}
\label{groupIIICOpoor}

The group-III pAGB objects are those with similar C23 colors as group-I objects but with much weaker 
CO\,2-1 line. They are probably the lowest mass pAGB stars, predominantly O-rich, 
with very high percentage of known binaries, which is in agreement
with their type-0, I and II SED (near-IR excess). Most of the RV\,Tau variables are in this group.

\subsubsection{Transition region}
\label{transitionregion}
The objects in this region have C23 colors intermediate between groups-I and II. They are predominantly O-rich as group-II objects, but have $R_{CO25}$ ratios similar as group-I objects. Their CSE expansion velocities are also intermediate between the two groups. 

\section{SUMMARY}
\label{summary}

A survey of CO\,2-1 line has been performed toward 58 known high Galactic latitude
pAGB stars. Circumstellar CO lines were detected only
toward five of these objects, with one new detection. The detected line profiles show various features
such as line wings, absorption features and triangular peak. 

To complete our survey, we performed a compilation of literature reporting single dish CO\,2-1 line observation for all 393 pAGB stars (likely, RCrB-eHe-LTP and RV\,Tau types 
from the Torun Catalog of pAGB stars). We found observations for
133 pAGB stars (34 per cent of all objects in the Torun Catalog).
The CO\,2-1 line has been detected in 44 objects among them.
The CO\,2-1 line data for AGB stars and PNe
were also compiled for comparison. 

CO-IR flux ratio
$R_{\rm CO25}$ is defined using the integrated CO\,2-1 line flux and {\it IRAS} $25\micron$ flux density. 
This ratio is compared with IR color C23 that is defined with the {\it IRAS} $60$ to $25\micron$ flux ratio. So constructed CO-IR diagram is used to investigate the pAGB phase of stellar evolution.

Post-AGB stars segregate into three groups on the CO-IR diagram: 
group-I pAGB stars show a narrow range of $R_{\rm CO25}$ ratios that are independent of the C23 colors (Eq.~\ref{eq-I-25}); 
groups-II pAGB stars have redder C23 colors and usually larger $R_{\rm CO25}$ ratios; 
group-III pAGB stars have significantly smaller $R_{\rm CO25}$ ratios (weak CO lines). 

Comparison of the pAGB stars with AGB stars and PNe on the CO-IR diagram reveals 
that post AGB objects are really located between the AGB stars and PNe. Planetary nebulae show a profound trend on the CO-IR diagram which agrees well with the trend of group-II pAGB stars. 

Combining these features with various properties such as chemical types, central star spectral types, 
binarity, SED types, CSE expansion velocities and pAGB sub-types 
(defined in the Torun Catalog) of the pAGB stars, the three pAGB CO-IR groups are found to have distinctive characteristics related to mass, evolutionary stage and binarity.

The CO-IR diagram is proven to be a powerful tool to 
discriminate the different effects of stellar mass, evolution and binarity of pAGB stars 
and to investigate the co-evolution of circumstellar gas and dust 
during the fast post-AGB stage. 



\acknowledgments

This research has intensively made use of NASA's Astrophysics Data System Bibliographic Services and 
SIMBAD database, operated at CDS, Strasbourg, France.
We thank ARO telescope operators for the assistance in remote
observations. The SMT is
operated by the Arizona Radio Observatory (ARO), Steward Observatory,
University of Arizona. J.H. thanks Dr. T.M., Dame for kindly offering
additional information from their Galactic CO\,1-0 line survey to help us identify possible 
interstellar contamination                                                                                                                . 

This work benefited from the collaboration among the coauthors funded by Marie Curie Actions
International Research Staff Exchange Scheme (project No. 269193) of
European Union. J.H. also thanks the projects (No. 11173056 and 11033008) of
the National Natural Science Foundation of China. R.Sz. and M.S. acknowledge the support from grants 2011/01/B/ST9/02031
and 2011/01/B/ST9/02229. T.I. acknowledges
the support from NSC grant NSC 96-2112-M-001-018-MY3. 



{\it Facilities:} \facility{HHT ()}.



\appendix

\section{COMPILATION OF LITERATURE CO\,2-1 LINE DATA FOR AGB, pAGB STARS AND PNe}
\label{app-lit-data}

\begin{sidewaystable*}[htbp]
\scriptsize
\centering
\caption{ Literature CO\,2-1 line data of post-AGB stars. }
\label{lit-pAGB-CO}
\begin{tabular}{l@{ }l@{}c@{}r@{ }c@{ }l@{ }l@{ }llllll}
\hline\hline
IRAS &
Other name &
Chem &
Tp($\sigma$) &
NoteT$^{a}$ &
I($\sigma$) &
$V_{\rm sys}(\sigma)$ &
$V_{\rm exp}(\sigma)$ &
Tel$^{b}$ &
Tscale&
Obs.mode$^{c}$ &
Obs.Date &
Literature \\
 & & & mK & & K\,km\,s$^{-1}$ & km\,s$^{-1}$ & km\,s$^{-1}$ & & & & yyyy-mm-dd & \\
\hline
             & CRL 2688          &      & 1300       &      &             & -28  (3)   & 31        & OVRO    & TA* & S    & 1978-03,04                  & 1979ApJ...230..149W \\
             & CRL 2688          & C    & 2300 (500) &      &             &            &           & OVRO    & TA* & MapC & 1978-11 to 12,1979-11 to 12 & 1982ApJ...252..616K \\
             & CRL 2688          &      & 4200       &      &             &            &           & ARO12m  & TA* & S    & 1985-02,04                  & 1987ApJ...319..367W \\
             & CRL 2688          &      &            &      & 185  (2)    & -36.0      & 17.9(2.3) & IRAM30m & Tmb & S    & 1988-08                     & 1992A\&A...256..235 \\
             & CRL 2688          &      &            &      & 312.5(1.3)  & -35.6      & 60        & IRAM30m & Tmb & S    & 1991-06                     & 1997A\&A...324.1123 \\
             & CRL 2688          &      & 5890 (20)  &      & 151.7       & -35.7      & 31        & JCMT    & TA* & S    & 1995-05                     & 1997A\&A...327..342 \\
             & CRL 2688          &      &            &      & 312.5       &            &           & IRAM30m & Tmb & S    & 1996-11,1997-05             & 2000A\&A...355...69 \\
             & CRL 2688          & C    & 16670      & Jy/K &             &            &           & IRAM30m & Tmb & S    & 2000-09                     & 2002ApJ...577..961H \\
             & HD 107369         &      &      (200) &      &             &            &           & SEST    & Tmb & S    & 1989-05,1990-07             & 1993A\&A...269..231 \\
             & V4334 Sgr         &      &      (23)  &      &             &            &           & JCMT    & TA* & S    & 1997-06-29                  & 1998A\&A...335..292 \\
01005+7910   &                   &      &      (50)  &      &             &            &           & IRAM30m & Tmb & S    & 1989-01                     & 1993A\&A...267..515 \\
01005+7910   &                   & C    &      (16)  &      &             &            &           & SMT10m  & Tmb & S    & 1999-12-03 to 2000-02-04    & 2005ApJ...624..331H \\
02143+5852   &                   &      &      (248) & cont &             &            &           & IRAM30m & Tmb & S    & 1989-01                     & 1993A\&A...267..515 \\
02143+5852   &                   &      &      (14)  & cont &             &            &           & SMT10m  & Tmb & S    & 1999-12-03 to 2000-02-04    & 2005ApJ...624..331H \\
04166+5719   & TW Cam            & O    &      (50)  &      &             &            &           & IRAM30m & Tmb & S    & 1989-04 to 06               & 1991A\&A...245..499 \\
04296+3429   &                   &      & 400        &      &             & -62.3      & 15.6      & JCMT    & TR* & S    & 1988-12                     & 1990A\&A...228..503 \\
04296+3429   &                   & C    & 450  (128) &      & 8.3         & -66        & 12.0      & IRAM30m & Tmb & MapC & 1989-01                     & 1993A\&A...267..515 \\
04296+3429   &                   & C    & 70   (17)  &      & 1.12        & -64.8      & 10.8      & SMT10m  & Tmb & S    & 1999-12-03 to 2000-02-04    & 2005ApJ...624..331H \\
04395+3601   & CRL 618           & C    & 2100 (800) &      &             & -29.0(3)   & 17.0(2)   & OVRO    & TA* & MapC & 1978-11 to 12,1979-11 to 12 & 1982ApJ...252..616K \\
04395+3601   & CRL 618           &      & 2000       &      &             &            &           & ARO12m  & TA* & S    & 1985-02,04                  & 1987ApJ...319..367W \\
04395+3601   & CRL 618           &      & 2200 (150) &      & 53.9        & -22        & 18        & ARO12m  & Tmb & S    & 1985-12 to 1988-01          & 1989ApJ...346..201H \\
04395+3601   & CRL 618           &      & 3500 (90)  &      &             & -21.7(0.3) & 18.6(0.4) & ARO12m  & TA* & S    & 1987-01-13 to 16            & 1990ApJ...358..251W \\
04395+3601   & CRL 618           &      & 9300 (160) &      & 250.0       & -21.9      & 11.6      & IRAM30m & Tmb & MapC & 1987-02                     & 1988A\&A...196L...5 \\
04395+3601   & CRL 618           &      &            &      & 161  (1)    & -21.4      & 18.5(1.7) & IRAM30m & Tmb & S    & 1988-08                     & 1992A\&A...256..235 \\
04395+3601   & CRL 618           &      & 8640       & 2C   & 270.0(3.4)  & -21.9(0.2) & 18.2(0.5) & IRAM30m & Tmb & S    & 1988-08                     & 1989A\&A...222L...1 \\
04395+3601   & CRL 618           &      & 5160 (200) & 2C   &             & -22.6(1.0) & 19.1(0.8) & CSO     & Tmb & S    & 1988-08,1989-11             & 1989ApJ...345L..87G \\
04395+3601   & CRL 618           &      &            &      & 182.2(1.1)  & -21.5      & 40        & IRAM30m & Tmb & S    & 1991-06                     & 1997A\&A...324.1123 \\
04395+3601   & CRL 618           &      &            &      & 46.2 (5.1)  &            &           & ARO12m  & Tmb & MapC & 1994-11-30                  & 1998ApJ...509..392M \\
04395+3601   & CRL 618           &      &            &      & 182.2       &            &           & IRAM30m & Tmb & S    & 1996-11,1997-05             & 2000A\&A...355...69 \\
04395+3601   & CRL 618           & C    & 2400 (90)  &      & 60.0 (0.8)  & -22.0      & 23.9      & ARO12m  & Tmb & S    & 1997-05,1997-12             & 2002ApJ...572..326B \\
04395+3601   & CRL 618           & C    & 11190      & Jy/K &             &            &           & IRAM30m & Tmb & S    & 2000-09                     & 2002ApJ...577..961H \\
04440+2605   & RV Tau            &      &      (37)  &      &             &            &           & ARO12m  & TA* & S    & 1987-01-13 to 16            & 1990ApJ...358..251W \\
04440+2605   & RV Tau            &      &      (60)  &      &             &            &           & IRAM30m & Tmb & S    & 1988-06                     & 1988A\&A...206L..17 \\
04440+2605   & RV Tau            & O    &      (30)  &      &             &            &           & IRAM30m & Tmb & S    & 1989-04 to 06               & 1991A\&A...245..499 \\
05338-3051   & RV Col            &      &      (200) &      &             &            &           & SEST    & Tmb & S    & 1989-05,1990-07             & 1993A\&A...269..231 \\
06034+1354   & DY Ori            &      &      (30)  &      &             &            &           & IRAM30m & Tmb & S    & 1989-04 to 06               & 1991A\&A...245..499 \\
06072+0953   & CT Ori            &      &      (40)  &      &             &            &           & IRAM30m & Tmb & S    & 1989-04 to 06               & 1991A\&A...245..499 \\
06108+2743   & SU Gem            &      &      (26)  &      &             &            &           & ARO12m  & TA* & S    & 1987-01-13 to 16            & 1990ApJ...358..251W \\
06108+2743   & SU Gem            & O    &      (100) &      &             &            &           & IRAM30m & Tmb & S    & 1989-04 to 06               & 1991A\&A...245..499 \\
06176-1036   & Red Rectangle     &      &      (160) &      &             &            &           & ARO12m  & Tmb & S    & 1984-12                     & 1986ApJ...304..394Z \\
06176-1036   & Red Rectangle     &      & 1500       & hand & 6.6         & +0.5       & 2.2       & IRAM30m & Tmb & S    & 1993-04                     & 1995ApJ...453..721J \\
06176-1036   & Red Rectangle     &      &      (18)  &      & 1.36        & -1         & 11        & ARO12m  & Tmb & S    & 1994-01                     & 1996ApJ...472..703D \\
06176-1036   & Red Rectangle     &      & 260  (20)  &      & 1.80        & +0.2       & 9         & JCMT    & TA* & S    & 1995-05                     & 1997A\&A...327..342 \\
06530-0213   &                   &      & 610  (110) &      & 7.0         & +33        & 10        & SEST    & Tmb & S    & 1991-04                     & 1994A\&AS..103..301 \\
06530-0213   &                   & C    & 120  (12)  &      & 2.47        & +31.4      & 12.5      & SMT10m  & Tmb & S    & 1999-12-03 to 2000-02-04    & 2005ApJ...624..331H \\
07134+1005   & HD 56126          & C    & 3210 (97)  &      & 43.9        & +73        & 10.2      & IRAM30m & Tmb & S    & 1989-01                     & 1993A\&A...267..515 \\
07134+1005   & HD 56126          & C    & 2200 (100) &      & 28          & +73        &           & IRAM30m & Tmb & MapC & 1990-04 to 06,1991-01       & 1992A\&A...257..701 \\
07134+1005   & HD 56126          & C    & 640  (70)  &      & 9.2  (0.9)  & +73.0(0.8) & 10.7(1.1) & CSO     & Tmb & S    & 1996-12-28 to 1997-01-02    & 1998ApJS..117..209K \\
07134+1005   & HD 56126          & C    & 630  (5)   &      & 9.03        & +73.5      & 10.6      & SMT10m  & Tmb & S    & 1999-12-03 to 2000-02-04    & 2005ApJ...624..331H \\
07284-0940   & U Mon             &      &      (33)  &      &             &            &           & ARO12m  & TA* & S    & 1987-01-13 to 16            & 1990ApJ...358..251W \\
\hline
\end{tabular}
\begin{tablenotes}                                                                                            
\item  [1] 
Note: Repeated observations of the same object are listed as separate entries. Our observations of high Galactic latitutde pAGB objects are not included.                                               
\end{tablenotes}                                                                                              
\clearpage
\end{sidewaystable*}
\begin{sidewaystable*}[htbp]
\scriptsize
\centering
\addtocounter{table}{-1}
\caption{ (continued)}
\begin{tabular}{l@{ }l@{}c@{}r@{ }c@{ }l@{ }l@{ }llllll}
\hline\hline
IRAS &
Other name &
Chem &
Tp($\sigma$) &
NoteT$^{a}$ &
I($\sigma$) &
$V_{\rm sys}(\sigma)$ &
$V_{\rm exp}(\sigma)$ &
Tel$^{b}$ &
Tscale&
Obs.mode$^{c}$ &
Obs.Date &
Literature \\
 & & & mK & & K\,km\,s$^{-1}$ & km\,s$^{-1}$ & km\,s$^{-1}$ & & & & yyyy-mm-dd & \\
\hline
07284-0940   & U Mon             &      &      (150) &      &             &            &           & IRAM30m & Tmb & S    & 1988-06                     & 1988A\&A...206L..17 \\
07284-0940   & U Mon             & O    &      (70)  &      &             &            &           & IRAM30m & Tmb & S    & 1989-04 to 06               & 1991A\&A...245..499 \\
07331+0021   & AI CMi            & O    &      (56)  &      &             &            &           & IRAM30m & Tmb & S    & 1989-01                     & 1993A\&A...267..515 \\
07399-1435   & OH231.8+4.2       & O    & 3600       & hand &             & +33        & 30  (5)   & SEST    & Tmb & MapC & 1996-11                     & 1997A\&A...327..689 \\
07399-1435   & OH231.8+4.2       & O    & 580  (20)  &      &             & +36.2(2.6) & 41.6(2.6) & SMT10m  & TA* & S    & 2003-10 to 2007-06          & 2009ApJ...690..837M \\
07430+1115   &                   &      &      (90)  &      &             &            &           & IRAM30m & Tmb & S    & 1989-01                     & 1993A\&A...267..515 \\
07430+1115   &                   & C    &      (13)  &      &             &            &           & SMT10m  & Tmb & S    & 1999-12-03 to 2000-02-04    & 2005ApJ...624..331H \\
08005-2356   &                   &      & 120  (60)  &      & 14          & +50        & 100       & SEST    & Tmb & S    & 1991-04                     & 1994A\&AS..103..301 \\
08011-3627   & AR Pup            & O    &      (200) &      &             &            &           & IRAM30m & Tmb & S    & 1989-04 to 06               & 1991A\&A...245..499 \\
08187-1905   & HD 70379          &      &      (18)  &      &             &            &           & SMT10m  & Tmb & S    & 1999-12-03 to 2000-02-04    & 2005ApJ...624..331H \\
08544-4431   & V390 Vel          & C    & 140  (10)  &      & 1.2         & +45.6      & 8.0       & SEST    & Tmb & S    & 1999-10,2001-03             & 2002A\&A...390..501 \\
08544-4431   & V390 Vel          &      &            &      & 3.0         & +61.3      & 6.4       & SEST    & Tmb & S    & 1999-10-02,2000-07-28       & 2003A\&A...405..271 \\
09032-3953   &                   &      & 180  (60)  &      & 6.7         & +36        & 25        & SEST    & Tmb & S    & 1991-04                     & 1994A\&AS..103..301 \\
09370-4826   &                   &      &      (160) &      &             &            &           & SEST    & Tmb & S    & 1991-04                     & 1994A\&AS..103..301 \\
09371+1212   & FROSTY LEONIS     &      & 176  (7)   &      &             & -5.6 (0.4) & 36.6(0.5) & ARO12m  & TA* & S    & 1987-01-13 to 16            & 1990ApJ...358..251W \\
10158-2844   & HR 4049           &      &      (200) &      &             &            &           & SEST    & Tmb & S    & 1989-05,1990-07             & 1993A\&A...269..231 \\
10158-2844   & HR 4049           & C    &      (40)  &      &             &            &           & IRAM30m & Tmb & S    & 1990-04 to 06,1991-01       & 1992A\&A...257..701 \\
10178-5958   & Hen 3-401         &      & 190  (30)  &      & 5.2         & -29.0      & 15        & SEST    & Tmb & S    & 19??-09-01 to 04            & 1991A\&A...242..247 \\
10194-5625   &                   &      &      (160) &      &             &            &           & SEST    & Tmb & S    & 1991-04                     & 1994A\&AS..103..301 \\
10197-5750   & Roberts 22        &      & 150  (40)  &      & 9.0         & -0.1       &           & SEST    & Tmb & S    & 19??-09-01 to 04            & 1991A\&A...242..247 \\
11000-6153   & HD 95767          &      &      (200) &      &             &            &           & SEST    & Tmb & S    & 1989-05,1990-07             & 1993A\&A...269..231 \\
11381-6401   &                   &      &      (190) &      &             &            &           & SEST    & Tmb & S    & 1991-04                     & 1994A\&AS..103..301 \\
11385-5517   & HD 101584         & O    & 300        &      & 41.3        & +44        & 145.      & SEST    & Tmb & MapC & 1997-08                     & 1999A\&A...347..194 \\
11544-6408   &                   &      &      (140) &      &             &            &           & SEST    & Tmb & S    & 1991-04                     & 1994A\&AS..103..301 \\
12067-4508   & HD 105578         &      &      (200) &      &             &            &           & SEST    & Tmb & S    & 1989-05,1990-07             & 1993A\&A...269..231 \\
12222-4652   & HD 108015         &      &      (200) &      &             &            &           & SEST    & Tmb & S    & 1989-05,1990-07             & 1993A\&A...269..231 \\
12419-5414   & Boomerang Nebula  &      & 160        & hand &             & -10        & 35        & SEST    & Tmb & MapC & 1994-08,1995-08 to 1995-10  & 1997ApJ...487L.155S \\
12419-5414   & Boomerang Nebula  &      & 150  (20)  &      & 5.0         & -3.5       &           & SEST    & Tmb & S    & 19??-09-01 to 04            & 1991A\&A...242..247 \\
14488-5405   & CPD -53 5736      &      & 90         &      & 2.07        & -10        & 10.0      & SEST    & Tmb & S    & 1998 to 2002                & 2005A\&A...429..977 \\
14524-6838   & HD 131356         &      &      (200) &      &             &            &           & SEST    & Tmb & S    & 1989-05,1990-07             & 1993A\&A...269..231 \\
15465+2818   & R CrB             &      &      (100) &      &             &            &           & ARO12m  & Tmb & S    & 1984-12                     & 1986ApJ...304..394Z \\
15465+2818   & R CrB             &      &      (41)  &      &             &            &           & ARO12m  & TA* & S    & 1987-01-13 to 16            & 1990ApJ...358..251W \\
16342-3814   & OH 344.1 +5.8     & O    & 29   (7)   &      & 2.14 (0.14) & +44  (1)   & 46  (1)   & SMT10m  & Tmb & S    & 2008-05-09                  & 2008A\&A...488L..21 \\
16594-4656   &                   &      & 1750       &      & 32.60       & -25        & 10.0      & SEST    & Tmb & S    & 1998 to 2002                & 2005A\&A...429..977 \\
17106-3046   &                   &      & 300        &      & 1.84        & 0          & 2         & SEST    & Tmb & S    & 1998 to 2002                & 2005A\&A...429..977 \\
17150-3224   & RAFGL 6815        &      & 1000 (300) &      & 18.6        & +15        & 15        & JCMT    & Tmb & S    & 1990-11,12                  & 1994A\&AS..103..301 \\
17150-3224   & RAFGL 6815        & O    & 1000       &      &             & +15        & 15        & JCMT    & Tmb & S    & 1990-12                     & 1993A\&A...273..185 \\
17150-3224   & RAFGL 6815        &      & 650        &      & 12.84       & +15        &           & SEST    & Tmb & S    & 1998 to 2002                & 2005A\&A...429..977 \\
17245-3951   & OH348.8-2.8       &      & 40         &      & 0.58        & 0          & 15        & SEST    & Tmb & S    & 1998 to 2002                & 2005A\&A...429..977 \\
17423-1755   & Hen 3-1475        &      & 280  (40)  &      & 15.9 (0.5)  & +47.5(0.9) & 53  (2)   & IRAM30m & Tmb & S    & 1997-09                     & 2004A\&A...414..581 \\
17436+5003   & V814 Her          &      & 370        &      &             & -37.5      & 7.7       & ARO12m  & Tmb & S    & 1986-04,1987-04             & 1989A\&A...209..119 \\
17436+5003   & V814 Her          &      & 1800       &      & 33          & -35.5      & 13.2      & IRAM30m & Tmb & S    & 1987-10                     & 1991A\&A...246..153 \\
17436+5003   & V814 Her          & O    & 2000 (40)  &      & 36.1        & -35        &           & IRAM30m & Tmb & MapC & 1990-04 to 06,1991-01       & 1992A\&A...257..701 \\
17436+5003   & V814 Her          & O    & 1500       &      & 29.3        & -34.7      & 11.7      & IRAM30m & Tmb & S    & 1993-08                     & 1998A\&AS..130....1 \\
17436+5003   & V814 Her          & O    & 230  (6)   &      & 4.24        & -34.8      & 13.8      & SMT10m  & Tmb & S    & 1999-12-03 to 2000-02-04    & 2005ApJ...624..331H \\
17441-2411   & AFGL 5385         &      & 670  (160) &      & 8.3         & +5         & 15        & SEST    & Tmb & S    & 1991-04                     & 1994A\&AS..103..301 \\
17441-2411   & AFGL 5385         &      & 570        &      & 6.24        & +110       &           & SEST    & Tmb & S    & 1998 to 2002                & 2005A\&A...429..977 \\
17530-3348   & AI Sco            &      &      (600) &      &             &            &           & IRAM30m & Tmb & S    & 1988-06                     & 1988A\&A...206L..17 \\
17530-3348   & AI Sco            & O    & 300  (100) &      & 1.8         & -37        &           & IRAM30m & Tmb & S    & 1989-04 to 06               & 1991A\&A...245..499 \\
17534+2603   & 89 Her            &      & 600        &      & 4.1         & -7.9       & 3.2       & IRAM30m & Tmb & S    & 1987-10                     & 1991A\&A...246..153 \\
\hline
\end{tabular}
\clearpage
\end{sidewaystable*}
\begin{sidewaystable*}[htbp]
\scriptsize
\centering
\addtocounter{table}{-1}
\caption{ (continued)}
\begin{tabular}{l@{ }l@{}c@{}r@{ }c@{ }l@{ }l@{ }llllll}
\hline\hline
IRAS &
Other name &
Chem &
Tp($\sigma$) &
NoteT$^{a}$ &
I($\sigma$) &
$V_{\rm sys}(\sigma)$ &
$V_{\rm exp}(\sigma)$ &
Tel$^{b}$ &
Tscale&
Obs.mode$^{c}$ &
Obs.Date &
Literature \\
 & & & mK & & K\,km\,s$^{-1}$ & km\,s$^{-1}$ & km\,s$^{-1}$ & & & & yyyy-mm-dd & \\
\hline
17534+2603   & 89 Her            &      & 1000 (70)  &      & 4.6         & -7.9       & 4         & IRAM30m & Tmb & S    & 1989-04 to 06               & 1991A\&A...245..499 \\
17534+2603   & 89 Her            &      & 200  (50)  &      &             & -8.0 (1.3) & 8.8 (1.3) & SMT10m  & TA* & S    & 2003-10 to 2007-06          & 2009ApJ...690..837M \\
18025-3906   &                   &      &      (130) &      &             &            &           & SEST    & Tmb & S    & 1991-04                     & 1994A\&AS..103..301 \\
18095+2704   &                   & O    &      (30)  &      &             &            &           & IRAM30m & Tmb & S    & 1990-04 to 06,1991-01       & 1992A\&A...257..701 \\
18276-1431   & OH17.7-2.0        &      &      (100) &      &             &            &           & CSO     & TA* & S    & 1987-07                     & 1989ApJ...336..822K \\
18276-1431   & OH17.7-2.0        & O    & 1000 (170) &      & 19.2 (1.0)  & +61.6      & 12.2      & IRAM30m & Tmb & S    & 1988-02                     & 1990A\&A...239..173 \\
18281+2149   & AC Her            &      &      (20)  &      &             &            &           & ARO12m  & TA* & S    & 1987-01-13 to 16            & 1990ApJ...358..251W \\
18281+2149   & AC Her            &      & 200        &      & 0.7         &            &           & IRAM30m & Tmb & S    & 1988-06                     & 1988A\&A...206L..17 \\
18281+2149   & AC Her            & C    & 100  (300) &      & 0.5         & -10        &           & IRAM30m & Tmb & S    & 1989-04 to 06               & 1991A\&A...245..499 \\
18372-2257   & V348 Sgr          &      &      (150) &      &             &            &           & SEST    & Tmb & S    & 1989-05,1990-07             & 1993A\&A...269..231 \\
18415-2100   & MV Sgr            &      &      (200) &      &             &            &           & SEST    & Tmb & S    & 1989-05,1990-07             & 1993A\&A...269..231 \\
19114+0002   & AFGL 2343         &      & 380        &      &             & +94.3      & 33.7      & ARO12m  & Tmb & S    & 1986-04,1987-04             & 1989A\&A...209..119 \\
19114+0002   & AFGL 2343         & O    & 2520 (107) &      & 126.8       & +99        & 33.7      & IRAM30m & Tmb & S    & 1989-01                     & 1993A\&A...267..515 \\
19114+0002   & AFGL 2343         &      & 470  (70)  &      & 22.8 (0.5)  & +99.0(0.2) & 28.1(0.2) & SEST    & Tmb & S    & 1989-05,1990-07             & 1993A\&A...269..231 \\
19114+0002   & AFGL 2343         & O    & 3400 (100) &      & 171         & +99        &           & IRAM30m & Tmb & MapC & 1990-04 to 06,1991-01       & 1992A\&A...257..701 \\
19114+0002   & AFGL 2343         & O    & 3450 (80)  &      & 172.4       & +97        & 37        & IRAM30m & Tmb & MapC & 1999-08                     & 2001A\&A...367..826 \\
19132-3336   & RY Sgr            &      &      (300) &      &             &            &           & SEST    & Tmb & S    & 1989-05,1990-07             & 1993A\&A...269..231 \\
19306+1407   &                   &      &      (77)  &      &             &            &           & IRAM30m & Tmb & S    & 1989-01                     & 1993A\&A...267..515 \\
19343+2926   & PN M1-92          & O    & 1100       &      & 56.6        & +0.3       & 32.5      & IRAM30m & Tmb & S    & 1990-10                     & 1998A\&AS..130....1 \\
19374+2359   &                   &      &      (123) & cont &             &            &           & IRAM30m & Tmb & S    & 1989-01                     & 1993A\&A...267..515 \\
19374+2359   &                   &      &      (200) &      &             &            &           & JCMT    & Tmb & S    & 1990-11,12                  & 1994A\&AS..103..301 \\
19386+0155   &                   &      &      (63)  &      &             &            &           & IRAM30m & Tmb & S    & 1990-07                     & 1993A\&A...267..515 \\
19454+2920   &                   &      & 730        &      & 14.3        & +21.4      & 14.5      & IRAM30m & Tmb & S    & 1988-01                     & 1991A\&A...246..153 \\
19454+2920   &                   & C    & 730  (41)  &      & 14.3        & +21        & 14.5      & IRAM30m & Tmb & MapC & 1989-01                     & 1993A\&A...267..515 \\
19454+2920   &                   &      & 210        &      & 3.8         & +19.7      & 17.8      & JCMT    & TA* & S    & 1991-10-25 to 27            & 1993ApJ...402..292V \\
19475+3119   & HD 331319         &      & 770        &      & 15          & +17.7      & 14.6      & IRAM30m & Tmb & S    & 1989-01                     & 1991A\&A...246..153 \\
19475+3119   & HD 331319         &      & 720  (62)  &      & 14.9        & +18        & 14.5      & IRAM30m & Tmb & MapC & 1989-01                     & 1993A\&A...267..515 \\
19475+3119   & HD 331319         & O    & 160  (7)   &      & 3.51        & +18.4      & 16.2      & SMT10m  & Tmb & S    & 1999-12-03 to 2000-02-04    & 2005ApJ...624..331H \\
19477+2401   &                   &      &      (10)  & cont &             &            &           & SMT10m  & Tmb & S    & 1999-12-03 to 2000-02-04    & 2005ApJ...624..331H \\
19480+2504   &                   &      & 600        &      & 13          & +41.7      & 15.4      & IRAM30m & Tmb & S    & 1989-01                     & 1991A\&A...246..153 \\
19480+2504   &                   & C    & 610  (54)  &      & 12.8        & +42        & 12.3      & IRAM30m & Tmb & S    & 1989-01                     & 1993A\&A...267..515 \\
19480+2504   &                   & C    & 470  (94)  &      & 9.8         & +41        & 15.4      & IRAM30m & Tmb & MapC & 1990-07                     & 1993A\&A...267..515 \\
19480+2504   &                   &      & 170        &      & 5.8         & +42.0      & 27.0      & JCMT    & TA* & S    & 1991-10-25 to 27            & 1993ApJ...402..292V \\
19486+1350   & TW Aql            & O    &      (60)  &      &             &            &           & IRAM30m & Tmb & S    & 1989-04 to 06               & 1991A\&A...245..499 \\
19500-1709   & HD 187885         &      & 300  (70)  &      & 3.8  (0.3)  & +25        & 8         & CSO     & TA* & S    & 1987-07                     & 1989ApJ...336..822K \\
19500-1709   & HD 187885         &      & 1230 (45)  &      & 27.7        & +25        & 12.0      & IRAM30m & Tmb & MapC & 1989-01                     & 1993A\&A...267..515 \\
19500-1709   & HD 187885         &      & 390  (40)  &      & 7.0  (0.2)  & +26.0(0.3) & 13.2(0.5) & SEST    & Tmb & S    & 1989-05,1990-07             & 1993A\&A...269..231 \\
19500-1709   & HD 187885         & C    & 1900 (100) &      & 48          & +24        &           & IRAM30m & Tmb & MapC & 1990-04 to 06,1991-01       & 1992A\&A...257..701 \\
19500-1709   & HD 187885         &      & 1360 (51)  &      & 23.1        & +24        & 12.8      & IRAM30m & Tmb & S    & 1990-07                     & 1993A\&A...267..515 \\
19500-1709   & HD 187885         &      & 1100       &      & 22.0        & +25.0      &           & IRAM30m & Tmb & S    & 1990-10                     & 1998A\&AS..130....1 \\
19500-1709   & HD 187885         & C    & 190  (6)   &      & 4.41        & +25.0      & 17.2      & SMT10m  & Tmb & S    & 1999-12-03 to 2000-02-04    & 2005ApJ...624..331H \\
19590-1249   & LS IV -12 111     & O    &      (17)  &      &             &            &           & SMT10m  & Tmb & S    & 1999-12-03 to 2000-02-04    & 2005ApJ...624..331H \\
20000+3239   &                   &      & 700        &      & 9.4         & +13.5      & 12.3      & IRAM30m & Tmb & S    & 1989-01                     & 1991A\&A...246..153 \\
20000+3239   &                   & C    & 550  (64)  & cont & 9.5         & +14        & 12.0      & IRAM30m & Tmb & MapC & 1989-01                     & 1993A\&A...267..515 \\
20004+2955   & V1027 Cyg         & O    &      (50)  &      &             &            &           & IRAM30m & Tmb & S    & 1990-04 to 06,1991-01       & 1992A\&A...257..701 \\
20028+3910   & PN PM 2-43        &      & 160        &      &             & +4.2       & 10.5      & ARO12m  & Tmb & S    & 1984-12                     & 1986ApJ...304..394Z \\
20028+3910   & PN PM 2-43        &      & 750        &      & 14.9        & +5.9       & 16.0      & IRAM30m & Tmb & S    & 1989-01                     & 1991A\&A...246..153 \\
20028+3910   & PN PM 2-43        &      & 700  (40)  &      & 15.1        & +6         & 17.5      & IRAM30m & Tmb & MapC & 1989-01                     & 1993A\&A...267..515 \\
20028+3910   & PN PM 2-43        &      & 900        &      & 17.7        & +4.0       & 15.6      & IRAM30m & Tmb & S    & 1993-08                     & 1998A\&AS..130....1 \\
20028+3910   & PN PM 2-43        &      & 130  (11)  &      & 2.53        & +6.2       & 13.0      & SMT10m  & Tmb & S    & 1999-12-03 to 2000-02-04    & 2005ApJ...624..331H \\
\hline
\end{tabular}
\clearpage
\end{sidewaystable*}
\begin{sidewaystable*}[htbp]
\scriptsize
\centering
\addtocounter{table}{-1}
\caption{ (continued)}
\begin{tabular}{l@{ }l@{}c@{}r@{ }c@{ }l@{ }l@{ }llllll}
\hline\hline
IRAS &
Other name &
Chem &
Tp($\sigma$) &
NoteT$^{a}$ &
I($\sigma$) &
$V_{\rm sys}(\sigma)$ &
$V_{\rm exp}(\sigma)$ &
Tel$^{b}$ &
Tscale&
Obs.mode$^{c}$ &
Obs.Date &
Literature \\
 & & & mK & & K\,km\,s$^{-1}$ & km\,s$^{-1}$ & km\,s$^{-1}$ & & & & yyyy-mm-dd & \\
\hline
20117+1634   & R Sge             & O    &      (100) &      &             &            &           & IRAM30m & Tmb & S    & 1989-04 to 06               & 1991A\&A...245..499 \\
20343+2625   & V Vul             & O    &      (30)  &      &             &            &           & IRAM30m & Tmb & S    & 1989-04 to 06               & 1991A\&A...245..499 \\
20462+3416   & LS II +34 26      &      &      (15)  &      &             &            &           & SMT10m  & Tmb & S    & 1999-12-03 to 2000-02-04    & 2005ApJ...624..331H \\
20572+4919   &                   &      &      (14)  & cont &             &            &           & SMT10m  & Tmb & S    & 1999-12-03 to 2000-02-04    & 2005ApJ...624..331H \\
21537+6435   &                   &      &      (180) &      &             &            &           & JCMT    & Tmb & S    & 1990-11,12                  & 1994A\&AS..103..301 \\
22023+5249   &                   &      &      (80)  &      &             &            &           & IRAM30m & Tmb & S    & 1989-01                     & 1993A\&A...267..515 \\
22223+4327   &                   &      &      (300) &      &             &            &           & IRAM30m & Tmb & S    & 1988-01                     & 1991A\&A...246..153 \\
22223+4327   &                   & C    & 640  (32)  &      & 12.3        & -30        & 14.0      & IRAM30m & Tmb & MapC & 1989-01                     & 1993A\&A...267..515 \\
22272+5435   & HD 235858         &      & 3100       &      &             & -28.0      & 9.6       & JCMT    & TR* & S    & 1988-12                     & 1990A\&A...228..503 \\
22272+5435   & HD 235858         & C    & 6270 (55)  &      & 80.5        & -28        & 11.8      & IRAM30m & Tmb & MapC & 1989-01                     & 1993A\&A...267..515 \\
22272+5435   & HD 235858         & C    & 6030 (119) &      & 76.3        & -29        & 9.5       & IRAM30m & Tmb & MapC & 1990-07                     & 1993A\&A...267..515 \\
22272+5435   & HD 235858         & C    & 5300       &      & 125.6       & -28.4      & 10.4      & IRAM30m & Tmb & S    & 1991-05                     & 1998A\&AS..130....1 \\
22272+5435   & HD 235858         & C    & 1120 (8)   &      & 13.6        & -28.1      & 9.1       & SMT10m  & Tmb & S    & 1999-12-03 to 2000-02-04    & 2005ApJ...624..331H \\
22327-1731   & HD213985          & C    &      (70)  &      &             &            &           & IRAM30m & Tmb & S    & 1990-04 to 06,1991-01       & 1992A\&A...257..701 \\
22574+6609   &                   &      & 500        &      & 11          & -64.0      & 15        & IRAM30m & Tmb & S    & 1988-01                     & 1991A\&A...246..153 \\
22574+6609   &                   & C    & 60   (18)  &      & 1.95        & -61.7      & 24.6      & SMT10m  & Tmb & S    & 1999-12-03 to 2000-02-04    & 2005ApJ...624..331H \\
23304+6147   &                   &      & 1600       &      & 25          & -16.1      & 11.5      & IRAM30m & Tmb & S    & 1988-01                     & 1991A\&A...246..153 \\
23304+6147   &                   &      & 700        &      &             & -15.9      & 15.5      & JCMT    & TR* & S    & 1988-12                     & 1990A\&A...228..503 \\
23304+6147   &                   & C    & 200  (8)   &      & 2.92        & -16.6      & 9.2       & SMT10m  & Tmb & S    & 1999-12-03 to 2000-02-04    & 2005ApJ...624..331H \\
23321+6545   & PN PM 2-48        &      & 270  (110) &      & 7.5  (0.6)  & -59        & 15        & CSO     & TA* & S    & 1987-07                     & 1989ApJ...336..822K \\
23321+6545   & PN PM 2-48        &      & 1200       &      & 28          & -55.2      & 17.0      & IRAM30m & Tmb & S    & 1989-01                     & 1991A\&A...246..153 \\
23321+6545   & PN PM 2-48        & C    & 1400       &      & 27.0        & -53.7      & 14.3      & IRAM30m & Tmb & S    & 1991-05                     & 1998A\&AS..130....1 \\
23321+6545   & PN PM 2-48        &      & 370        &      & 7.1         & -55.3      & 18.2      & JCMT    & TA* & S    & 1991-10-25 to 27            & 1993ApJ...402..292V \\
23541+7031   & PK 118+08.1       &      & 470        &      & 25.3        & -27        & 54        & IRAM30m & Tmb & S    & 1989-10 to 1993-05          & 1996ApJ...467..341H \\
Z02229+6208  &                   & C    & 480  (6)   &      & 6.96        & +24.1      & 10.7      & SMT10m  & Tmb & S    & 1999-12-03 to 2000-02-04    & 2005ApJ...624..331H \\
\hline
\end{tabular}
\begin{tablenotes}                                                                                            
\item [1]  $^{a}$                                                                                             
 cont = interstellar contamination;                                                                           
 Jy/K = line peak temperature converted back from line peak flux in Jy;                                       
 hand = line peak temperature measured by hand from published spectral plots;                                 
 2C = line peak temperature as the sum of a two components fit.                                               
\item [2]  $^{b}$                                                                                             
Telescopes for CO observations:                                                                               
 ARO12m = ARO 12\,m telescope at Kitt Peak;                                                                   
 CSO = the 10.4\,m Robert B. Leighton telescope of the Caltech Submillimeter Observatory on Mauna Kea, Hawaii;
 IRAM30m = IRAM 30m telescope at Pico Veleta;                                                                 
 JCMT = 15\,m James Clerk Maxwell Telescope on Mauna Kea, Hawaii;                                             
 OVRO = the Caltech 10m telescope at the Owens Valley Radio Observatory;                                      
 SEST = the 15\,m Swedish-ESO Submillimetre Telescope on La Silla, Chile;                                     
 SMT10m = the 10\,m Sub-Millimeter Telescope on Mt. Graham, AZ.                                               
\item [3]  $^{c}$                                                                                             
Observation modes:                                                                                            
 S = single point;                                                                                            
 Map = single dish map center.                                                                                
\end{tablenotes}                                                                                              
\clearpage
\end{sidewaystable*}
%

\begin{table*}
\scriptsize
\centering
\begin{minipage}{140mm}
\caption{List of literature for Table~\ref{lit-pAGB-CO}.}
\label{lit-pAGB}
\begin{tabular*}{\textwidth}{llll}
\hline\hline
1979ApJ...230..149W  & \citet{1979ApJ...230..149W} & 1995ApJ...453..721J  & \citet{1995ApJ...453..721J}\\
1982ApJ...252..616K  & \citet{1982ApJ...252..616K} & 1996ApJ...467..341H  & \citet{1996ApJ...467..341H} \\
1986ApJ...304..394Z  & \citet{1986ApJ...304..394Z} & 1996ApJ...472..703D  & \citet{1996ApJ...472..703D} \\
1987ApJ...319..367W  & \citet{1987ApJ...319..367W} & 1997A\&A...324.1123B & \citet{1997AA...324.1123B}  \\
1988A\&A...196L...5B & \citet{1988AA...196L...5B}  & 1997A\&A...327..342G & \citet{1997AA...327..342G} \\
1988A\&A...206L..17B & \citet{1988AA...206L..17B}  & 1997A\&A...327..689S & \citet{1997AA...327..689S} \\
1989A\&A...209..119Z & \citet{1989AA...209..119Z}  & 1997ApJ...487L.155S  & \citet{1997ApJ...487L.155S}\\
1989A\&A...222L...1C & \citet{1989AA...222L...1C}  & 1998A\&A...335..292E & \citet{1998AA...335..292E}  \\
1989ApJ...336..822K  & \citet{1989ApJ...336..822K} & 1998A\&AS..130....1N & \citet{1998AAS..130....1N} \\
1989ApJ...345L..87G  & \citet{1989ApJ...345L..87G} & 1998ApJ...509..392M  & \citet{1998ApJ...509..392M}\\
1989ApJ...346..201H  & \citet{1989ApJ...346..201H} & 1998ApJS..117..209K  & \citet{1998ApJS..117..209K} \\
1990A\&A...228..503W & \citet{1990AA...228..503W}  & 1999A\&A...347..194O & \citet{1999AA...347..194O}  \\
1990A\&A...239..173H & \citet{1990AA...239..173H}  & 2000A\&A...355...69P & \citet{2000AA...355...69P} \\
1990ApJ...358..251W  & \citet{1990ApJ...358..251W} & 2001A\&A...367..826J & \citet{2001AA...367..826J} \\
1991A\&A...242..247B & \citet{1991AA...242..247B}  & 2002A\&A...390..501G & \citet{2002AA...390..501G} \\
1991A\&A...245..499A & \citet{1991AA...245..499A}  & 2002ApJ...572..326B  & \citet{2002ApJ...572..326B}\\
1991A\&A...246..153L & \citet{1991AA...246..153L}  & 2002ApJ...577..961H  & \citet{2002ApJ...577..961H} \\
1992A\&A...256..235K & \citet{1992AA...256..235K}  & 2003A\&A...405..271M & \citet{2003AA...405..271M}  \\
1992A\&A...257..701B & \citet{1992AA...257..701B}  & 2004A\&A...414..581H & \citet{2004AA...414..581H} \\
1993A\&A...267..515O & \citet{1993AA...267..515O}  & 2005A\&A...429..977W & \citet{2005AA...429..977W} \\
1993A\&A...269..231V & \citet{1993AA...269..231V}  & 2005ApJ...624..331H  & \citet{2005ApJ...624..331H}\\
1993A\&A...273..185H & \citet{1993AA...273..185H}  & 2008A\&A...488L..21H & \citet{2008AA...488L..21H}  \\
1993ApJ...402..292V  & \citet{1993ApJ...402..292V} & 2009ApJ...690..837M  & \citet{2009ApJ...690..837M}\\
1994A\&AS..103..301H & \citet{1994AAS..103..301H}  &                                                   \\
\hline
\end{tabular*}
\end{minipage}
\end{table*}

\begin{deluxetable}{l@{ }l@{ }r@{ }l@{ }l@{ }l@{ }l@{ }r@{ }r@{ }r@{  }r}
\tablecolumns{11} 
\tabletypesize{\tiny}
\tablewidth{0pc}
\tablecaption{Observed properties from literature for all the 87 considered post-AGB stars.
\label{lit-pAGB-obj}}
\tablehead{
\colhead{Object}                    &
\colhead{Other name}                &
\colhead{$F_{\rm CO2-1}(\sigma)$ }  &
\colhead{NoteCO$^{a}$}              &
\colhead{$V{\rm exp}$}              &
\colhead{F12}                       &
\colhead{F25}                       &
\colhead{F60}                       &
\colhead{C12}                       &
\colhead{C23}                       &
\colhead{RCO25($\sigma$)}           \\
\colhead{(or IRAS)}                 &
\colhead{ }                         &
\colhead{Jy\,MHz}                   &
\colhead{km\,s$^{-1}$}              &
\colhead{ }                         &
\colhead{Jy}                        &
\colhead{Jy}                        &
\colhead{Jy}                        &
\colhead{ }                         &
\colhead{ }                         &
\colhead{MHz}                       }
\startdata
CRL 2688    & EGG NEBULA         & 2220.96(964.73) &      & 17.6  &         &         &         &       &       &                  \\
HD 107369   & CD-31 9638         &        (81.86)  & Tp   &       &         &         &         &       &       &                  \\
V4334 Sgr   & SAKURAI'S OBJECT   &        (11.63)  & Tp   &       &         &         &         &       &       &                  \\
01005+7910  &                    &        (6.00)   & Tp   &       & 3.905   & 24.23   & 10.07   &  1.98 & -0.95 &        (  0.248) \\
02143+5852  & GLMP 26            &        (10.18)  &      &       & 5.897   & 18.06   & 5.395   &  1.22 & -1.31 &        (  0.564) \\
04166+5719  & TW Cam             &        (6.00)   & Tp   &       & 8.252   & 5.602   & 1.84    & -0.42 & -1.21 &        (  1.071) \\
04296+3429  & GLMP 74            & 38.24           &      & 12.8  & 12.74   & 45.94   & 15.45   &  1.39 & -1.18 &   0.832          \\
04395+3601  & CRL 618            & 1354.12(255.78) &      & 16.7  & 470.8   & 1106.   & 1036.   &  0.93 & -0.07 &   1.224(  0.231) \\
04440+2605  & RV Tau             &        (3.60)   &      &       & 22.52   & 18.05   & 6.502   & -0.24 & -1.11 &        (  0.199) \\
05338-3051  & RV Col             &        (81.86)  & Tp   &       & 0.349   &         &         &       &       &                  \\
06034+1354  & DY Ori             &        (3.60)   & Tp   &       & 12.43   & 14.89   & 4.182   &  0.20 & -1.38 &        (  0.242) \\
06072+0953  & CT Ori             &        (4.80)   & Tp   &       & 6.145   & 5.545   & 1.24    & -0.11 & -1.63 &        (  0.866) \\
06108+2743  & SU Gem             &        (12.00)  &      &       & 7.905   & 5.684   & 2.193   & -0.36 & -1.03 &        (  2.111) \\
06176-1036  & Red Rectangle      & 40.25  (2.30)   &      & 4.9   & 421.6   & 456.1   & 173.1   &  0.09 & -1.05 &   0.088(  0.005) \\
06530-0213  & PN PM 1-24         & 109.45 (25.12)  &      & 11.3  & 6.114   & 27.41   & 15.05   &  1.63 & -0.65 &   3.993(  0.916) \\
07134+1005  & HD 56126           & 229.47 (61.58)  &      & 9.9   & 24.51   & 116.7   & 50.13   &  1.69 & -0.92 &   1.966(  0.528) \\
07284-0940  & U Mon              &        (8.40)   &      &       & 124.3   & 88.43   & 26.59   & -0.37 & -1.30 &        (  0.095) \\
07331+0021  & AI CMi             &        (6.72)   & Tp   &       & 15.32   & 68.11   & 18.51   &  1.62 & -1.41 &        (  0.099) \\
07399-1435  & OH231.8+4.2        & 2248.50         &      & 21.4  & 18.98   & 226.3   & 548.3   &  2.69 &  0.96 &   9.936          \\
07430+1115  & GLMP 192           &        (9.45)   &      &       & 7.685   & 29.93   & 10.67   &  1.48 & -1.12 &        (  0.316) \\
08005-2356  & V510 Pup           & 269.15          &      & 100.0 & 17.96   & 51.8    & 29.83   &  1.15 & -0.60 &   5.196          \\
08011-3627  & AR Pup             &        (24.00)  & Tp   &       & 131.3   & 94.32   & 26.12   & -0.36 & -1.39 &        (  0.254) \\
08187-1905  & HD 70379           &        (13.08)  & Tp   &       & 0.714   & 17.62   & 12.31   &  3.48 & -0.39 &        (  0.742) \\
08544-4431  & V390 Vel           & 40.37  (17.30)  &      & 5.6   & 180.3   & 158.8   & 56.25   & -0.14 & -1.13 &   0.254(  0.109) \\
09032-3953  & PN PM 2-9          & 128.81          &      & 25.0  & 20.38   & 125.9   & 92.31   &  1.98 & -0.34 &   1.023          \\
09370-4826  & GLMP 255           &        (65.49)  & Tp   &       & 10.82   & 30.14   & 14.16   &  1.11 & -0.82 &        (  2.173) \\
09371+1212  & FROSTY LEONIS      & 875.07          & Tp   & 36.6  &         & 4.594   & 70.7    &       &  2.97 & 190.481          \\
10158-2844  & HR 4049            &        (4.80)   &      &       & 48.25   & 9.553   & 1.77    & -1.76 & -1.83 &        (  0.502) \\
10178-5958  & Hen 3-401          & 99.97           &      & 13.0  & 4.124   & 38.33   & 76.14   &  2.42 &  0.75 &   2.608          \\
10194-5625  & GLMP 271           &        (65.49)  & Tp   &       & 16.75   & 57.02   & 27.78   &  1.33 & -0.78 &        (  1.149) \\
10197-5750  & Roberts 22         & 173.02          &      & 31.0  & 200.    & 1092.   & 588.3   &  1.84 & -0.67 &   0.158          \\
11000-6153  & HD 95767           &        (81.86)  & Tp   &       & 22.13   & 15.65   & 10.9    & -0.38 & -0.39 &        (  5.231) \\
11381-6401  &                    &        (77.76)  & Tp   &       & 5.41    & 27.76   & 17.02   &  1.78 & -0.53 &        (  2.801) \\
11385-5517  & HD 101584          & 793.98          &      & 45.0  & 92.6    & 138.3   & 193.    &  0.44 &  0.36 &   5.741          \\
11544-6408  & GLMP 315           &        (57.30)  & Tp   &       & 12.49   & 29.81   & 12.52   &  0.94 & -0.94 &        (  1.922) \\
12067-4508  & HD 105578          &        (81.86)  & Tp   &       & 5.321   & 10.96   & 5.652   &  0.78 & -0.72 &        (  7.469) \\
12222-4652  & HD 108015          &        (81.86)  & Tp   &       & 32.46   & 33.23   & 7.993   &  0.03 & -1.55 &        (  2.463) \\
12419-5414  & Boomerang Nebula   & 184.84 (62.73)  &      & 25.5  & 4.173   & 5.502   & 13.65   &  0.30 &  0.99 &  33.595( 11.401) \\
14488-5405  & CPD -53 5736       & 39.80           &      & 15.0  & 6.587   & 63.74   & 30.54   &  2.46 & -0.80 &   0.624          \\
14524-6838  & HD 131356          &        (81.86)  & Tp   &       & 13.23   & 10.26   & 4.107   & -0.28 & -0.99 &        (  7.979) \\
15465+2818  & R CrB              &        (55.70)  &      &       & 38.86   & 17.06   & 3.936   & -0.89 & -1.59 &        (  3.265) \\
16342-3814  & OH 344.1 +5.8      & 73.07  (4.78)   &      & 46.0  & 16.2    & 199.8   & 290.2   &  2.73 &  0.41 &   0.366(  0.024) \\
16594-4656  & WATER LILY NEBULA  & 626.73          &      & 14.0  & 44.92   & 298.    & 131.4   &  2.05 & -0.89 &   2.103          \\
17106-3046  & PN PM 2-23         & 35.37           &      & 1.0   & 4.007   & 62.38   & 51.19   &  2.98 & -0.21 &   0.567          \\
17150-3224  & RAFGL 6815         & 273.61 (26.76)  &      & 14.8  & 57.92   & 322.3   & 268.3   &  1.86 & -0.20 &   0.849(  0.083) \\
17245-3951  & OH348.8-2.8        & 11.15           &      & 15.0  & 3.356   & 44.73   & 38.23   &  2.81 & -0.17 &   0.249          \\
17423-1755  & Hen 3-1475         & 89.62  (2.82)   &      & 26.5  & 7.052   & 28.31   & 63.68   &  1.51 &  0.88 &   3.166(  0.100) \\
17436+5003  & V814 Her           & 180.62 (22.78)  &      & 11.5  & 6.122   & 183.5   & 151.7   &  3.69 & -0.21 &   0.984(  0.124) \\
17441-2411  & AFGL 5385          & 139.76 (19.80)  &      & 12.5  & 42.78   & 191.1   & 106.5   &  1.63 & -0.63 &   0.731(  0.104) \\
17530-3348  & AI Sco             & 10.15           &      & 2.8   & 17.59   & 11.38   & 2.95    & -0.47 & -1.47 &   0.892          \\
17534+2603  & 89 Her             & 24.52  (1.41)   &      & 3.9   & 97.52   & 54.49   & 13.42   & -0.63 & -1.52 &   0.450(  0.026) \\
18025-3906  & PN PM 2-34         &        (53.21)  & Tp   &       & 4.295   & 41.2    & 30.11   &  2.45 & -0.34 &        (  1.292) \\
18095+2704  & V887 Her           &        (3.60)   & Tp   &       & 45.09   & 125.7   & 27.83   &  1.11 & -1.64 &        (  0.029) \\
18276-1431  & OH17.7-2.0         & 108.22 (5.64)   &      & 12.2  & 22.65   & 132.    & 120.    &  1.91 & -0.10 &   0.820(  0.043) \\
18281+2149  & AC Her             & 3.38   (0.56)   &      & 1.5   & 41.43   & 65.33   & 21.37   &  0.49 & -1.21 &   0.052(  0.009) \\
18372-2257  & V348 Sgr           &        (61.39)  & Tp   &       & 5.533   & 2.999   & 2.876   & -0.66 & -0.05 &        ( 20.470) \\
18415-2100  & MV Sgr             &        (81.86)  & Tp   &       & 0.597   & 1.565   & 0.777   &  1.05 & -0.76 &        ( 52.307) \\
19114+0002  & AFGL 2343          & 837.39 (201.00) &      & 31.5  & 31.33   & 648.3   & 515.9   &  3.29 & -0.25 &   1.292(  0.310) \\
19132-3336  & RY Sgr             &        (122.78) & Tp   &       & 77.17   & 26.25   & 5.433   & -1.17 & -1.71 &        (  4.677) \\
19306+1407  & GLMP 923           &        (9.24)   & Tp   &       & 3.584   & 58.65   & 31.83   &  3.03 & -0.66 &        (  0.158) \\
19343+2926  & PN M1-92           & 319.13          &      & 32.5  & 17.52   & 59.76   & 118.4   &  1.33 &  0.74 &   5.340          \\
19374+2359  &                    &        (14.76)  & Tp   &       & 23.62   & 98.18   & 70.87   &  1.55 & -0.35 &        (  0.150) \\
19386+0155  & V1648 Aql          &        (7.56)   & Tp   &       & 17.38   & 47.44   & 18.56   &  1.09 & -1.02 &        (  0.159) \\
19454+2920  & GLMP 950           & 91.14  (18.24)  &      & 15.6  & 17.27   & 89.56   & 54.43   &  1.79 & -0.54 &   1.018(  0.204) \\
19475+3119  & HD 331319          & 93.09  (15.45)  &      & 15.1  & 0.537   & 37.99   & 55.83   &  4.62 &  0.42 &   2.450(  0.407) \\
19477+2401  & CLOVERLEAF NEBULA  &        (7.27)   & Tp   &       & 11.24   & 54.92   & 27.13   &  1.72 & -0.77 &        (  0.132) \\
19480+2504  & GLMP 953           & 88.65  (49.96)  &      & 17.5  & 20.81   & 67.89   & 43.16   &  1.28 & -0.49 &   1.306(  0.736) \\
19486+1350  & TW Aql             &        (7.20)   & Tp   &       &         & 5.232   & 11.25   &       &  0.83 &        (  1.376) \\
19500-1709  & HD 187885          & 173.63 (53.58)  &      & 11.3  & 27.82   & 165.    & 73.4    &  1.93 & -0.88 &   1.052(  0.325) \\
19590-1249  & LS IV -12 111      &        (12.36)  & Tp   &       & 0.293   & 10.26   & 6.45    &  3.86 & -0.50 &        (  1.205) \\
20000+3239  & GLMP 963           & 53.36  (0.27)   &      & 12.2  & 15.03   & 70.97   & 29.99   &  1.69 & -0.94 &   0.752(  0.004) \\
20004+2955  & V1027 Cyg          &        (6.00)   & Tp   &       & 31.72   & 36.96   & 4.656   &  0.17 & -2.25 &        (  0.162) \\
20028+3910  & PN PM 2-43         & 88.11  (5.98)   &      & 14.5  & 41.78   & 210.8   & 143.1   &  1.76 & -0.42 &   0.418(  0.028) \\
20117+1634  & R Sge              &        (12.00)  & Tp   &       & 10.6    & 7.543   & 2.123   & -0.37 & -1.38 &        (  1.591) \\
20343+2625  & V Vul              &        (3.60)   & Tp   &       & 12.35   & 5.688   & 1.294   & -0.84 & -1.61 &        (  0.633) \\
20462+3416  & LS II +34 26       &        (10.90)  & Tp   &       & 0.287   & 13.68   & 12.12   &  4.20 & -0.13 &        (  0.797) \\
20572+4919  & V2324 Cyg          &        (10.18)  & Tp   &       & 4.335   & 10.97   & 9.696   &  1.01 & -0.13 &        (  0.928) \\
21537+6435  & PN PM 1-334        &        (61.88)  & Tp   &       & 6.913   & 26.1    & 13.34   &  1.44 & -0.73 &        (  2.371) \\
22023+5249  & GLMP 1051          &        (9.60)   & Tp   &       & 1.017   & 24.69   & 14.52   &  3.46 & -0.58 &        (  0.389) \\
22223+4327  & V448 Lac           & 69.33           &      & 14.0  & 2.121   & 37.1    & 22.4    &  3.11 & -0.55 &   1.869          \\
22272+5435  & HD 235858          & 490.04 (98.40)  &      & 10.1  & 73.88   & 302.4   & 96.59   &  1.53 & -1.24 &   1.621(  0.325) \\
22327-1731  & HD213985           &        (8.40)   & Tp   &       & 5.569   & 4.664   & 2.107   & -0.19 & -0.86 &        (  1.801) \\
22574+6609  &                    & 64.29  (2.29)   &      & 19.8  & 9.003   & 29.47   & 20.64   &  1.29 & -0.39 &   2.182(  0.078) \\
23304+6147  & PN PM 2-47         & 120.31 (20.61)  &      & 12.1  & 11.36   & 59.07   & 26.6    &  1.79 & -0.87 &   2.037(  0.349) \\
23321+6545  & PN PM 2-48         & 218.78 (74.04)  &      & 16.1  & 13.66   & 85.61   & 63.96   &  1.99 & -0.32 &   2.556(  0.865) \\
23541+7031  & PK 118+08.1        & 142.61          &      & 54.0  & 91.36   & 92.28   & 55.7    &  0.01 & -0.55 &   1.545          \\
Z02229+6208 &                    & 237.64          &      & 10.7  &         &         &         &       &       &                  \\
\enddata
\tablecomments{The CO\,2-1 line quantities are the average values from literature. New observations from this work are not included.}
\tablenotetext{a}{'Tp' means the line area is estimated from line peak temperature                         
 and line width and thus is not as reliable as those values directly given in literature.}
\end{deluxetable}

Refereed papers that contain original single dish observations of CO\,2-1 line have 
been searched for in the ADS database using object position 
(with a search radius of $1\,\arcmin$) for all 
393 known pAGB stars in the version 2 of the Torun Catalog of pAGB stars \citep{szcz11}.
The search was done by Oct. 26, 2011. Both single pointing observations and single dish mapping 
are considered. Interferometer data are not considered due to the missing flux issue.

CO line parameters such as line peak temperature 
($T_{\rm mb}$ or $T_{\rm A}^*$ or $T_{\rm A}'$), line area, line center velocity, and CSE expansion 
velocity are compiled, if available. The RMS noise level in the baseline are 
used as the 1\,$\sigma$ upper limit of undetected lines. The information about the telescopes, 
e.g., conversion factor to obtain main beam temperature from other antenna temperature scales, 
the telescope response (Jy/K) for conversion of main beam temperature into line flux (Jy), 
and the beam size of the telescopes at CO 2-1 line frequency are also collected. 
The telescope responses are adopted as
nominal values for all involved telescopes:
7.33\,Jy/K for the $13\arcsec$ beam of the IRAM 30-m, 
21\,Jy/K for the $22\arcsec$ beam of the JCMT 15-m, 
25\,Jy/K for the $24\arcsec$ beam of the SEST 15-m, 
44.4\,Jy/K for the $32\arcsec$ beam of the AROSMT 10-m, 
39\,Jy/K for the $30\arcsec$ beam of both the NRAO 12-m and the CSO 10.4-m telescopes, and 
29.4\,Jy/K for the $26\arcsec$ beam of the old 10-m Caltech OVRO telescope.

Here we give the list of all 175 CO\,2-1 data records for 87 pAGB stars
in Table~\ref{lit-pAGB-CO}, in which data records are sorted in increasing alphabetic
order of object names and increasing observation dates. All the involved literature
is collected in Table~\ref{lit-pAGB}. When available, we also collect the information about the chemistry of the observed circumstellar shells (see Table~\ref{lit-pAGB-CO}). 

Duplicated CO 2-1 observations were averaged to give a mean line strength. 
Before averaging, we convert all antenna temperature scales into velocity-integrated line 
flux in Jy\,MHz, so that the measured line strengths by different telescopes can be directly 
compared to check consistency. In the case when only line peak temperature was given in the papers, 
the line width $V_{\rm exp}$ is used to estimate the line area by assuming a Gaussian line profile. 
In very few cases when the $V_{\rm exp}$ is also not given, a fixed value of 10\,km\,s$^{-1}$ is assumed. 
The repeated observations usually agree with each other within a factor of 2-3, 
which is significantly larger than the usually accepted flux calibration uncertainty of 20 per cent. 
The reason of the large variation could be attributed to pointing error, 
bad weather, or technical problem during the observations.
Weights are used in the averaging, which are set as follows: 
a weight of unity is set for most of the data entries, while a weight of 0.5 is given to those line area data 
estimated from line peak temperature and $V_{\rm exp}$, and a weight of 2 is given to those 
single dish mapping data because these observations have less problems 
with pointing and are more reliable for nearby extended objects. 

The average quantities for all the 87 pAGB objects (the velocity-integrated CO\,2-1 line flux, and CSE expansion velocity $V_{\rm exp}$), as well as IRAS flux densities, IRAS colors (defined in Sect.~\ref{pagb}), and CO-IR flux ratio $R_{\rm CO25}$ (defined by Eq.~\ref{eq1} in Sect.~\ref{results}) are collected in Table~\ref{lit-pAGB-obj}.

\section{STATISTICAL PROPERTIES OF THE pAGB STAR SAMPLE}
\label{app-statistics}

Statistical properties of the pAGB star sample are discussed here to assess
if any selection effects would have biased the conclusions
drawn from the sample. We discuss in this section the completeness of the Torun Catalog of pAGB stars, 
our compilation of CO\,2-1 observations, and the detection rates of CO\,2-1 lines among 
the observed objects.

\begin{figure*}
\epsscale{1.0}
\plotone{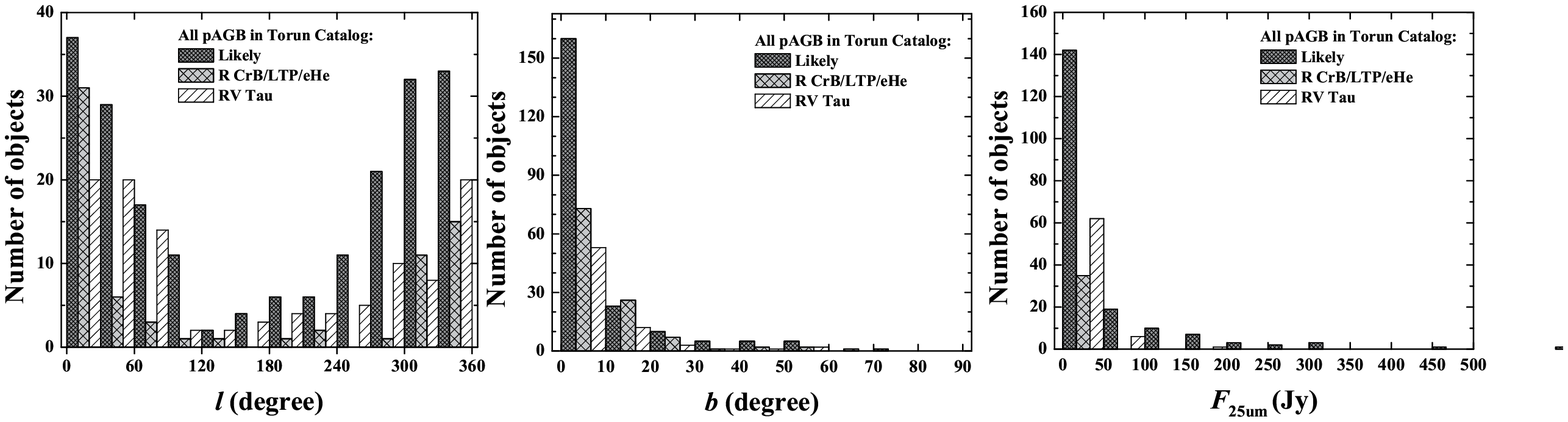}
\caption{The histograms of the Galactic positions and $IRAS\,25\,\micron$ flux density distributions of the pAGB stars in the Torun Catalog.
Only the three most important sub-catalogs: Likely, R CrB/LTP/eHe, and RV Tau, are considered.}
\label{app-hist-torun-cat}
\end{figure*}

First of all, the Torun Catalog of pAGB stars was created with various object selection 
criteria \citep{szcz07}. It is still unknown if the objects in this catalog are representative 
to all pAGB stars in the Galaxy. The version 2 of the catalog \citep{szcz11} is now divided into 
several sub-catalogs: likely, possible, RV Tau, R CrB/LTP/eHe, and unlikely. Here we plot 
the histograms of the Galactic position and {\it IRAS} $25\micron$ flux densities 
of the three most important sub-catalogs: likely, RV\,Tau and R\,CrB/LTP/eHe, to check if the catalogs have any obvious bias.
As shown in Fig.~\ref{app-hist-torun-cat}, all three sub-catalogs show decreasing trends from Galactic Center (GC) to 
anti-GC and from Galactic disc to high latitude and from $25\micron$-weak to $25\micron$-strong objects. Although 
the detailed comparison of these distributions with that of pAGB star model prediction for 
the Milky Way Galaxy is beyond the scope of this work, the trends roughly agree to the distribution of Galactic zero age main sequence stars.
Thus, it is concluded that no severe bias can be seen in the three sub-catalogs from their 
Galactic position and IR flux density distributions.

\begin{figure*}
\epsscale{1.0}
\plotone{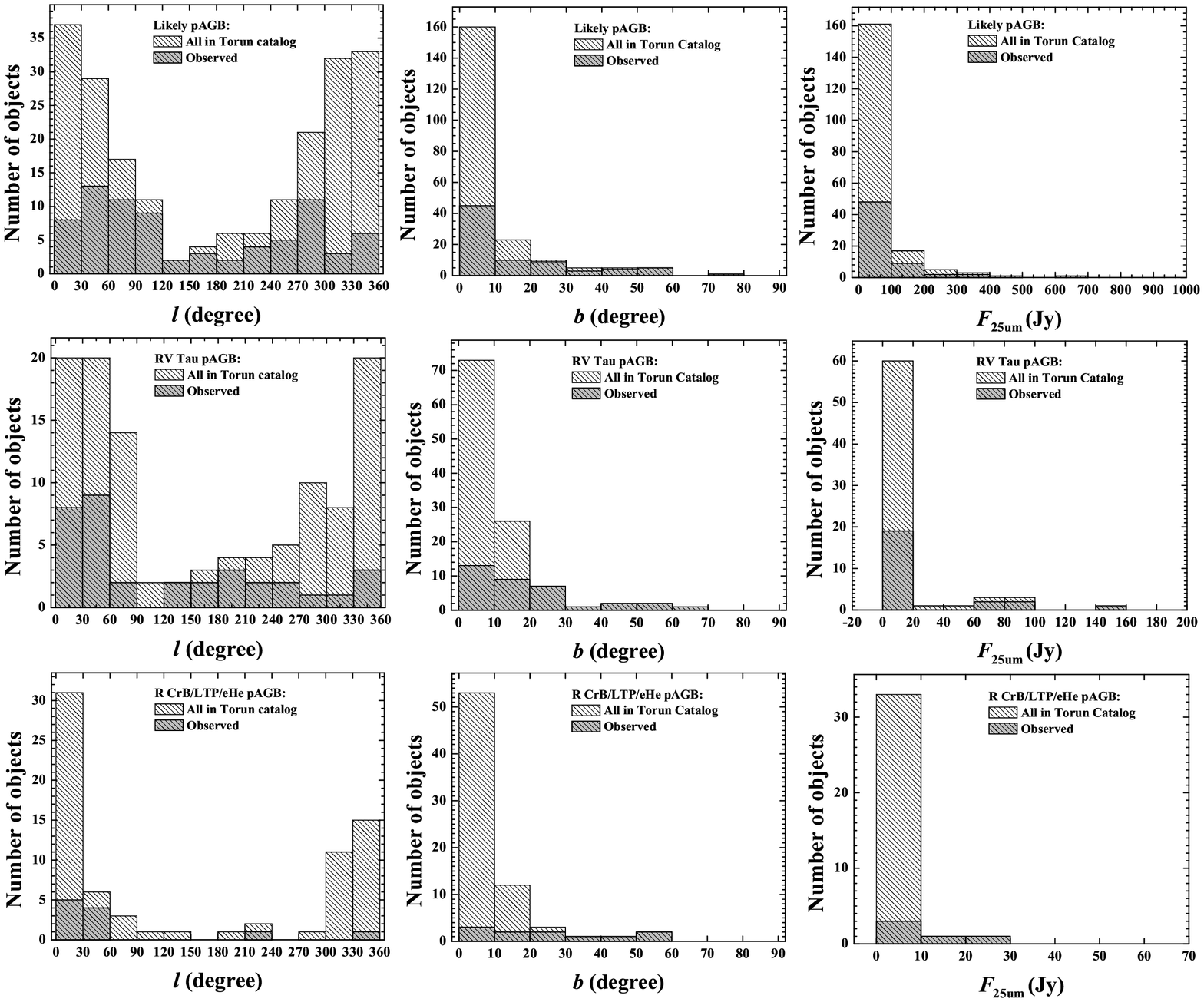}
\caption{Comparison of the histograms of the pAGB stars observed in CO\,2-1 with that of all pAGB stars in the Torun Catalog.
Galactic positions and $IRAS\,25\,\micron$ flux density distributions are shown in different columns of panels, while the three most 
important sub-groups of pAGB stars: Likely, R CrB/LTP/eHe, and RV Tau, are shown in different rows of panels.}
\label{app-hist-pAGB-obs}
\end{figure*}

Secondly, the pAGB stars with CO\,2-1 observations in literature and in our survey (this work) are compared 
with all pAGB stars in the Torun Catalog in Fig.~\ref{app-hist-pAGB-obs}. The left and middle column of panels 
show that pAGB stars toward the GC and the Galactic disc have the lowest percentage of CO\,2-1 observations,
while most of the high Galactic latitude pAGB stars have been observed in CO\,2-1 line. 
In the right column of Fig.~\ref{app-hist-pAGB-obs},
the CO observations of all three sub-groups of pAGB stars are clearly biased toward IR strong objects.
Conversely, the CO observations of the R CrB/LTP/eHe types of pAGB stars in the third row of the figure 
are the most incomplete, because observers usually do not expected them to show CO line emission. 

\begin{figure*}
\epsscale{1.0}
\plotone{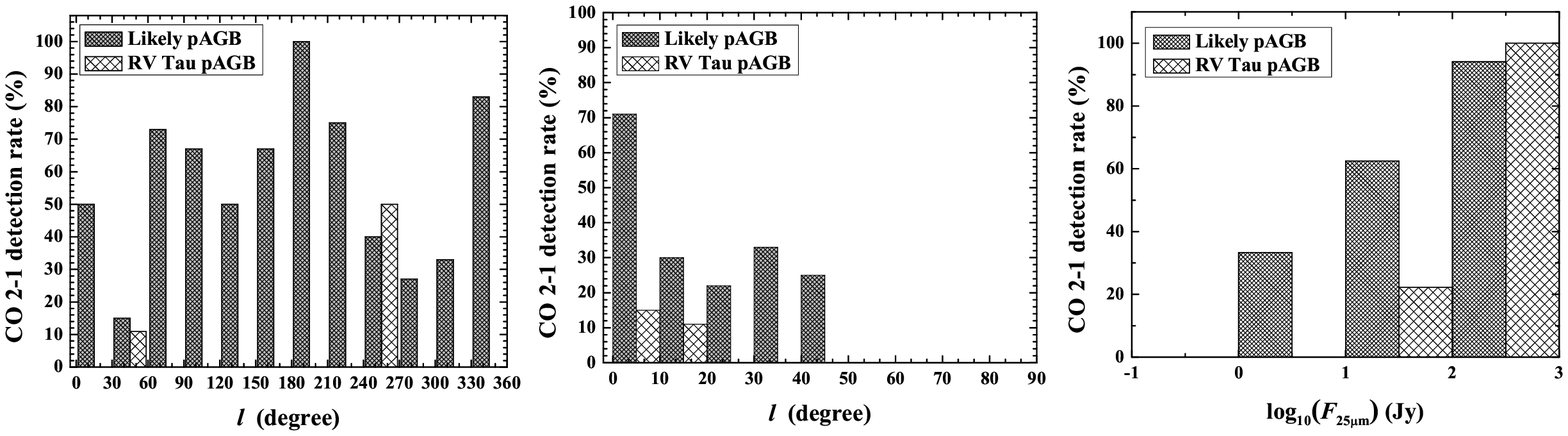}
\caption{Comparison of the CO\,2-1 line detection rates of `Likely' and RV\,Tau type pAGB stars as a function of 
Galactic positions and $IRAS\,25\,\micron$ flux densities.}
\label{app-hist-det-rate}
\end{figure*}

Thirdly, the detection rates of the CO\,2-1 line among the `Likely' and RV\,Tau type pAGB stars are plotted in Fig.~\ref{app-hist-det-rate}.
The R\,CrB/LTP/eHe type pAGB stars are not shown, because none of them have been detected in CO\,2-1. 
Although there is no obvious trend along the Galactic longitude in the left panel of the figure, 
one can see in the middle panel that the CO\,2-1 line detection rate in the Galactic disc is 
more than two times higher than at higher Galactic latitudes. This could be due to the higher masses of the pAGB stars in the Galactic disc, 
because more massive pAGB stars have thicker relic circumstellar envelope and thus have their CO lines more easily detected. In the right 
panel of the figure, there is a clear decreasing trend of CO\,2-1 line detection rates toward IR-weaker objects, which indicates that the CO\,2-1 
line observations are sensitivity limited.

In a summary, the major bias effects found in the CO observations of pAGB stars are 
1) PAGB stars in the GC direction are not adequately observed in CO\,2-1 line; 
2) The CO line observations in literature and our work are 
biased to IR-strong sources and are sensitivity limited.
3) The CO line observations of R CrB/LTP/eHe type pAGB stars are very incomplete in any sense; 

\section{COMPARING AGB STARS ON THE CO-IR DIAGRAMS}
\label{app-AGB-stars}

Because this work is concentrated on pAGB stars, we only briefly mention some interesting points about 
the distribution of AGB stars and PNe on the new CO-IR diagram. Given below are more details not mentioned in 
Sect.~\ref{AGB-PN} of the main text.

\subsection{O- and C-rich AGB stars}
\label{O-stars}

Although the majority of the O-rich AGB stars (empty 
blue circles in the second row of Fig.~\ref{figcompare}) concentrate 
in a small region on both the CO-IR and IRAS color-color diagrams, 
some objects do show significantly smaller $R_{\rm CO25}$ ratios (CO-weak) or significantly redder C23 colors. 
The redder objects could be either post-thermal pulse objects with detached CSEs
\citep[see, e.g., ][]{stef98}, OH/IR star candidates, or mis-identified pAGB stars. 
Many of the CO-weak objects are semi-regular variables of which the CSEs are probably optically thinner and thus 
CO molecules could have been partial destroyed by photodissociation. 

The 28 OH/IR stars (half-shaded black circles) scatter in a larger region with similar or smaller CO-IR flux ratios 
but redder C12 and C23 colors than the other O-rich AGB stars. 
This can be explained by cooler CO gas and lower average dust temperatures 
in the very thick CSEs of the OH/IR stars \citep{1990AA...239..173H,kast92}.

The O-rich AGB stars have slightly bluer C23 colors and about 3 times smaller $R_{\rm CO25}$ 
ratios than the C stars on average. 
The redder C23 color of C stars had been known \citep[e.g., in the study of IRAS color-color diagram by ][]{vand88} as 
due to the shallower emissivity of carbon rich dust in the 12-100\,$\micron$ region \citep{zuck86}. The stronger CO line in C stars 
had long been recognized since the CO line survey of \citet{nyma92}.
The most possible explanation 
is higher CO abundance in carbon stars than in O-rich AGB stars, because only part of the 
oxygen atoms are used to make CO molecules in the latter.

\subsection{S stars against O- and C-rich AGB stars}
\label{app-C-S-stars}

S stars on the $R_{\rm CO25}$-C23 diagram 
(red filled circles in the top row of Fig.~\ref{figcompare}) 
scatter in the similar regions as the C stars, 
indicating that the gas and dust properties of the CSE of S stars are closer to 
that of C stars than O-rich AGB stars. However, because S stars are expected to have much less 
dust in their CSEs due to the lock of most C and O atoms into gaseous CO molecules, 
their CO-IR flux ratios are intuitively expected to be even larger than that of typical C stars, 
which is not true in our Fig.~\ref{figcompare}. The not high enough CO-IR ratios of S stars could be 
explained by assuming that the CO molecules are not efficiently formed through ideal equilibrium chemistry 
\citep[as discussed for C stars by ][]{Papo08} and by assuming dust grains such as the recently proposed solid SiO dust 
\citep{Wetz12} may still be efficiently formed around S stars.

Comparing the IRAS C-C diagrams in the upper right and middle right panels of Fig.~\ref{figcompare}, one can see that 
the S stars also have different IRAS colors than the C- and O-rich AGB stars. Their C12 colors are generally bluer than the latter, while 
their C23 colors are similar to C stars but redder than O-rich AGB stars.
Again, the solid SiO dust proposed by \citet{Wetz12} has the potential to naturally explain 
these color differences. The solid SiO grains have the $10\micron$ feature as normal silicates, 
but not the $18\micron$ feature. Comparing with the Silicates dust in O-rich AGB stars, the lack of the $18\micron$ feature of SiO dust 
in S stars just results in a weaker IRAS $25\micron$ flux density, and thus bluer C12 and redder C23 colors of S stars than O-rich AGB stars, as showed above. 
Comparing with the C-rich dust in C stars, the most salient feature of the SiO dust is the appearance of the emission feature at $10\micron$ which naturally causes 
the enhancement of the IRAS $12\micron$ flux density, and thus bluer C12 and comparable C23 colors of the S stars against C stars.




\bibliographystyle{apj}  

\begin{thebibliography}{137}
\expandafter\ifx\csname natexlab\endcsname\relax\def\natexlab#1{#1}\fi

\bibitem[{{Alcolea} \& {Bujarrabal}(1991)}]{1991AA...245..499A}
{Alcolea}, J., \& {Bujarrabal}, V. 1991, A\&A, 245, 499

\bibitem[{{Alksnis} {et~al.}(2001){Alksnis}, {Balklavs}, {Dzervitis},
  {Eglitis}, {Paupers}, \& {Pundure}}]{GCGCS}
{Alksnis}, A., {Balklavs}, A., {Dzervitis}, U., {Eglitis}, I., {Paupers}, O.,
  \& {Pundure}, I. 2001, Baltic Astronomy, 10, 1

\bibitem[{{Arkhipova} {et~al.}(1999){Arkhipova}, {Ikonnikova}, {Noskova},
  {Sokol}, {Esipov}, \& {Klochkova}}]{Arkhipova:1999nx}
{Arkhipova}, V.~P., {Ikonnikova}, N.~P., {Noskova}, R.~I., {Sokol}, G.~V.,
  {Esipov}, V.~F., \& {Klochkova}, V.~G. 1999, Astronomy Letters, 25, 25

\bibitem[{{Bachiller} {et~al.}(1997){Bachiller}, {Forveille}, {Huggins}, \&
  {Cox}}]{1997AA...324.1123B}
{Bachiller}, R., {Forveille}, T., {Huggins}, P.~J., \& {Cox}, P. 1997, A\&A,
  324, 1123

\bibitem[{{Bachiller} {et~al.}(1988){Bachiller}, {Gomez-Gonzalez},
  {Bujarrabal}, \& {Martin-Pintado}}]{1988AA...196L...5B}
{Bachiller}, R., {Gomez-Gonzalez}, J., {Bujarrabal}, V., \& {Martin-Pintado},
  J. 1988, A\&A, 196, L5

\bibitem[{{Bakker} {et~al.}(1997){Bakker}, {van Dishoeck}, {Waters}, \&
  {Schoenmaker}}]{Bak97}
{Bakker}, E.~J., {van Dishoeck}, E.~F., {Waters}, L.~B.~F.~M., \&
  {Schoenmaker}, T. 1997, \aap, 323, 469

\bibitem[{{Balser} {et~al.}(2002){Balser}, {McMullin}, \&
  {Wilson}}]{2002ApJ...572..326B}
{Balser}, D.~S., {McMullin}, J.~P., \& {Wilson}, T.~L. 2002, ApJ, 572, 326

\bibitem[{{Bujarrabal} {et~al.}(1992){Bujarrabal}, {Alcolea}, \&
  {Planesas}}]{1992AA...257..701B}
{Bujarrabal}, V., {Alcolea}, J., \& {Planesas}, P. 1992, A\&A, 257, 701

\bibitem[{{Bujarrabal} {et~al.}(2013){Bujarrabal}, {Alcolea}, {Van Winckel},
  {Santander-Garc{\'{\i}}a}, \& {Castro-Carrizo}}]{Buja13}
{Bujarrabal}, V., {Alcolea}, J., {Van Winckel}, H., {Santander-Garc{\'{\i}}a},
  M., \& {Castro-Carrizo}, A. 2013, \aap, 557, A104

\bibitem[{{Bujarrabal} \& {Bachiller}(1991)}]{1991AA...242..247B}
{Bujarrabal}, V., \& {Bachiller}, R. 1991, A\&A, 242, 247

\bibitem[{{Bujarrabal} {et~al.}(1988){Bujarrabal}, {Bachiller}, {Alcolea}, \&
  {Martin-Pintado}}]{1988AA...206L..17B}
{Bujarrabal}, V., {Bachiller}, R., {Alcolea}, J., \& {Martin-Pintado}, J. 1988,
  A\&A, 206, L17

\bibitem[{{Bujarrabal} {et~al.}(2005){Bujarrabal}, {Castro-Carrizo}, {Alcolea},
  \& {Neri}}]{Bujarrabal:2005yq}
{Bujarrabal}, V., {Castro-Carrizo}, A., {Alcolea}, J., \& {Neri}, R. 2005,
  \aap, 441, 1031

\bibitem[{{Bujarrabal} {et~al.}(2001){Bujarrabal}, {Castro-Carrizo}, {Alcolea},
  \& {S{\'a}nchez Contreras}}]{buja01}
{Bujarrabal}, V., {Castro-Carrizo}, A., {Alcolea}, J., \& {S{\'a}nchez
  Contreras}, C. 2001, A\&A, 377, 868

\bibitem[{{Bujarrabal} {et~al.}(2007){Bujarrabal}, {van Winckel}, {Neri},
  {Alcolea}, {Castro-Carrizo}, \& {Deroo}}]{buja07}
{Bujarrabal}, V., {van Winckel}, H., {Neri}, R., {Alcolea}, J.,
  {Castro-Carrizo}, A., \& {Deroo}, P. 2007, A\&A, 468, L45

\bibitem[{{{\c S}ahin} {et~al.}(2011){{\c S}ahin}, {Lambert}, {Klochkova}, \&
  {Tavolganskaya}}]{Sahin11}
{{\c S}ahin}, T., {Lambert}, D.~L., {Klochkova}, V.~G., \& {Tavolganskaya},
  N.~S. 2011, \mnras, 410, 612

\bibitem[{{Castro-Carrizo} {et~al.}(2002){Castro-Carrizo}, {Bujarrabal},
  {S{\'a}nchez Contreras}, {Alcolea}, \& {Neri}}]{Castro-Carrizo:2002ly}
{Castro-Carrizo}, A., {Bujarrabal}, V., {S{\'a}nchez Contreras}, C., {Alcolea},
  J., \& {Neri}, R. 2002, \aap, 386, 633

\bibitem[{{Castro-Carrizo} {et~al.}(2005){Castro-Carrizo}, {Bujarrabal},
  {S{\'a}nchez Contreras}, {Sahai}, \& {Alcolea}}]{cast05}
{Castro-Carrizo}, A., {Bujarrabal}, V., {S{\'a}nchez Contreras}, C., {Sahai},
  R., \& {Alcolea}, J. 2005, A\&A, 431, 979

\bibitem[{{Castro-Carrizo} {et~al.}(2004){Castro-Carrizo}, {Neri}, \&
  {Winters}}]{cast04}
{Castro-Carrizo}, A., {Neri}, R., \& {Winters}, J.~M. 2004, in Astronomical
  Society of the Pacific Conference Series, Vol. 313, Asymmetrical Planetary
  Nebulae III: Winds, Structure and the Thunderbird, ed. {M.~Meixner,
  J.~H.~Kastner, B.~Balick, \& N.~Soker}, 314

\bibitem[{{Castro-Carrizo} {et~al.}(2010){Castro-Carrizo}, {Quintana-Lacaci},
  {Neri}, {Bujarrabal}, {Sch{\"o}ier}, {Winters}, {Olofsson}, {Lindqvist},
  {Alcolea}, {Lucas}, \& {Grewing}}]{cast10}
{Castro-Carrizo}, A., {et~al.} 2010, A\&A, 523, A59

\bibitem[{{Cernicharo} {et~al.}(1989){Cernicharo}, {Guelin}, {Penalver},
  {Martin-Pintado}, \& {Mauersberger}}]{1989AA...222L...1C}
{Cernicharo}, J., {Guelin}, M., {Penalver}, J., {Martin-Pintado}, J., \&
  {Mauersberger}, R. 1989, A\&A, 222, L1

\bibitem[{{Cerrigone} {et~al.}(2009){Cerrigone}, {Hora}, {Umana}, \&
  {Trigilio}}]{Cerr09}
{Cerrigone}, L., {Hora}, J.~L., {Umana}, G., \& {Trigilio}, C. 2009, \apj, 703,
  585

\bibitem[{{Cox} {et~al.}(2000){Cox}, {Lucas}, {Huggins}, {Forveille},
  {Bachiller}, {Guilloteau}, {Maillard}, \& {Omont}}]{cox00}
{Cox}, P., {Lucas}, R., {Huggins}, P.~J., {Forveille}, T., {Bachiller}, R.,
  {Guilloteau}, S., {Maillard}, J.~P., \& {Omont}, A. 2000, A\&A, 353, L25

\bibitem[{{Dame} {et~al.}(2001){Dame}, {Hartmann}, \& {Thaddeus}}]{dame01}
{Dame}, T.~M., {Hartmann}, D., \& {Thaddeus}, P. 2001, ApJ, 547, 792

\bibitem[{{Dayal} \& {Bieging}(1996)}]{1996ApJ...472..703D}
{Dayal}, A., \& {Bieging}, J.~H. 1996, ApJ, 472, 703

\bibitem[{{de Ruyter} {et~al.}(2005){de Ruyter}, {van Winckel}, {Dominik},
  {Waters}, \& {Dejonghe}}]{de-Ruyter:2005dq}
{de Ruyter}, S., {van Winckel}, H., {Dominik}, C., {Waters}, L.~B.~F.~M., \&
  {Dejonghe}, H. 2005, \aap, 435, 161

\bibitem[{{de Ruyter} {et~al.}(2006){de Ruyter}, {van Winckel}, {Maas}, {Lloyd
  Evans}, {Waters}, \& {Dejonghe}}]{deru06}
{de Ruyter}, S., {van Winckel}, H., {Maas}, T., {Lloyd Evans}, T., {Waters},
  L.~B.~F.~M., \& {Dejonghe}, H. 2006, A\&A, 448, 641

\bibitem[{{Dougados} {et~al.}(1990){Dougados}, {Rouan}, {Lacombe}, {Tiphene},
  \& {Forveille}}]{Doug90}
{Dougados}, C., {Rouan}, D., {Lacombe}, F., {Tiphene}, D., \& {Forveille}, T.
  1990, A\&A, 227, 437

\bibitem[{{Eder} {et~al.}(1988){Eder}, {Lewis}, \& {Terzian}}]{Eder88}
{Eder}, J., {Lewis}, B.~M., \& {Terzian}, Y. 1988, \apjs, 66, 183

\bibitem[{{Evans} {et~al.}(1998){Evans}, {Eyres}, {Naylor}, \&
  {Salama}}]{1998AA...335..292E}
{Evans}, A., {Eyres}, S.~P.~S., {Naylor}, T., \& {Salama}, A. 1998, A\&A, 335,
  292

\bibitem[{{Forveille} {et~al.}(1987){Forveille}, {Morris}, {Omont}, \&
  {Likkel}}]{forv87}
{Forveille}, T., {Morris}, M., {Omont}, A., \& {Likkel}, L. 1987, A\&A, 176,
  L13

\bibitem[{{Gammie} {et~al.}(1989){Gammie}, {Knapp}, {Young}, {Phillips}, \&
  {Falgarone}}]{1989ApJ...345L..87G}
{Gammie}, C.~F., {Knapp}, G.~R., {Young}, K., {Phillips}, T.~G., \&
  {Falgarone}, E. 1989, ApJL, 345, L87

\bibitem[{{G{\'e}rard} \& {Le Bertre}(2003)}]{gera03}
{G{\'e}rard}, E., \& {Le Bertre}, T. 2003, A\&A, 397, L17

\bibitem[{{Gielen} {et~al.}(2011){Gielen}, {Bouwman}, {van Winckel}, {Lloyd
  Evans}, {Woods}, {Kemper}, {Marengo}, {Meixner}, {Sloan}, \&
  {Tielens}}]{Giel11}
{Gielen}, C., {et~al.} 2011, \aap, 533, A99

\bibitem[{{Giridhar} {et~al.}(2005){Giridhar}, {Lambert}, {Reddy}, {Gonzalez},
  \& {Yong}}]{Giri05}
{Giridhar}, S., {Lambert}, D.~L., {Reddy}, B.~E., {Gonzalez}, G., \& {Yong}, D.
  2005, \apj, 627, 432

\bibitem[{{Greaves} \& {Holland}(1997)}]{1997AA...327..342G}
{Greaves}, J.~S., \& {Holland}, W.~S. 1997, A\&A, 327, 342

\bibitem[{{Groenewegen} {et~al.}(2002){Groenewegen}, {Sevenster}, {Spoon}, \&
  {P{\'e}rez}}]{2002AA...390..501G}
{Groenewegen}, M.~A.~T., {Sevenster}, M., {Spoon}, H.~W.~W., \& {P{\'e}rez}, I.
  2002, A\&A, 390, 501

\bibitem[{{Hajian} {et~al.}(1996){Hajian}, {Phillips}, \&
  {Terzian}}]{1996ApJ...467..341H}
{Hajian}, A.~R., {Phillips}, J.~A., \& {Terzian}, Y. 1996, ApJ, 467, 341

\bibitem[{{He} {et~al.}(2008){He}, {Imai}, {Hasegawa}, {Campbell}, \&
  {Nakashima}}]{2008AA...488L..21H}
{He}, J.~H., {Imai}, H., {Hasegawa}, T.~I., {Campbell}, S.~W., \& {Nakashima},
  J. 2008, A\&A, 488, L21

\bibitem[{{Herpin} {et~al.}(2002){Herpin}, {Goicoechea}, {Pardo}, \&
  {Cernicharo}}]{2002ApJ...577..961H}
{Herpin}, F., {Goicoechea}, J.~R., {Pardo}, J.~R., \& {Cernicharo}, J. 2002,
  ApJ, 577, 961

\bibitem[{{Heske} {et~al.}(1990){Heske}, {Forveille}, {Omont}, {van der Veen},
  \& {Habing}}]{1990AA...239..173H}
{Heske}, A., {Forveille}, T., {Omont}, A., {van der Veen}, W.~E.~C.~J., \&
  {Habing}, H.~J. 1990, A\&A, 239, 173

\bibitem[{{Hodge} {et~al.}(2004){Hodge}, {Kraemer}, {Price}, \&
  {Walker}}]{Hodg04}
{Hodge}, T.~M., {Kraemer}, K.~E., {Price}, S.~D., \& {Walker}, H.~J. 2004,
  \apjs, 151, 299

\bibitem[{{Hony} {et~al.}(2002){Hony}, {Waters}, \& {Tielens}}]{Hony02}
{Hony}, S., {Waters}, L.~B.~F.~M., \& {Tielens}, A.~G.~G.~M. 2002, \aap, 390,
  533

\bibitem[{{Hrivnak} \& {Bieging}(2005)}]{2005ApJ...624..331H}
{Hrivnak}, B.~J., \& {Bieging}, J.~H. 2005, ApJ, 624, 331

\bibitem[{{Hrivnak} {et~al.}(2010){Hrivnak}, {Lu}, {Maupin}, \&
  {Spitzbart}}]{Hrivnak:2010pd}
{Hrivnak}, B.~J., {Lu}, W., {Maupin}, R.~E., \& {Spitzbart}, B.~D. 2010, \apj,
  709, 1042

\bibitem[{{Hu} {et~al.}(1993){Hu}, {Slijkhuis}, {Nguyen-Q-Rieu}, \& {de
  Jong}}]{1993AA...273..185H}
{Hu}, J.~Y., {Slijkhuis}, S., {Nguyen-Q-Rieu}, \& {de Jong}, T. 1993, A\&A,
  273, 185

\bibitem[{{Hu} {et~al.}(1994){Hu}, {te Lintel Hekkert}, {Slijkhuis}, {Baas},
  {Sahai}, \& {Wood}}]{1994AAS..103..301H}
{Hu}, J.~Y., {te Lintel Hekkert}, P., {Slijkhuis}, F., {Baas}, F., {Sahai}, R.,
  \& {Wood}, P.~R. 1994, A\&AS, 103, 301

\bibitem[{{Huggins} {et~al.}(1996){Huggins}, {Bachiller}, {Cox}, \&
  {Forveille}}]{hugg96}
{Huggins}, P.~J., {Bachiller}, R., {Cox}, P., \& {Forveille}, T. 1996, A\&A,
  315, 284

\bibitem[{{Huggins} {et~al.}(2005){Huggins}, {Bachiller}, {Planesas},
  {Forveille}, \& {Cox}}]{hugg05}
{Huggins}, P.~J., {Bachiller}, R., {Planesas}, P., {Forveille}, T., \& {Cox},
  P. 2005, ApJS, 160, 272

\bibitem[{{Huggins} \& {Healy}(1989)}]{1989ApJ...346..201H}
{Huggins}, P.~J., \& {Healy}, A.~P. 1989, ApJ, 346, 201

\bibitem[{{Huggins} {et~al.}(2004){Huggins}, {Muthu}, {Bachiller}, {Forveille},
  \& {Cox}}]{2004AA...414..581H}
{Huggins}, P.~J., {Muthu}, C., {Bachiller}, R., {Forveille}, T., \& {Cox}, P.
  2004, A\&A, 414, 581

\bibitem[{{Jorissen} \& {Knapp}(1998)}]{jori98}
{Jorissen}, A., \& {Knapp}, G.~R. 1998, A\&AS, 129, 363

\bibitem[{{Josselin} \& {L{\`e}bre}(2001)}]{2001AA...367..826J}
{Josselin}, E., \& {L{\`e}bre}, A. 2001, A\&A, 367, 826

\bibitem[{{Jura}(1986)}]{Jura:1986lq}
{Jura}, M. 1986, \apj, 309, 732

\bibitem[{{Jura} {et~al.}(1995){Jura}, {Balm}, \&
  {Kahane}}]{1995ApJ...453..721J}
{Jura}, M., {Balm}, S.~P., \& {Kahane}, C. 1995, ApJ, 453, 721

\bibitem[{{Kahane} {et~al.}(1992){Kahane}, {Cernicharo}, {Gomez-Gonzalez}, \&
  {Guelin}}]{1992AA...256..235K}
{Kahane}, C., {Cernicharo}, J., {Gomez-Gonzalez}, J., \& {Guelin}, M. 1992,
  A\&A, 256, 235

\bibitem[{{Kastner}(1992)}]{kast92}
{Kastner}, J.~H. 1992, ApJ, 401, 337

\bibitem[{{Kawabe} {et~al.}(1987){Kawabe}, {Ishiguro}, {Kasuga}, {Morita},
  {Ukita}, {Kobayashi}, {Okumura}, {Fomalont}, \& {Kaifu}}]{kawa87}
{Kawabe}, R., {et~al.} 1987, ApJ, 314, 322

\bibitem[{{Kerschbaum} \& {Olofsson}(1999)}]{kers99}
{Kerschbaum}, F., \& {Olofsson}, H. 1999, A\&AS, 138, 299

\bibitem[{{Knapp} {et~al.}(1982){Knapp}, {Phillips}, {Leighton}, {Lo},
  {Wannier}, {Wootten}, \& {Huggins}}]{1982ApJ...252..616K}
{Knapp}, G.~R., {Phillips}, T.~G., {Leighton}, R.~B., {Lo}, K.~Y., {Wannier},
  P.~G., {Wootten}, H.~A., \& {Huggins}, P.~J. 1982, ApJ, 252, 616

\bibitem[{{Knapp} {et~al.}(1998){Knapp}, {Young}, {Lee}, \&
  {Jorissen}}]{1998ApJS..117..209K}
{Knapp}, G.~R., {Young}, K., {Lee}, E., \& {Jorissen}, A. 1998, ApJS, 117, 209

\bibitem[{{Knapp} {et~al.}(1989){Knapp}, {Sutin}, {Phillips}, {Ellison},
  {Keene}, {Leighton}, {Masson}, {Steiger}, {Veidt}, \&
  {Young}}]{1989ApJ...336..822K}
{Knapp}, G.~R., {et~al.} 1989, ApJ, 336, 822

\bibitem[{{Kwok}(1993)}]{kowk93}
{Kwok}, S. 1993, ARA\&A, 31, 63

\bibitem[{{Kwok} {et~al.}(1987){Kwok}, {Boreiko}, \& {Hrivnak}}]{Kwok87}
{Kwok}, S., {Boreiko}, R.~T., \& {Hrivnak}, B.~J. 1987, \apj, 312, 303

\bibitem[{{Kwok} {et~al.}(1989){Kwok}, {Volk}, \& {Hrivnak}}]{Kwok:1989ul}
{Kwok}, S., {Volk}, K.~M., \& {Hrivnak}, B.~J. 1989, \apjl, 345, L51

\bibitem[{{Libert} {et~al.}(2010){Libert}, {Winters}, {Le Bertre},
  {G{\'e}rard}, \& {Matthews}}]{libe10}
{Libert}, Y., {Winters}, J.~M., {Le Bertre}, T., {G{\'e}rard}, E., \&
  {Matthews}, L.~D. 2010, A\&A, 515, A112

\bibitem[{{Likkel}(1989)}]{Lik89}
{Likkel}, L. 1989, \apj, 344, 350

\bibitem[{{Likkel} {et~al.}(1991){Likkel}, {Forveille}, {Omont}, \&
  {Morris}}]{1991AA...246..153L}
{Likkel}, L., {Forveille}, T., {Omont}, A., \& {Morris}, M. 1991, A\&A, 246,
  153

\bibitem[{{Likkel} \& {Morris}(1988)}]{Lik88}
{Likkel}, L., \& {Morris}, M. 1988, \apj, 329, 914

\bibitem[{{Loup} {et~al.}(1993){Loup}, {Forveille}, {Omont}, \&
  {Paul}}]{loup93}
{Loup}, C., {Forveille}, T., {Omont}, A., \& {Paul}, J.~F. 1993, A\&AS, 99, 291

\bibitem[{{Maas} {et~al.}(2003){Maas}, {Van Winckel}, {Lloyd Evans}, {Nyman},
  {Kilkenny}, {Martinez}, {Marang}, \& {van Wyk}}]{2003AA...405..271M}
{Maas}, T., {Van Winckel}, H., {Lloyd Evans}, T., {Nyman}, L.-{\AA}.,
  {Kilkenny}, D., {Martinez}, P., {Marang}, F., \& {van Wyk}, F. 2003, A\&A,
  405, 271

\bibitem[{{Mamon} {et~al.}(1988){Mamon}, {Glassgold}, \& {Huggins}}]{mamo88}
{Mamon}, G.~A., {Glassgold}, A.~E., \& {Huggins}, P.~J. 1988, ApJ, 328, 797

\bibitem[{{Manteiga} {et~al.}(2011){Manteiga}, {Garc{\'{\i}}a-Hern{\'a}ndez},
  {Ulla}, {Manchado}, \& {Garc{\'{\i}}a-Lario}}]{Mant11}
{Manteiga}, M., {Garc{\'{\i}}a-Hern{\'a}ndez}, D.~A., {Ulla}, A., {Manchado},
  A., \& {Garc{\'{\i}}a-Lario}, P. 2011, \aj, 141, 80

\bibitem[{{Meixner} {et~al.}(1998){Meixner}, {Campbell}, {Welch}, \&
  {Likkel}}]{1998ApJ...509..392M}
{Meixner}, M., {Campbell}, M.~T., {Welch}, W.~J., \& {Likkel}, L. 1998, ApJ,
  509, 392

\bibitem[{{Men'shchikov} {et~al.}(2002){Men'shchikov}, {Schertl}, {Tuthill},
  {Weigelt}, \& {Yungelson}}]{mens02}
{Men'shchikov}, A.~B., {Schertl}, D., {Tuthill}, P.~G., {Weigelt}, G., \&
  {Yungelson}, L.~R. 2002, A\&A, 393, 867

\bibitem[{{Milam} {et~al.}(2009){Milam}, {Woolf}, \&
  {Ziurys}}]{2009ApJ...690..837M}
{Milam}, S.~N., {Woolf}, N.~J., \& {Ziurys}, L.~M. 2009, ApJ, 690, 837

\bibitem[{{Morris} \& {Bowers}(1980)}]{Morr80}
{Morris}, M., \& {Bowers}, P.~F. 1980, \aj, 85, 724

\bibitem[{{Murakawa} {et~al.}(2008){Murakawa}, {Ohnaka}, {Driebe}, {Hofmann},
  {Oya}, {Schertl}, \& {Weigelt}}]{mura08}
{Murakawa}, K., {Ohnaka}, K., {Driebe}, T., {Hofmann}, K., {Oya}, S.,
  {Schertl}, D., \& {Weigelt}, G. 2008, A\&A, 489, 195

\bibitem[{{Neri} {et~al.}(1998){Neri}, {Kahane}, {Lucas}, {Bujarrabal}, \&
  {Loup}}]{1998AAS..130....1N}
{Neri}, R., {Kahane}, C., {Lucas}, R., {Bujarrabal}, V., \& {Loup}, C. 1998,
  A\&AS, 130, 1

\bibitem[{{Nyman} {et~al.}(1992){Nyman}, {Booth}, {Carlstrom}, {Habing},
  {Heske}, {Sahai}, {Stark}, {van der Veen}, \& {Winnberg}}]{nyma92}
{Nyman}, L.-{\AA}., {et~al.} 1992, A\&AS, 93, 121

\bibitem[{{Olofsson} \& {Nyman}(1999)}]{1999AA...347..194O}
{Olofsson}, H., \& {Nyman}, L.-{\AA}. 1999, A\&A, 347, 194

\bibitem[{{Omont} {et~al.}(1993){Omont}, {Loup}, {Forveille}, {te Lintel
  Hekkert}, {Habing}, \& {Sivagnanam}}]{1993AA...267..515O}
{Omont}, A., {Loup}, C., {Forveille}, T., {te Lintel Hekkert}, P., {Habing},
  H., \& {Sivagnanam}, P. 1993, A\&A, 267, 515

\bibitem[{{Palla} {et~al.}(2000){Palla}, {Bachiller}, {Stanghellini}, {Tosi},
  \& {Galli}}]{2000AA...355...69P}
{Palla}, F., {Bachiller}, R., {Stanghellini}, L., {Tosi}, M., \& {Galli}, D.
  2000, A\&A, 355, 69

\bibitem[{{Papoular}(2008)}]{Papo08}
{Papoular}, R. 2008, MNRAS, 390, 1727

\bibitem[{{Parthasarathy} {et~al.}(2001){Parthasarathy}, {Garc{\'{\i}}a-Lario},
  {Gauba}, {de Martino}, {Nakada}, {Fujii}, {Pottasch}, \& {San Fern{\'a}ndez
  de C{\'o}rdoba}}]{Part01}
{Parthasarathy}, M., {Garc{\'{\i}}a-Lario}, P., {Gauba}, G., {de Martino}, D.,
  {Nakada}, Y., {Fujii}, T., {Pottasch}, S.~R., \& {San Fern{\'a}ndez de
  C{\'o}rdoba}, L. 2001, \aap, 376, 941

\bibitem[{{Peeters} {et~al.}(2002){Peeters}, {Hony}, {Van Kerckhoven},
  {Tielens}, {Allamandola}, {Hudgins}, \& {Bauschlicher}}]{Pee02}
{Peeters}, E., {Hony}, S., {Van Kerckhoven}, C., {Tielens}, A.~G.~G.~M.,
  {Allamandola}, L.~J., {Hudgins}, D.~M., \& {Bauschlicher}, C.~W. 2002, \aap,
  390, 1089

\bibitem[{{Pereira} {et~al.}(2004){Pereira}, {Lorenz-Martins}, \&
  {Machado}}]{Pere04}
{Pereira}, C.~B., {Lorenz-Martins}, S., \& {Machado}, M. 2004, \aap, 422, 637

\bibitem[{{Pereira} \& {Miranda}(2007)}]{Pere07}
{Pereira}, C.~B., \& {Miranda}, L.~F. 2007, \aap, 462, 231

\bibitem[{{Piovan} {et~al.}(2003){Piovan}, {Tantalo}, \& {Chiosi}}]{Piovan03}
{Piovan}, L., {Tantalo}, R., \& {Chiosi}, C. 2003, \aap, 408, 559

\bibitem[{{Preston} {et~al.}(1963){Preston}, {Krzeminski}, {Smak}, \&
  {Williams}}]{Preston:1963bh}
{Preston}, G.~W., {Krzeminski}, W., {Smak}, J., \& {Williams}, J.~A. 1963,
  \apj, 137, 401

\bibitem[{{Ramstedt} {et~al.}(2009){Ramstedt}, {Sch{\"o}ier}, \&
  {Olofsson}}]{rams09}
{Ramstedt}, S., {Sch{\"o}ier}, F.~L., \& {Olofsson}, H. 2009, A\&A, 499, 515

\bibitem[{{Rao} {et~al.}(2012){Rao}, {Giridhar}, \& {Lambert}}]{Rao12}
{Rao}, S.~S., {Giridhar}, S., \& {Lambert}, D.~L. 2012, \mnras, 419, 1254

\bibitem[{{Roberts}(1962)}]{robe62}
{Roberts}, M.~S. 1962, \aj, 67, 79

\bibitem[{{Roddier} {et~al.}(1995){Roddier}, {Roddier}, {Graves}, \&
  {Northcott}}]{rodd95}
{Roddier}, F., {Roddier}, C., {Graves}, J.~E., \& {Northcott}, M.~J. 1995, ApJ,
  443, 249

\bibitem[{{Ryans} {et~al.}(2003){Ryans}, {Dufton}, {Mooney}, {Rolleston},
  {Keenan}, {Hubeny}, \& {Lanz}}]{Ryans:2003eu}
{Ryans}, R.~S.~I., {Dufton}, P.~L., {Mooney}, C.~J., {Rolleston}, W.~R.~J.,
  {Keenan}, F.~P., {Hubeny}, I., \& {Lanz}, T. 2003, \aap, 401, 1119

\bibitem[{{Sahai} {et~al.}(2000){Sahai}, {Bujarrabal}, {Castro-Carrizo}, \&
  {Zijlstra}}]{saha00}
{Sahai}, R., {Bujarrabal}, V., {Castro-Carrizo}, A., \& {Zijlstra}, A. 2000,
  A\&A, 360, L9

\bibitem[{{Sahai} \& {Nyman}(1997)}]{1997ApJ...487L.155S}
{Sahai}, R., \& {Nyman}, L.-{\AA}. 1997, ApJL, 487, L155

\bibitem[{{S{\'a}nchez Contreras} {et~al.}(1997){S{\'a}nchez Contreras},
  {Bujarrabal}, \& {Alcolea}}]{1997AA...327..689S}
{S{\'a}nchez Contreras}, C., {Bujarrabal}, V., \& {Alcolea}, J. 1997, A\&A,
  327, 689

\bibitem[{{S{\'a}nchez Contreras} {et~al.}(2004){S{\'a}nchez Contreras},
  {Bujarrabal}, {Castro-Carrizo}, {Alcolea}, \& {Sargent}}]{sanc04}
{S{\'a}nchez Contreras}, C., {Bujarrabal}, V., {Castro-Carrizo}, A., {Alcolea},
  J., \& {Sargent}, A. 2004, ApJ, 617, 1142

\bibitem[{{S{\'a}nchez Contreras} {et~al.}(2010){S{\'a}nchez Contreras},
  {Cortijo-Ferrero}, {Miranda}, {Castro-Carrizo}, \&
  {Bujarrabal}}]{Sanchez-Contreras:2010gf}
{S{\'a}nchez Contreras}, C., {Cortijo-Ferrero}, C., {Miranda}, L.~F.,
  {Castro-Carrizo}, A., \& {Bujarrabal}, V. 2010, \apj, 715, 143

\bibitem[{{S{\'a}nchez Contreras} {et~al.}(2008){S{\'a}nchez Contreras},
  {Sahai}, {Gil de Paz}, \& {Goodrich}}]{Sanc08}
{S{\'a}nchez Contreras}, C., {Sahai}, R., {Gil de Paz}, A., \& {Goodrich}, R.
  2008, \apjs, 179, 166

\bibitem[{{Seaquist} {et~al.}(1991){Seaquist}, {Plume}, \& {Davis}}]{Seaq91}
{Seaquist}, E.~R., {Plume}, R., \& {Davis}, L.~E. 1991, \apj, 367, 200

\bibitem[{{Sevenster}(2002)}]{Sev02}
{Sevenster}, M.~N. 2002, \aj, 123, 2772

\bibitem[{{Silva} {et~al.}(1993){Silva}, {Azcarate}, {Poppel}, \&
  {Likkel}}]{Silva93}
{Silva}, A.~M., {Azcarate}, I.~N., {Poppel}, W.~G.~L., \& {Likkel}, L. 1993,
  \aap, 275, 510

\bibitem[{{Slijkhuis} {et~al.}(1991){Slijkhuis}, {de Jong}, \& {Hu}}]{slij91}
{Slijkhuis}, S., {de Jong}, T., \& {Hu}, J.~Y. 1991, A\&A, 248, 547

\bibitem[{{Stasi{\'n}ska} {et~al.}(2006){Stasi{\'n}ska}, {Szczerba}, {Schmidt},
  \& {Si{\'o}dmiak}}]{SSSS}
{Stasi{\'n}ska}, G., {Szczerba}, R., {Schmidt}, M., \& {Si{\'o}dmiak}, N. 2006,
  \aap, 450, 701

\bibitem[{{Steffen} {et~al.}(1998){Steffen}, {Szczerba}, \&
  {Schoenberner}}]{stef98}
{Steffen}, M., {Szczerba}, R., \& {Schoenberner}, D. 1998, A\&A, 337, 149

\bibitem[{{Stephenson}(1989)}]{CGCS}
{Stephenson}, C.~B. 1989, Publications of the Warner \& Swasey Observatory, 3,
  53

\bibitem[{{Su{\'a}rez} {et~al.}(2006){Su{\'a}rez}, {Garc{\'{\i}}a-Lario},
  {Manchado}, {Manteiga}, {Ulla}, \& {Pottasch}}]{Suar06}
{Su{\'a}rez}, O., {Garc{\'{\i}}a-Lario}, P., {Manchado}, A., {Manteiga}, M.,
  {Ulla}, A., \& {Pottasch}, S.~R. 2006, \aap, 458, 173

\bibitem[{{Szczerba} \& {Marten}(1993)}]{Szczerba:1993fk}
{Szczerba}, R., \& {Marten}, H. 1993, in European Southern Observatory
  Conference and Workshop Proceedings, Vol.~46, European Southern Observatory
  Conference and Workshop Proceedings, ed. H.~E. {Schwarz}, 90

\bibitem[{{Szczerba} {et~al.}(2007){Szczerba}, {Si{\'o}dmiak}, {Stasi{\'n}ska},
  \& {Borkowski}}]{szcz07}
{Szczerba}, R., {Si{\'o}dmiak}, N., {Stasi{\'n}ska}, G., \& {Borkowski}, J.
  2007, A\&A, 469, 799

\bibitem[{{Szczerba} {et~al.}(2012){Szczerba}, {Si{\'o}dmiak}, {Stasi{\'n}ska},
  {Borkowski}, {Garc{\'{\i}}a-Lario}, {Su{\'a}rez}, {Hajduk}, \&
  {Garc{\'{\i}}a-Hern{\'a}ndez}}]{szcz11}
{Szczerba}, R., {Si{\'o}dmiak}, N., {Stasi{\'n}ska}, G., {Borkowski}, J.,
  {Garc{\'{\i}}a-Lario}, P., {Su{\'a}rez}, O., {Hajduk}, M., \&
  {Garc{\'{\i}}a-Hern{\'a}ndez}, D.~A. 2012, in IAU Symposium, Vol. 283, IAU
  Symposium, 506--507

\bibitem[{{te Lintel Hekkert}(1991)}]{teLin91}
{te Lintel Hekkert}, P. 1991, \aap, 248, 209

\bibitem[{{te Lintel Hekkert} \& {Chapman}(1996)}]{teLin96}
{te Lintel Hekkert}, P., \& {Chapman}, J.~M. 1996, \aaps, 119, 459

\bibitem[{{te Lintel Hekkert} {et~al.}(1992){te Lintel Hekkert}, {Chapman}, \&
  {Zijlstra}}]{teLin92}
{te Lintel Hekkert}, P.~T.~L., {Chapman}, J.~M., \& {Zijlstra}, A.~A. 1992,
  \apjl, 390, L23

\bibitem[{{van der Veen} \& {Habing}(1988)}]{vand88}
{van der Veen}, W.~E.~C.~J., \& {Habing}, H.~J. 1988, A\&A, 194, 125

\bibitem[{{van der Veen} {et~al.}(1989){van der Veen}, {Habing}, \&
  {Geballe}}]{vand89}
{van der Veen}, W.~E.~C.~J., {Habing}, H.~J., \& {Geballe}, T.~R. 1989, A\&A,
  226, 108

\bibitem[{{van der Veen} {et~al.}(1993){van der Veen}, {Trams}, \&
  {Waters}}]{1993AA...269..231V}
{van der Veen}, W.~E.~C.~J., {Trams}, N.~R., \& {Waters}, L.~B.~F.~M. 1993,
  A\&A, 269, 231

\bibitem[{{van Winckel}(2003)}]{van-Winckel:2003rt}
{van Winckel}, H. 2003, \araa, 41, 391

\bibitem[{{van Winckel} {et~al.}(2009){van Winckel}, {Lloyd Evans}, {Briquet},
  {De Cat}, {Degroote}, {De Meester}, {De Ridder}, \& et~al.}]{vanw09}
{van Winckel}, H., {Lloyd Evans}, T., {Briquet}, M., {De Cat}, P., {Degroote},
  P., {De Meester}, W., {De Ridder}, J., \& et~al. 2009, A\&A, 505, 1221

\bibitem[{{Van Winckel} \& {Reyniers}(2000)}]{Van-Winckel:2000qf}
{Van Winckel}, H., \& {Reyniers}, M. 2000, \aap, 354, 135

\bibitem[{{Vandenbussche} {et~al.}(2002){Vandenbussche}, {Beintema}, {de
  Graauw}, {Decin}, {Feuchtgruber}, {Heras}, {Kester}, {Lahuis}, {Lenorzer},
  {Lorente}, {Salama}, {Waelkens}, {Waters}, \& {Wieprecht}}]{vand02}
{Vandenbussche}, B., {et~al.} 2002, \aap, 390, 1033

\bibitem[{{Volk} \& {Kwok}(1987)}]{Volk87}
{Volk}, K., \& {Kwok}, S. 1987, \apj, 315, 654

\bibitem[{{Volk} {et~al.}(1993){Volk}, {Kwok}, \&
  {Woodsworth}}]{1993ApJ...402..292V}
{Volk}, K., {Kwok}, S., \& {Woodsworth}, A.~W. 1993, ApJ, 402, 292

\bibitem[{{Wannier} {et~al.}(1979){Wannier}, {Leighton}, {Knapp}, {Redman},
  {Phillips}, \& {Huggins}}]{1979ApJ...230..149W}
{Wannier}, P.~G., {Leighton}, R.~B., {Knapp}, G.~R., {Redman}, R.~O.,
  {Phillips}, T.~G., \& {Huggins}, P.~J. 1979, ApJ, 230, 149

\bibitem[{{Wannier} \& {Sahai}(1987)}]{1987ApJ...319..367W}
{Wannier}, P.~G., \& {Sahai}, R. 1987, ApJ, 319, 367

\bibitem[{{Wannier} {et~al.}(1990){Wannier}, {Sahai}, {Andersson}, \&
  {Johnson}}]{1990ApJ...358..251W}
{Wannier}, P.~G., {Sahai}, R., {Andersson}, B.-G., \& {Johnson}, H.~R. 1990,
  ApJ, 358, 251

\bibitem[{{Waters} {et~al.}(1998){Waters}, {Cami}, {de Jong}, {Molster}, {van
  Loon}, {Bouwman}, {de Koter}, {Waelkens}, {Van Winckel}, {Morris}, {Cami},
  {Bouwman}, {de Koter}, {de Jong}, \& {de Graauw}}]{wat98}
{Waters}, L.~B.~F.~M., {et~al.} 1998, \nat, 391, 868

\bibitem[{{Wetzel} {et~al.}(2013){Wetzel}, {Klevenz}, {Gail}, {Pucci}, \&
  {Trieloff}}]{Wetz12}
{Wetzel}, S., {Klevenz}, M., {Gail}, H.-P., {Pucci}, A., \& {Trieloff}, M.
  2013, \aap, 553, A92

\bibitem[{{Winters} {et~al.}(2003){Winters}, {Le Bertre}, {Jeong}, {Nyman}, \&
  {Epchtein}}]{wint03}
{Winters}, J.~M., {Le Bertre}, T., {Jeong}, K.~S., {Nyman}, L., \& {Epchtein},
  N. 2003, A\&A, 409, 715

\bibitem[{{Winters} {et~al.}(2007){Winters}, {Le Bertre}, {Pety}, \&
  {Neri}}]{wint07}
{Winters}, J.~M., {Le Bertre}, T., {Pety}, J., \& {Neri}, R. 2007, A\&A, 475,
  559

\bibitem[{{Witt} {et~al.}(2009){Witt}, {Vijh}, {Hobbs}, {Aufdenberg},
  {Thorburn}, \& {York}}]{witt09}
{Witt}, A.~N., {Vijh}, U.~P., {Hobbs}, L.~M., {Aufdenberg}, J.~P., {Thorburn},
  J.~A., \& {York}, D.~G. 2009, ApJ, 693, 1946

\bibitem[{{Woods} {et~al.}(2005){Woods}, {Nyman}, {Sch{\"o}ier}, {Zijlstra},
  {Millar}, \& {Olofsson}}]{2005AA...429..977W}
{Woods}, P.~M., {Nyman}, L., {Sch{\"o}ier}, F.~L., {Zijlstra}, A.~A., {Millar},
  T.~J., \& {Olofsson}, H. 2005, A\&A, 429, 977

\bibitem[{{Woodsworth} {et~al.}(1990){Woodsworth}, {Kwok}, \&
  {Chan}}]{1990AA...228..503W}
{Woodsworth}, A.~W., {Kwok}, S., \& {Chan}, S.~J. 1990, A\&A, 228, 503

\bibitem[{{Zijlstra}(2007)}]{zijl07}
{Zijlstra}, A.~A. 2007, Baltic Astronomy, 16, 79

\bibitem[{{Zuckerman} \& {Dyck}(1986{\natexlab{a}})}]{1986ApJ...304..394Z}
{Zuckerman}, B., \& {Dyck}, H.~M. 1986{\natexlab{a}}, ApJ, 304, 394

\bibitem[{{Zuckerman} \& {Dyck}(1986{\natexlab{b}})}]{zuck86}
---. 1986{\natexlab{b}}, ApJ, 311, 345

\bibitem[{{Zuckerman} \& {Dyck}(1989)}]{1989AA...209..119Z}
---. 1989, A\&A, 209, 119

\end{thebibliography}


\end{document}